\documentclass[iop]{emulateapj}
\usepackage{epsfig}
\usepackage{graphicx,natbib}

\def\ltsima{$\; \buildrel < \over \sim \;$}
\def\lsim{\lower.5ex\hbox{\ltsima}}
\def\gtsima{$\; \buildrel > \over \sim \;$}
\def\gsim{\lower.5ex\hbox{\gtsima}}

\shorttitle{A Critical Study of the Stellar Mass of Galaxies}
\begin{document}

%% LaTeX will automatically break titles if they run longer than
%% one line. However, you may use \\ to force a line break if
%% you desire.

%%\title{CANDELS Investigation of Photometric Redshift Methods}
\title{A Critical Assessment of Stellar Mass Measurement Methods}
%% Use \author, \affil, and the \and command to format
%% author and affiliation information.
%% Note that \email has replaced the old \authoremail command
%% from AASTeX v4.0. You can use \email to mark an email address
%% anywhere in the paper, not just in the front matter.
%% As in the title, you can use \\ to force line breaks.

\author{
Bahram Mobasher \altaffilmark{1},
Tomas Dahlen \altaffilmark{2},
Henry C.~Ferguson \altaffilmark{2},
Viviana Acquaviva \altaffilmark{3},
Guillermo Barro \altaffilmark{4},
Steven L.~Finkelstein \altaffilmark{5},
Adriano Fontana \altaffilmark{6},
Ruth Gruetzbauch\altaffilmark{7},
Seth Johnson\altaffilmark{8},
Yu Lu \altaffilmark{9},
Casey J.~Papovich \altaffilmark{10},
Janine Pforr\altaffilmark{11},
Mara Salvato\altaffilmark{12},
Rachel S.~Somerville \altaffilmark{13},
Tommy Wiklind \altaffilmark{14},
Stijn Wuyts \altaffilmark{12},
Matthew L. N. Ashby \altaffilmark{15},
Eric Bell \altaffilmark{16},
Christopher J.~Conselice \altaffilmark{17},
Mark E. Dickinson \altaffilmark{11},
Sandra ~M.~Faber \altaffilmark{4},
Giovanni Fazio \altaffilmark{15},
Kristian Finlator\altaffilmark{18},
Audrey Galametz\altaffilmark{6},
Eric Gawiser \altaffilmark{13},
Mauro Giavalisco\altaffilmark{8},
Andrea Grazian \altaffilmark{6},
Norman A.~Grogin \altaffilmark{2},
Yicheng Guo \altaffilmark{4},
Nimish Hathi \altaffilmark{19},
Dale Kocevski\altaffilmark{20},
Anton M.~Koekemoer \altaffilmark{2},
David C. Koo \altaffilmark{4},
Jeffrey A.~Newman \altaffilmark{21},
Naveen Reddy \altaffilmark{1},
Paola Santini\altaffilmark{6},
Risa H.~Wechsler \altaffilmark{9}
%Tomas Dahlen\altaffilmark{1}, 
%Bahram Mobasher\altaffilmark{2},
%Sandra M. Faber\altaffilmark{3},
%Henry C Ferguson\altaffilmark{1}
}
%\email{dahlen@stsci.edu}

%\altaffiltext{1}{Space Telescope Science Institute, 3700 San Martin Drive, Baltimore, MD 21218}
%\altaffiltext{2}{Department of Physics and Astronomy, University of California, Riverside, CA 92521}

\begin{abstract}
This is the second paper in a series aimed at investigating the main sources of uncertainty in measuring the observable parameters in galaxies
from their Spectral Energy Distributions (SEDs). In the first paper (\citealt{dahlen}) we presented a detailed account of the photometric redshift measurements and an error analysis of this process.   
In this paper we perform a comprehensive study of the main sources of random and systematic error in stellar mass estimates for galaxies, and their relative contributions to the associated error budget. 
Since there is no prior knowledge of the stellar mass of galaxies 
(unlike their photometric redshifts), we use mock galaxy catalogs with simulated multi-waveband photometry and known redshift, stellar mass, age and extinction for individual galaxies. The multi-waveband photometry for the simulated galaxies were generated 
in 13 filters spanning from U-band to mid-infrared wavelengths.   
 Given different parameters affecting stellar mass measurement (photometric S/N ratios, SED fitting errors and systematic effects), the inherent degeneracies and correlated errors, we formulated 
different simulated galaxy catalogs to quantify these effects individually. 
 For comparison, we also generated catalogs based on observed photometric data of real galaxies in the GOODS-South field, spanning the same passbands. The simulated and observed 
 catalogs were provided to a number of 
teams within the Cosmic Assembly Near-infrared Deep Extragalactic Legacy 
Survey (CANDELS) collaboration to estimate the stellar masses for individual galaxies. A total of eleven teams participated, with   
different combinations of stellar mass measurement codes/methods, population synthesis models, star formation histories, extinction and age. 

For each simulated galaxy, the differences between the input stellar masses, $M_{input}$, and those
estimated by each team, $M_{est}$, is defined as
$\Delta\log(M) \equiv \log(M_{\rm estimated}) - \log(M_{\rm input})$, and used to identify the most fundamental parameters affecting stellar mass estimate in galaxies, with the following results:
 (1). no significant bias in $\Delta log(M)$ was found among different codes, with all having comparable scatter   
($\sigma (\Delta log(M))= 0.136$\,dex). The estimated stellar mass values are
seriously affected by low photometric $S/N$ ratios, with the {\it rms}
scatter increasing for galaxies with $H_{AB} > 26$ mag.; (2). A source of error contributing to the scatter in $\Delta log(M)$ is found to be due to photometric uncertainties ($0.136$\,dex) and low resolution in age and
extinction grids when generating the SED templates;(3). The median of stellar masses among different methods provides a stable measure of the 
mass associated with any given galaxy ($\sigma (\Delta log(M))= 0.142$\,dex); (4). The $\Delta log(M)$ values are
 strongly correlate with deviations in age (defined as the difference between the estimated and expected values), with a weaker correlation with extinction; (5). the {\it rms} scatter in the estimated stellar masses due to free parameters (after fixing redshifts and IMF) are quantified and found to be  $\sigma (\Delta log(M))= 0.110$\,dex; (6). Using the observed data, we studied the sensitivity of stellar masses to both the population synthesis codes and inclusion of nebular emission lines and found them to affect the stellar mass by $0.2$\,dex and $0.3$\,dex respectively. 

\end{abstract}

\altaffiltext{1}{Department of Physics and Astronomy, University of California, Riverside, CA 92521}
\altaffiltext{2}{Space Telescope Science Institute, 3700 San Martin Drive, Baltimore, MD 21218}
\altaffiltext{3}{Physics Department, CUNY NYC College of Technology, 300 Jay Street, Brooklyn, NY 11201}
\altaffiltext{4}{UCO/Lick Observatory, Department of Astronomy and Astrophysics, University of California, Santa Cruz, CA 95064}
\altaffiltext{5}{Department of Astronomy, The University of Texas at Austin, Austin, TX 78712}
\altaffiltext{6}{INAF, Osservatorio Astronomico di Roma, Via Frascati 33, I00040, Monteporzio, Italy}
\altaffiltext{7}{Center for Astronomy and Astrophysics, Observatorio Astronomico de Lisboa, Tapada da Ajuda, 1349-018, Lisboa, Portugal}
\altaffiltext{8}{Department of Astronomy, University of Massachusetts, 710 North Plesant Street, Amherst, MA 01003}
\altaffiltext{9}{Kavli Institute for Particle Physics and Cosmology, Stanford University, Stanford, CA 94305}
\altaffiltext{10}{Department of Physics and astronomy, Texas A\&M Research Foundation, College Station, TX 77843}
\altaffiltext{11}{National Optical Astronomy Observatories, 950 N Cherry Avenue, Tucson, AZ 85719}
\altaffiltext{12}{Max-Planck-Institut f\"ur extraterrestrische Physik, Giessenbachstrasse 1, D-85748 Garching bei M\"unchen, Germany}
\altaffiltext{13}{Department of Physics and Astronomy, Rutgers, The State University of New Jersey, 136 Frelinghuysen Road, Piscataway, NJ 08854}
\altaffiltext{14}{Joint ALMA Observatory, Alonso de Cordova 3107, Vitacura, Santiago, Chile}
\altaffiltext{15}{Harvard Smithsonian Center for Astrophysics, 60 Garden Street, MS-66, Cambridge, MA 02138-1516}
\altaffiltext{16}{Department of Astronomy, University of Michigan, 500 Church Street, Ann Arbor, MI 48109, USA}
\altaffiltext{17}{School of Physics and Astronomy, University of Nottingham, Nottingham, UK}
\altaffiltext{18}{Dark Cosmology Center, Niels Bohr Institute, University of Copenhagen, Denmark}
\altaffiltext{19}{LAM - Laboratoire d'Astrophysique de Marseille, 38, Rue R. Jollot-Curie, 13388 Marseille, Cedex 13, France}
\altaffiltext{20}{Department of Physics and Astronomy, University of Kentucky, Lexington, KY 40506}
\altaffiltext{21}{Department of Physics and Astronomy, University of Pittsburgh, Pittsburgh, PA 15260}
\altaffiltext{15}{Department of Astronomy, University of Michigan, 500 Chruch St., Ann Arbor, MI 48109}

%\email{mobasher@ucr.edu}

\section{\bf Introduction}
The questions of what governs the observed properties of galaxies, the reason
behind the correlations among these properties and how they change with
look-back time, are among the most fundamental in observational astronomy 
today.  This requires accurate measurement of redshifts as well as the rest-frame observables. In particular, detailed knowledge of the statistical properties of galaxies (i.e. luminosity and mass functions) at different redshifts is essential to constrain current hierarchical models for the formation of galaxies. This requires large and deep surveys with multi-waveband photometry, photometric redshifts and stellar mass estimates. 

The installation of wide field-of-view detectors with high optical and infrared quantum efficiency on space and ground-based observatories has now allowed construction of multi-waveband, large and deep galaxy surveys. These surveys occupy a large portion of the Area-Depth parameter space, from the very deep Hubble Ultra Deep Field (HUDF; \citealt{beckwith6}), designed for studies of very high redshift galaxies, to wide-area Cosmic Evolution Survey-COSMOS (\citealt{scoville}) formulated to study the large scale structure in the Universe and its evolution with redshift. These are complemented by the intermediate surveys (in terms of depth and area) such as the Great Observatories Origins Deep Survey (GOODS; \citealt{giavalisco}), designed specifically for studies of the evolution of galaxies to high redshifts. Recently, the wavelength range of these surveys has been extended to near-infrared bands in a Multi-cycle Treasury program, the Cosmic Assembly Near-infrared Deep Extragalactic Legacy Survey- CANDELS (\citealt{koekemoer}; \citealt{Grogin}). One important addition to these observations is the availability of deep mid-infrared data (3.6-8.0 $\mu$m) from the {\it Spitzer Space Telescope}, extending the observed wavelength range to 8 $\mu$m (\citealt{ashby}). This is essential in constraining the Spectral Energy Distributions (SEDs) of galaxies and in estimating accurate photometric redshifts and stellar masses spanning a range of redshifts.

The multi-waveband data from these surveys are extensively used to study the luminosity function and mass function of galaxies to very high redshifts, with often divergent results (\citealt{Ouchi}; \citealt{bouwens11}; \citealt{finkelstein}; \citealt{dahlen};\citealt{mcLure}; \citealt{schenker13a}). This is done through fitting of the observed SEDs of individual galaxies to model templates in order to estimate their photometric redshifts or measure rest-frame luminosities. However, there are a number of concerns regarding this process.  First, this requires accurate photometry for galaxies. Given that the photometric data points used for the SED fits are observed by different telescopes and instruments, with different point spread functions (PSFs), one needs to reduce them to the same scale (i.e. images with the highest resolution). This is to ensure that they are corrected so that the ratios of fluxes in different bands refer to the same regions of galaxies. Second,  this requires clear understanding of the accuracy and biases in photometric redshift and stellar mass measurement. Third, at the basic level, different investigators use different techniques, codes, templates and initial parameters to fit the observed SEDs and extract observable information from them. This alone introduces unknown differences in the photometric redshift and stellar mass estimates to the {\it same} galaxy. 
The first problem is generally addressed by degrading the data to a common PSF, or by fitting templates for galaxies from the higher resolution image convolved with a kernel to match the PSF of the lower resolution images, using the Template Fitting (TFIT) technique (\citealt{laidler}). This has successfully been used to generate self-consistent multiband dataset for individual galaxies across the wavelength range covered (\citealt{Guo}; \citealt{santini2015}; Nayyeri et al 2015 in preparation). However, there are still outstanding issues regarding the second and third points. 
In \cite {dahlen}, we addressed systematic uncertainties in photometric redshift estimation. In this paper, we focus on the stellar mass measurement. 

The most common method for measuring the stellar masses of galaxies is to fit
their observed SEDs, covering the wavelength range UV/optical/infrared, to templates generated 
from the population synthesis models. The templates consist of a large grid of 
model SEDs with a range of free parameters, including: redshift, 
Star Formation History (SFH), age, prescription for dust extinction and metallicity. For any galaxy, the parameters corresponding to the template SED which best fit its observed 
SED (minimum $\chi^2$) are associated to that galaxy. Having measured the $M/L$
ratio of the galaxy, and knowing its absolute luminosity, one could then estimate its stellar mass, as well as other physical parameters. However, there is 
significant degeneracy in this procedure. The fitting techniques do not necessarily yield unique models, with various combinations of 
free parameters leading to equally acceptable fits. Furthermore, the final
estimate of the stellar mass also depends on technical details such as the 
population synthesis models used to generate template SEDs, the 
fitting technique, the code used and the $S/N$ ratio of the observed 
photometric data. Therefore, it is important to understand the dependence
of the stellar mass on each of these parameters and to disentangle the 
interplay between them. 
 With this in mind, we have undertaken an extensive investigation of the sources of uncertainty in the stellar mass measurement from broadband photometry. The time is ripe for such a study, with the availability of the CANDELS data spanning a wide range in wavelength.

We perform two classes of tests: 1). comparison of estimated stellar
masses with ``true'' ones generated from theoretical mock catalogs and 
2). comparison
of estimated stellar masses from different codes and methods applied to
observational data, where the ``true'' masses are not known. This allows 
 a test of internal consistency between different
stellar mass methods, aiming to understand sources responsible for the 
observed divergencies between them. The CANDELS data used for this purpose are extremely deep, so the photometry has very low measurement errors. 

We generated simulated and real multi-waveband photometric catalogs of galaxies with known redshifts and stellar masses and asked a number of experts within the CANDELS team to independently estimate the observable quantities associated with them. We then compared the stellar masses with the ``true'' values in the mock catalogs and the measurements between different teams, aiming first to have a realistic estimate of the error budget and then, to develop a prescription to acquire the most accurate stellar mass for individual galaxies. Furthermore, we aim to understand parameters responsible for the observed divergencies between different algorithms used for stellar mass measurement. 

Several studies have recently undertaken similar investigation by addressing the accuracy of predicted physical parameters in galaxies using simulated catalogs (\citealt{wuyts2009}; \citealt{lee2012}; \citealt{pforr2012}). These studies often used one population synthesis codes to generate model templates (\citealt{wuyts2009}) and a single SED fitting technique (\citealt{long2012}). Furthermore, in the fitting process they fit all the free parameters simultaneously (i.e. age, metallicity, SFH, mass), causing serious degeneracies between the predicted parameters. Moreover, they either use a limited redshift range (\citealt{wuyts2009}) or are restricted to certain galaxy types (\citealt{lee2012}) and are hardly constrained by the observational data (\citealt{pforr2012}).  \cite{pforr2013} investigated the dependence of results on different population models, used a wider range in redshift, and explored the depndence of results on wavelength coverage and photometric filters. While they used \cite{maraston2005} as their population synthesis model, they also tested the results from other codes but used the same SED fitting method and procedure to estimate the parameters, showing serious degeneracies. None of these studies explores the dependence of the estimated stellar masses on the nebular emission lines, which is proved to be significant (\citealt{debarros}). 

This paper complements previous studies in various ways. It uses ten 
independent SED fitting techniques and codes from different 
teams and, at the same time, explores dependence of each of these results on a variety of 
population synthesis models. Furthermore, the mock catalogs generated for this purpose
 are selected to resemble observed galaxy surveys (i.e. CANDELS) in terms of 
redshift distribution, wavelength coverage and photometric uncertainty, 
so that the results would be directly applicable to the observed data. 
In addition to simulations, it also uses observed photometry and real data to 
explore the internal consistency of the stellar masses measured from different 
procedures. By fixing the parameters in the SED fitting process to those of the
input mock catalogs, we study the degeneracy amongst the parameters, estimating the errors contributed from each parameter to the final stellar mass. 
The present study also investigates dependence of stellar mass on nebular line emissions. 
Finally, the errors
associated with individual physical parameters are estimated and 
their contribution to the total error budget calculated. Results from this study are directly used to estimate stellar masses for the CANDELS galaxies by 
finding the technique which leads to the most accurate measurement.

In section 2, we present the procedures and the tests designed for this study. In Section 3 the participating teams are introduced, with a brief description of the methods and techniques used by each team. Sections 4-7 present different tests and explore sources of uncertainty and bias in stellar mass measurements from different teams. Comparison with other similar studies in literature is
performed in section 8, with 
the error budgests estimated and discussed in Section 9. Our conclusions are summarized in Section 10.  Throughout, we assume standard cosmology with $ H_0 =70$ Km/s/Mpc, $\Omega_M = 0.3$ and $\Omega_\Lambda =0.7$. Magnitudes are all in the AB system (Gunn  \& Oke 1983).

\section {\bf The Procedure}

In this investigation, we carry out four different tests, designed to explore different types of systematic errors in stellar mass measurement. 
We estimate stellar masses from different catalogs: an empirical mock catalog (TEST-1), a Semi-Analytic Mock catalog (SAM; TEST-2)  
and a ``real'' observational catalog (TEST-3 and TEST-4). The main parameters used to generate the mock catalogs and the input to stellar mass measurement codes (discussed in section 3) are listed in Table 1. In Appendix I, we summarize definitions of the stellar masses used in the SAMs in this study and most commonly used in literature.   
TEST-1, developed to evaluate different SED fitting codes, fits simulated data for galaxies with simple star-formation histories (SFH), using a limited number of free parameters (this test is strongly constrained). The mock catalogs are generated to have similar distribution of physical parameters as the observed 
catalogs (presented in Appendix II).  
TEST-2A and TEST-2B fit simulated data for galaxies with more complex SFHs drawn from a semi-analytic model. TEST-2A fits the mock data, simulated to mimic real galaxies as closely as possible.  TEST-2B is more constrained; there is no dust and fits are restricted to using the same evolutionary synthesis code for the fits. TEST-3A and TEST-3B compare masses when the same fitting parameters and techniques, used in TEST-2A and TEST-2B, are applied to real galaxies. TEST-4 repeats TEST-3A using somewhat shallower near-IR data typical of pre-CANDELS observations. The simulated multi-waveband mock catalogs used in TEST-2A and 2B were generated with halos extracted from Bolshoi N-body simulations (\citealt{klypin}; \citealt{behroozi}) and populated using semi-analytic models (\citealt{somerville08}; \citealt{somerville}). The bandpasses and quality of the photometry in all the tests approximate the observed data from the CANDELS. The stellar masses provided in the mock catalogs are defined
as the mass which is directly produced through SED fits. The age is defined as
the time since the on-set of star formation. One of the main sources of error in stellar mass estimates is
lack of knowledge of the SFHs (e.g. \citealt{lee2014}). 
Nearly all the methods make very simple assumptions about this 
and even when diverse SFHs are allowed, it is unclear
whether the methods can correctly select the right type of history based
on the photometry, given all the other uncertainties. The SAMs have a
semi-realistic mix of complex SFHs (including rising, ~constant, and
declining) however, they do not correctly reproduce the
observed trends between galaxy mass and Star Formation Rate (SFR) ie. downsizing.  The main characteristics of the tests in this section are listed in Table 1, 
with an overall comparison between different tests presented in Table 2 and detailed below:
\\
\noindent {\bf TEST-1: Test of the consistency of different SED fitting codes and techniques. } 
This test is designed to study how well different codes can measure the stellar masses and if there is any difference originating from the codes once we keep all the rest of the parameters fixed. To do this, we generate a mock catalog with known input parameters (redshift, stellar mass, age and extinction) using templates
produced from \cite{bruzual03} population synthesis models. To make the simulated galaxies comparable to the real data, we add noise to their photometry.  The parameters used to generate the mock SEDs are listed in Table 1. There are a total of 559 galaxies in TEST-1 mock catalog. The total number of simulated galaxies in TEST-1, and the distribution of their physical parameters are taken to be close to the real spectroscopic catalog in GOODS-S field, used to calibrate photometric redshifts and the SED fitting techniques. This allows results from TEST-1 to be directly applicable to observations. Details about the TEST-1 mock catalog and distribution of the observable parameters are given in Appendix II.

%($M_{int}=\int M\ dt $). The integrated mass corresponding to an exponentially 
%declining SFR, $SFR(t) = SFR_0 e^{-t/\tau}$, 
% is therefore calculated for each object as
%$$ M_{int} = \tau SFR_0 (e^{t/\tau} - 1) $$
%where $SFR_0$ is the SFR at $t=0$ and $\tau$ is the SFR time scale. 

The masses are derived by fixing the template SEDs to have the input values (listed in Table 1) and ONLY fitting for two quantities: the age of the star formation and color excess ($E(B-V)$). The age is defined as the time since the on-set
of star formation (assuming an exponentially declining SFR with a fixed $\tau$) and was constrained between 10 Myr and the age of the Universe at the redshift of the particular galaxy under consideration. The allowed range for the color excess, $E(B-V)$, is taken to be between 0 and 1.  The redshift for each galaxy was fixed to its input value.  Since the majority of the parameters affecting the stellar mass measurement are fixed, the only difference between the estimated stellar masses from the SED fits ($M_{est}/M_\odot$) and the expected stellar masses ($M_{input}/M_\odot$) is due to differences between the codes and the SED fitting techniques used between different teams.

TEST-1 is based on a set of 13 filters consisting of: U-band (VIMOS), optical- {\it F435W, F606W, F775W, F850LP} (ACS); near-infrared {\it F105W, F125W, F160W} (WFC3); 
HawkI {\it K-band} (VLT) and {\it Spitzer}/IRAC 3.6, 4.5, 5.8 and 8 $\mu$m. The selection criteria for 
galaxies in TEST-1
include: a) $S/N > 5$ in the H-band (the selection band in the simulated catalog); b) Detected with $S/N>1$ in at least six passbands: c) $0 <z<4$. 
\\
\noindent {\bf TEST-2: Test of the sensitivity of the stellar mass estimates to the free parameters.} 

This test is developed to study the effect of free parameters on the stellar mass measurement. It uses SAM catalogs containing 10,000 galaxies with known 
multi-waveband photometry, 
input mass, age, extinction and metallicity. The SEDs were constructed using \citealt{bruzual03}- BC03) models, with a modified
version of \cite {Charlot} prescription for 
 dust treatment as discussed in \cite{somerville}. Stellar mass and chemical evolution are calculated assuming instantaneous recycling. The SFHs are
diverse, consisting of exponentially declining, constant and rising.  Redshift distribution for galaxies in TEST-2 catalog closely follow the photometric redshift distribution in the GOODS-S field (Appendix II). 

For all the galaxies in TEST-2 mock catalogs, photometry is provided in 13 filters: U-band (VIMOS), optical- {\it F435W, F606W, F775W, F850LP} (ACS); near-infrared {\it F105W, F125W, F160W} (WFC3); HawkI {\it K-band} (VLT) and {\it Spitzer}/IRAC 3.6, 4.5, 5.8 and 8 $\mu$m.  
The selection criteria for the mock catalog here are: a) $S/N > 5$ in the H-band (the selection band in the simulated catalog); 
b) Detected with $S/N>1$ in at least six passbands; 
c) $z<6$.  TEST-2 is performed in two stages:

\noindent {\bf TEST-2A:} The mock catalog here is generated using a diversity of SFHs 
(exponentially declining, rising and constant) and metallicities. The data generated for this catalog have dust extinction applied to the photometry and hence, 
closely resemble the observations. To estimate the stellar masses, the participating teams were not restricted and were free to choose any template from any stellar population synthesis code, SFH, metallicity and extinction law to fit the mock 
SEDs. The only limitation was to use Chabrier IMF and to fix the redshift of the
galaxy to its input value in the mock catalog. 
 
\noindent {\bf TEST-2B:} Unlike TEST-2A, the mock catalog here is generated by constraining 
the free parameters. The SFH associated with template SEDs is fixed to an exponentially declining form, with the templates
produced from BC03 with solar metallicity. No dust extinction is applied to photometry in the mock catalog and therefore, TEST-2B is not 
representative of the ``real'' population of galaxies. Redshift is fixed to its 
input value and Chabrier IMF is used. 
The participating teams were asked to use the {\it same} input parameters as
the ones used to generate the mock data.

Comparison between the stellar mass estimates from TEST-2A and TEST-2B will
therefore reveal the effect of free parameters and degeneracy in the SED 
fitting process.

\noindent {\bf TEST-3: Comparison of the stellar mass measurements using real data}

Having estimated the sources of scatter in stellar mass measurements due to different codes (TEST-1) and due to degeneracy and interplay between the free parameters (TEST-2), we now apply the methods on a sample of observed SEDs with accurate multi-waveband data and available spectroscopic redshifts, selected from the TFIT catalog in the GOODS-S field (\citealt{guo2013}). This is the same sample used in \cite{dahlen} to calibrate the templates for measuring photometric redshifts. A total of 598 galaxies were used for this test. For the SED fits, the galaxy redshifts were fixed to their spectroscopic values. Unlike TEST-1 and TEST-2, where we used simulated photometric catalogs and hence, had estimates of the ``true'' stellar mass, here we do not have any absolute measure of the stellar mass and the comparison is only relative, measuring the consistency between different approaches. 

The photometry for TEST-3 is performed on the real data using the TFIT 
technique (\citealt{guo2013}) and consists of: U-band (VIMOS), 
optical- {\it F435W, F606W, F775W, F850LP} (ACS); near-infrared- {\it F098M, F105W, F125W, F160W } (WFC3); 
{\it Ks} (VLT/ISAAC) and mid-infrared {\it Spitzer}/IRAC 3.6, 4.5, 5.8 and 8 $\mu$m.  
 The {\it F098M} is only available for the {\it Early Release Survey} part of the GOODS-S, while {\it F105W} is only available for a sub-area of that field. TEST-3 is also done in two stages:

\noindent {\bf TEST-3A:} In this test no restriction was imposed on the free parameters when generating the template SEDs for the fits, except for the redshifts which were fixed to their spectroscopic values and the IMF which was chosen to be Chabrier (Table 1). The stellar masses were subsequently estimated from different methods (section 3) and compared with each other. 

\noindent {\bf TEST-3B:} This is the same as TEST-3A with additional restrictions imposed on the free parameters. This will show how close the results from different teams would be when some of the parameters in the SED fits are not allowed to vary. Therefore, it indicates the effect of the free parameters (and their possible interplay) in stellar mass measurements. 

\noindent {\bf TEST-4:} {\bf Tests the effect of selection wavelength and near-infrared photometric depth on the stellar mass measurement.} 
  
This is similar to TEST-3A with the only difference being the use of much shallower near-infrared data and selection in ACS z-band. This test is designed to examine the way different codes treat shallow infrared data and its effect on stellar mass measurement. As is often the case in galaxy surveys, due to smaller size of near-IR detectors, their lower sensitivity and the effect of sky brightness, the near-IR data are not as deep as their optical counterparts. We designed  TEST-4 to examine the sensitivity of stellar mass on these parameters. 

\begin{table*}
\caption{Details of different TESTs developed for stellar mass measurement. The listed parameters are used to generate the mock catalogs and as inputs in the codes discussed in section 3 to measure stellar masses}
\centering
\begin{tabular}{ll}
 & \\
\tableline 
\tableline

{\bf TEST-1:} & \\
    {\bf Fixed Parameters  }        & \\
               &  IMF: Chabrier (limits: 0.1 $< M/M_\odot <$ 100)\\
               & Redshift range $ 0 < z < 4$\\
               & Stellar population templates: Bruzual and Charlot 2003 (BC03).\\
               & SFH: Single burst Exponentially declining, $\tau$ fixed at 0.1 Gyr.\\
               &Gas recycling: no. \\
               & Dust extinction law: Calzetti.\\
               & IGM Absorption: Madau law, flux set to zero at $\lambda < 912\AA$ (restframe).\\
               &Metallicity: $Z = Z_\odot$\\
               & Nebular Emission: not included.\\
{\bf Free Parameters} & \\
               & Age between 10 Myr and the Age of the Universe at the redshift\\ 
               & of the galaxy.\\
               & $E(B-V)$ between 0 and 1.\\
               & \\
{\bf TEST-2A} &\\
{\bf Fixed parameters}\\
              & Chabrier IMF ($0.1 < M/M_\odot < 100$)\\
              & Redshift range $ 0 < z < 6$\\
              & Redshift is fixed to its input value.\\
              & Dust extinction is applied to the photometry in the mock catalog.\\
{\bf Free parameters} &\\
              & $^1$SFH, metallicity, extinction, population synthesis code\\
{\bf TEST-2B}\\
 {\bf Fixed Parameters} &\\
 & Templates: BC03 with Chabrier IMF ($0.1 < M/M_\odot < 100$)\\
 & Redshift range $ 0 < z < 6$\\
 & $^1$SFH: exponentially declining SFR\\
 & no extinction applied to the photometric points; E(B-V)=0\\
 & Metallicity: Solar \\
&Emission lines: not included\\
 & Redshift fixed to the provided value in the mock catalog\\
&\\
{\bf Free parameters}&\\
&The exponential time scale $\tau$ and the age of the stellar population.\\
{\bf TEST-3A}\\
{\bf Fixed Parameters}\\
& Observed F160W band selected multi-band TFIT photometric catalog for GOODS-S\\
& The objects are fixed at their spectroscopic redshifts\\
& Redshift range $0 < z < 6$\\
& Chabrier IMF ($0.1 < M/M_\odot < 100$) \\
{\bf Free Parameters}&\\
              & SFH, metallicity, extinction, population synthesis code, stellar mass, age\\
&\\
{\bf TEST-3B} &\\
{\bf Fixed Parameters}&\\
& Observed F160W band selected multi-band TFIT photometric catalog for GOODS-S\\
& Templates: BC03 [with Chabrier IMF ($0.1 < M/M_\odot < 100$)]\\
& Extinction: E(B-V)=Av=0, i.e., no extinction\\
 & $^1$SFH: Exponentially declining\\
 & Metallicity: Solar\\
&Redshift range $0 < z < 6$\\
{\bf Free Parameters}\\
& stellar mass, star formation time-scale, $\tau$, age\\
{\bf TEST-4} &\\
&\\
& The same as TEST-3A but selected in ACS z-band, with shallower observed near-infrared data \\
\tableline
\end{tabular}

\tablecomments{$^1$The SAMs use a diversity of SFHs depending on the host halo merger history. Therefore, the SFH of every mock galaxy is fixed. The forms of the SFHs adapted here are used to generate the SED templates for the stellar mass measurement methods.}

\end{table*}

\begin{table*}
\caption{Table shows the list of the parameters in the SED fitting methods which are kept fixed to values listed in Table 1 or are left free in the fit.}
\centering
\begin{tabular}{rllclll}
Parameters &				TEST-1 &	TEST-2A&TEST-2B	&	TEST-3A&TEST-3B&	TEST-4\\
&&&&&&\\
Star Formation History	&		Exp. Declining&	Free&	Exp. Declining&	Free&	Exp. Declining&	Free\\
&&&&&&\\
Population Synthesis Models	&	BC03 	&	Free&	BC03 	&	Free&	BC03 	&	Free\\
&&&&&&\\
$\tau$& Fixed&Free  &Fixed &Free &Fixed &Free\\
&&&&&&\\
IMF				&	Fixed	&	Fixed&	Fixed	&	Fixed& 	Fixed	&	Fixed\\
&&&&&&\\
Redshift 			&	Fixed	&	Fixed&	 Fixed	&	Fixed&	Fixed	&	Fixed\\
&&&&&&\\
Extinction			&	Fixed	&	Free&	 None	&	Free&	None	&	Free\\
&&&&&&\\
Age				&	Free	&	Free&	 Free	&	Free&	Free	&	Free\\
&&&&&&\\
Metallicity			&	Fixed	&	Free&	 Fixed	&	Free&	Fixed	&	Free\\
&&&&&&\\
Nebular Emission		&	No	&	No&	 No	&	No&	No	&	No\\
&&&&&&\\
IGM Absorption			&	Fixed	&	Fixed&	 Fixed	&	Fixed&	Fixed	&	Fixed\\
\tableline
\end{tabular}
\end{table*}

\section {\bf The SED Fitting Techniques and Stellar Mass Measurement}

The catalogs discussed in section 2 were provided to the CANDELS team members. Using the instructions for different tests (Table 1), the teams were asked to predict the stellar masses for galaxies in the catalogs, satisfying the requirements for each TEST. To perform this as objectively as possible, the $M_{input}$ values in the mock catalogs were not revealed to the participants.  

A total of ten teams participated in this exercise (not all the teams participated in all the tests). In many cases, the codes and templates used to measure the stellar masses were different from those used for the photometric redshifts in Dahlen et al (2013). Below, details of the codes and the assumptions when applied on TESTs 2A, 3A and 4 are described (in these tests the participants were free to choose templates from any population synthesis models or any SFH). For TESTs 1, 2B and 3B, all the methods used BC03 evolutionary synthesis models and exponentially declining star-formation histories. 
 Details are also listed in Table 2. Where possible, we use the same identification for the teams as in \cite{dahlen}.  

\noindent {\bf Acquaviva (1.A)}- {\it GalMC} code (\citealt{acquaviva11}): 
The algorithm is a Markov Chain
Monte Carlo (MCMC) sampler based on Bayesian statistics. In this approach,
the SED fitting parameters (age, mass, reddening, and e-folding time
for $\tau$ models) are treated as random variables. The parameter space
is explored with a random walk biased so that the frequency of visited
locations is proportional to the posterior Probability Density
Function (PDF). The desired intervals of the SED fitting
parameters are then obtained by marginalizing the PDF, which in MCMC
simply corresponds to summing over the points in the chains. Here, a new 
version of the {\it GalMC} is used ({\it SpeedyMC}; \citealt{Acquaviva12}), 
which is 20,000 times faster and at every step of the chain, the spectra are
generated through multi-linear interpolation of a library of
pre-computed models. The best-fit stellar masses and the 68\%
uncertainties are predicted from these marginalized distributions. 

For TESTs 2A, 3A and 4, this code used templates based on Charlot and Bruzual (2007- CB07) while for other tests it used BC03 population synthesis models. Two metallicities are used: Solar and 0.2 $Z_\odot$ with the one giving the optimum $\chi^2$ value chosen.  

\noindent {\bf Finkelstein (4.B)}- {\it own code}:
This uses $\chi^2$ fitting method with CB07 population synthesis model. It uses a hybrid SFH (exponentially declining + rising star formation rate). 

\noindent {\bf Fontana (6.C)}- {\it own code}:
This uses $\chi^2$ fitting method with the SED templates generated from BC03 and exponentially declining SFR. 
The templates are generated with both Calzetti and SMC dust models and hence, the code can choose between the two dust extinction scenarios, whichever gives a better fit. 

\noindent {\bf Gruetzbauch (7.D)}- {\it EAZY code}- \cite{brammer}:
Uses $\chi^2$ fitting method with BC03 and an exponentially declining SFR with a large set of $\tau$ values. Also uses a large set of metallicity and extinction values.

\noindent {\bf Johnson (8.E)}- {\it SATMC code}- \cite{Johnson}:
Uses the {\it MCMC} to fit the SEDs, similar to method 1.A. BC03 templates are used with instantaneous burst of star formation. This is the only experiment which uses this SFH. For the fit, all the parameters in the code are varied.

\noindent {\bf Papovich (9.F)}- {\it own code}: This is a $\chi^2$ minimization code. It uses an exponentially declining SFR. Solar metallicity is assumed. The code uses templates based on BC03 models. 

\noindent {\bf Pforr (10.G)}- {\it HYPERZ} code \cite{bolzonella}: This is a $\chi^2$ minimization code. It uses hybrid SFH consisting of exponentially declining, truncated and constant SFRs. In this respect, 10.G is different from many of the methods listed in Table 3 but is similar to others (eg. 4.B). This is the only method which uses \cite{maraston2005} population synthesis code to generate templates. 

\noindent {\bf Salvato (11.H)}- {\it La Phare} code \cite{arnauts}: 
This uses a $\chi^2$ minimization technique and BC03 code to generate templates. Exponentially declining star formation rate is used. The prior $E(B-V) < 0.15$ is applied if the ratio $t/\tau > 4$ (i.e. significant extinction is only allowed for galaxies with high SFR). 

\noindent {\bf Wiklind (12.I)}- {\it own code}- \cite{wiklind}- Uses $\chi^2$ minimization of the SEDs. The errors in stellar mass are estimated from Monte Carlo simulations. Exponentially declining star formation rates are used with $\tau=0$ representing an instantaneous burst. The template SEDs are based on BC03. 

\noindent {\bf Wuyts (13.J)}- {\it FAST} code \cite{kriek}:
Uses $\chi^2$ fitting technique with exponentially declining SFR. The templates are from BC03 with solar metallicity. 

\begin{table*}
\caption{Details of the methods and parameters used for stellar mass measurement}
\centering
\begin{tabular}{ll}
 & \\
\tableline 
\tableline

{\bf Method 1.A}\\
Team ID: 1 &\\
PI: Acquaviva &\\
Code ID: A &\\
Code: {\it GalMC} \citep{acquaviva11}&\\
Fitting Method: MCMC&\\
Stellar Population Synthesis Templates: CB07 or BC03 (see the text)&\\
Star Formation History: Constant&\\
Extinction law: Calzetti, $E(B-V)=0.0-1.0$&\\
Ages: $10^6$-$1.4\times 10^{10}$ yrs&\\
Nebular emission: yes&\\
Metallicity: $Z_\odot$ and $0.2 Z_\odot$&\\
&\\
{\bf Method 4.B}\\
Team ID: 4 &\\
PI: Finkelstein &\\
Code ID: B &\\
Code: {\it own code} &\\
Fitting Method: $\chi^2$&\\
Stellar Population Synthesis Templates: CB07 &\\
Star Formation History: Exponentially declining, rising ($\tau= 0.0001, 0.01, 0.1, 1.0, 100.0, -0.3, -1.0, -10.0 $) Gyrs &\\
(the negative values correspond to a rising SFR)&\\
Extinction law: Calzetti, $E(B-V)=0.0-0.51$&\\
Ages: 1Myr - 13Gyrs&\\
Nebular emission: yes&\\
Metallicity: $Z_\odot$ and $0.2 Z_\odot$&\\
&\\
{\bf Method 6.C}\\
Team ID: 6 &\\
PI: Fontana &\\
Code ID: C &\\
Code: {\it own code} &\\
Fitting Method: $\chi^2$&\\
Stellar Population Synthesis Templates: BC03 &\\
Star Formation History: Exponentially declining with $\tau = 0.1, 0.3, 0.6, 1, 2, 3, 5, 9, 15$ Gyrs &\\
Extinction law: Calzetti+SMC, $E(B-V)=0.0-1.1$ in increments of 0.05&\\
Ages: $log(age)= 7-7.35$ (in 0.05 steps), $7.4-8.9$ (0.1 steps), $9-10.3$ (0.05 steps)  &\\
Nebular emission: no&\\
Metallicity: $0.2 Z_\odot$, $0.4 Z_\odot$, $ Z_\odot$, $2.5 Z_\odot$ also a subset of models with age $<$ 1 Gyrs and $Z_\odot =0.02$  &\\
&\\
{\bf Method 7.D}\\
Team ID: 7 &\\
PI: Gruetzbauch &\\
Code ID: D &\\
Code: {\it EAZY} \cite{brammer}&\\
Fitting Method: $\chi^2$&\\
Stellar Population Synthesis Templates: BC03 &\\
Star Formation History: Exponentially declining with $\tau=$ 0.01, 0.03, 0.06, 0.1, 0.25, 0.5, 0.75, 1.0, 1.3, 1.7, 2.2, 2.7,  &\\
3.25, 3.75, 4.25, 4.75, 5.25, 5.75, 6.25, 6.75, 7.25, 7.75, 8.25, 8.75, 9.25, 9.75, 10.25, 10.75 Gyrs &\\
Extinction law: Calzetti, $A_v=0.,0.2,0.4,0.6,0.8,1.0,1.33,1.66,2,2.5$&\\
Ages: 1Myr - 13 Gyrs- ages required to be smaller than the age of the Universe at each redshift  &\\
Nebular emission: no&\\
Metallicity: (X, Y, Z): (0.7696, 0.2303, 0.0001), (0.7686, 0.231, 0.0004), (0.756, 0.24, 0.004), \\
(0.742, 0.25, 0.008), (0.70, 0.28, 0.02), (0.5980, 0.352, 0.0500), (0.4250, 0.475, 0.1000)&\\
&\\
\end{tabular}
\end {table*}

\begin{table*}
\begin{tabular}{ll}
 & \\
\tableline 
\tableline

{\bf Method 8.E}\\
Team ID: 8 &\\
PI: Johnson &\\
Code ID: E &\\
Code: {\it SATMC} (Johnson et al 2013)&\\
Fitting Method: MCMC&\\
Stellar Population Synthesis Templates: BC03 &\\
Star Formation History: Instantaneous burst  &\\
Extinction law: Calzetti, $E(B-V)=0.0-4.5$ &\\
Ages: $0.01-10$ Gyr (unequally spaced and taken directly from BC03 library)&\\
Nebular emission: no&\\
Metallicity: $0.0001 Z_\odot$, $0.0004 Z_\odot$, $0.004 Z_\odot$, $0.02 Z_\odot$ $0.05 Z_\odot$  &\\
&\\

{\bf Method 9.F}\\
Team ID: 9 &\\
PI: Papovich &\\
Code ID: F &\\
Code: {\it Own Code} &\\
Fitting Method: $\chi^2$&\\
Stellar Population Synthesis Templates: BC03 &\\
Star Formation History: Exponentially declining ($\tau=0.001, 0.01,0.03, 0.1, 0.3, 1.0, 3.0, 10.0, 100.0$ Gyr)  &\\
Extinction law: Calzetti &\\
Ages: 0.0251, 0.04, 0.064, 0.1015, 0.161, 0.255, 0.6405, 1.0152, 1.609, 2.5, 4.0, 6.25 and 10.0 Gyrs &\\
Nebular emission: no&\\
Metallicity: Solar, except for TEST-2A for which the following are used: 0.02, 0.2, 0.4, 1.0, 2.5 $ Z_\odot$  &\\
&\\
{\bf Method 10.G}\\
Team ID: 10 &\\
PI: Pforr &\\
Code ID: G &\\
Code: {\it HyperZ} (\cite{bolzonella}&\\
Fitting Method: $\chi^2$&\\
Stellar Population Synthesis Templates: M05 &\\
Star Formation History: Exponentially declining ($\tau=0.1, 0.3,1.0$ Gyr), Constant SF at $t=0.1, 0.3, 1$ Gyr, &\\ 
 zero SF afterwards, Constant star formation  &\\
Extinction law: none &\\
Ages: $0-20$ Gyr (221 in total, grid as in BC03 templates)&\\
Nebular emission: no&\\
Metallicity: $0.2 Z_\odot$, $0.5 Z_\odot$, $1.0 Z_\odot$, $2 Z_\odot$   &\\
&\\
{\bf Method 11.H}\\
Team ID: 11 &\\
PI: Salvato &\\
Code ID: H &\\
Code: {\it Le Phare} \cite{arnauts}&\\
Fitting Method: $\chi^2$&\\
Stellar Population Synthesis Templates: BC03 &\\
Star Formation History: Exponentially declining ($\tau=0.1, 0.3,1, 2, 3, 5, 10, 15, 30$ Gyr)  &\\
Extinction law: Calzetti &\\
Ages: $0.01-13.5$ Gyr &\\
Nebular emission: yes&\\
Metallicity: 0.02 $Z_\odot$, 0.008 $Z_\odot$   &\\
&\\
{\bf Method 12.I}\\
Team ID: 12 &\\
PI: Wiklind &\\
Code ID: I &\\
Code: {\it Own Code} \cite{wiklind}&\\
Fitting Method: $\chi^2$&\\
Stellar Population Synthesis Templates: BC03 &\\
Star Formation History: Exponentially declining ($\tau=0.1,0.2,0.3,0.4, 0.6, 0.8, 1.0$ Gyr) and instantaneous burst ($\tau=0$)  &\\
Extinction law: Calzetti &\\
Ages: 0.05, 0.1, 0.2, 0.3, 0.4, 0.5, 0.6, 0.7, 0.8, 1.0, 1.2, 1.4, 1.6, 1.8, 2.0, 2.5, 3.0, 3.5, 4.0, 5.0, 6.0, 7.0 Gyrs &\\
Nebular emission: no&\\
Metallicity: $0.2 Z_\odot$, $0.4 Z_\odot$, $1.0 Z_\odot$, $2.5 Z_\odot$   &\\
\tableline
\end{tabular}
\end{table*}

\begin{table*}
\begin{tabular}{ll}
 & \\
\tableline 
\tableline

&\\
{\bf Method 13.J}\\
Team ID: 13 &\\
PI: Wuyts &\\
Code ID: J &\\
Code: {\it FAST} \cite{kriek}&\\
Fitting Method: $\chi^2$&\\
Stellar Population Synthesis Templates: BC03 &\\
Star Formation History: Exponentially declining (log($\tau$)=8.5-10 in increments of 0.1  &\\
Extinction law: Calzetti &\\
Ages: log($age$)=7.7 to 10.1 in increments of 0.1 &\\
Nebular emission: no&\\
Metallicity: solar   &\\
&\\

\tableline
\end{tabular}
\end{table*}

Details of individual methods are listed in Table 3. In the next section we 
 compare the input mass with the stellar mass estimates independently measured from different methods (Tables 2 and 3) to explore differences as a function of the method (TEST-1), free parameters (extinction, star formation history, age)-(TEST-2), templates used and internal consistency (TEST-3) and the photometric depth and selection wavelength (TEST-4). This allows a study of the absolute consistency (i.e. how well each code produces the expected mass) and relative consistency (how the estimated masses between different codes agree). In the following sections, we perform a step-by-step study of the above, using the information in Tables 1 and 3.

\section {\bf TEST-1: Comparison of Stellar Masses from Different Methods}

\subsection{\bf Dependence on the SED Fitting Codes}

The participating teams, listed in Table 3, used the mock catalog generated for TEST-1 and independently estimated the stellar mass for individual galaxies. For each code, Figure 1 shows changes between the input mass, log($M_{input}$), and $\Delta(log M)$,  
defined as the difference between the input mass 
and stellar mass estimated from that code, $M_{est}$: $\Delta(log M) = log(M_{input}) - log(M_{est})$. The very small scatter in the case of 1.A is to be expected because TEST-1 mock catalog was generated based on this method and therefore it confirms the consistency between the input and estimated masses. As a result, the observed scatter in the stellar mass from method 1.A is likely due to the
effect of photometric errors added to the mock data. 
This is supported by the results from Figure 2a which shows an increase in 
the scatter in $\Delta log (M)$ based on method 1.A from bright to faint
magnitudes (see below).  
 Figure 1 also confirms that all the methods used in this experiment recover the input mass values to good accuracy.  There are no systematic effects or mass-dependent biases, indicating that none of the methods in Table 3 is significantly biased. 

 It is clear from Figure 1 that for most of the methods, the scatter reduces towards the higher mass end ($M > 10^{10} \ M_\odot$).  
This is because these galaxies are often brighter with a higher photometric $S/N$ ratios. This is demonstrated in Figure 2a, where we study changes in $\Delta log(M)$ as a function of H-band ({\it F160W}) magnitude, showing an increase in the scatter at $H_{AB} > 26$ mag. This indicates that the main source of inconsistency between the stellar mass estimates among different codes, keeping everything else the same, is for the relatively fainter galaxies (and those with lower photometric $S/N$ ratios) and due to different ways the photometric errors are handled in the SED fitting process. Figure 2b shows the change in $\Delta log(M)$ as a function of redshift where, for most of the methods, we find {\it no} correlation and hence, no redshift-dependent biases. The exception is 7.D where shows a bias at $z < 1$.  The reason for the observed redshift-dependent bias here is not clear but is likely due to degeneracy caused by using a wide range of metallicities. The redshift distribution in TEST-1 catalog, presented in Appendix II, is fixed to be the same as the observed (spectroscopic) distribution for the GOODS-S field. Therefore, the results from this study are directly applicable to the observed samples. 

\begin{figure*}
\epsscale{0.8}
\plotone{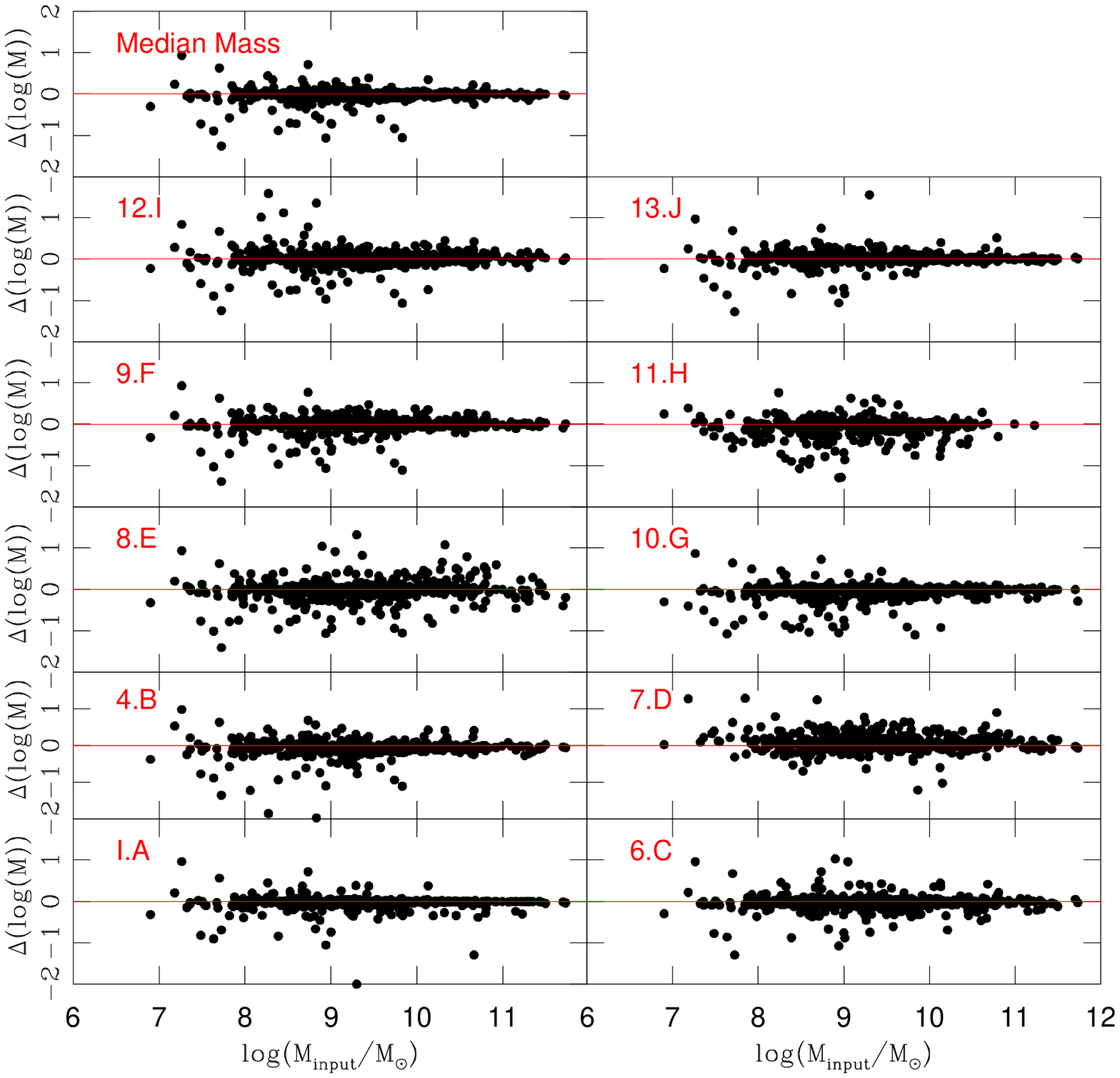}
\figcaption[TEST4_Stell_mass_comp_1n.eps]{The stellar mass difference ($\Delta log(M)$) as a function of log($M_{input}$) 
measured from all the participating methods, using TEST-1. 
$\Delta log (M)$ is defined as $\Delta log (M) = log(M_{input}) - log(M_{est})$ where $M_{est}$ is the estimated stellar mass.  
The red horizontal line shows the expected relation if the input stellar mass is exactly produced. A total of 559 simulated galaxies are used. This test examines the sensitivity of the stellar mass to the methods/codes listed in Table 3.}
\end{figure*}
\begin{figure*}
\epsscale{0.75}
\plotone{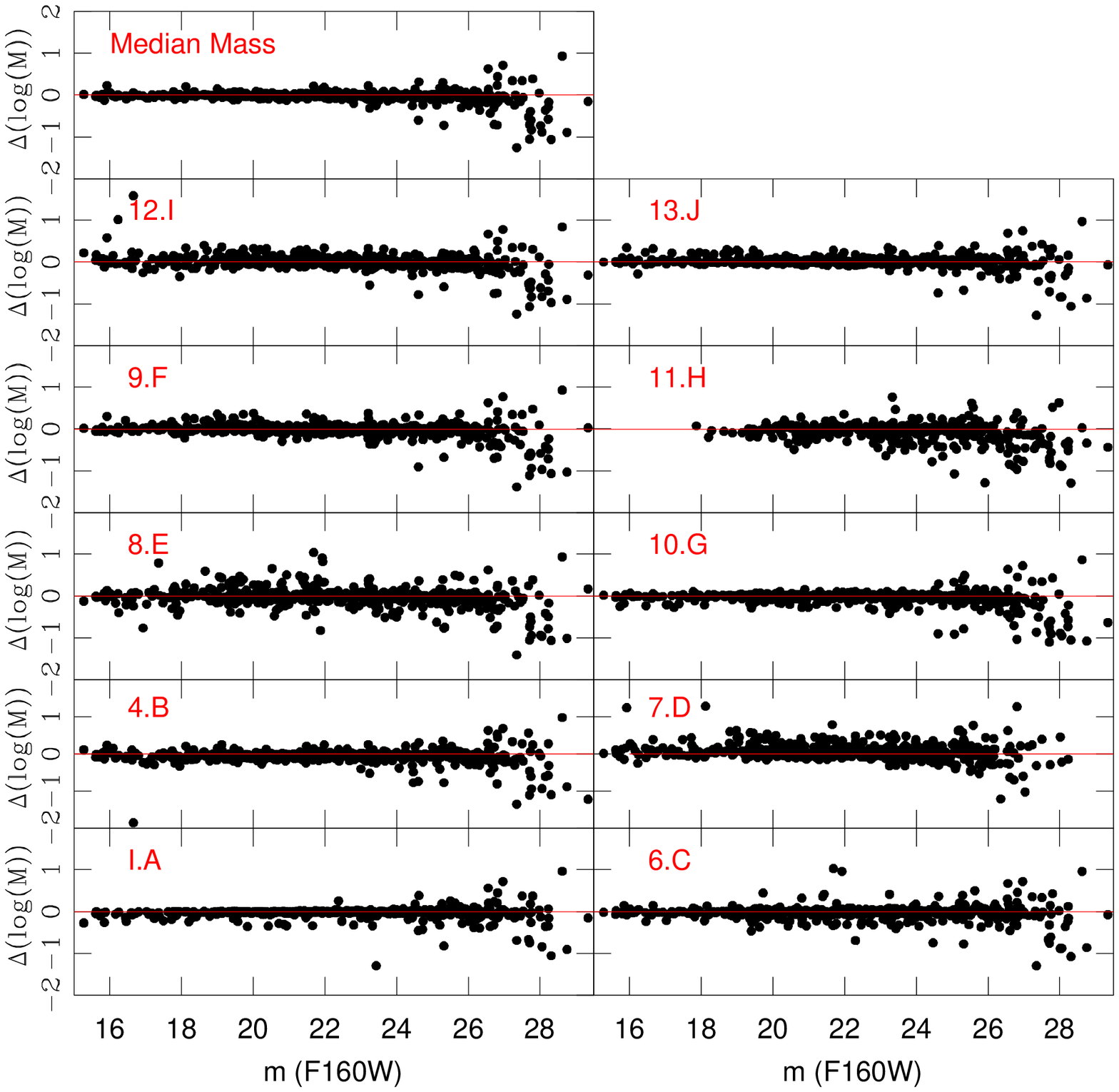}
\plotone{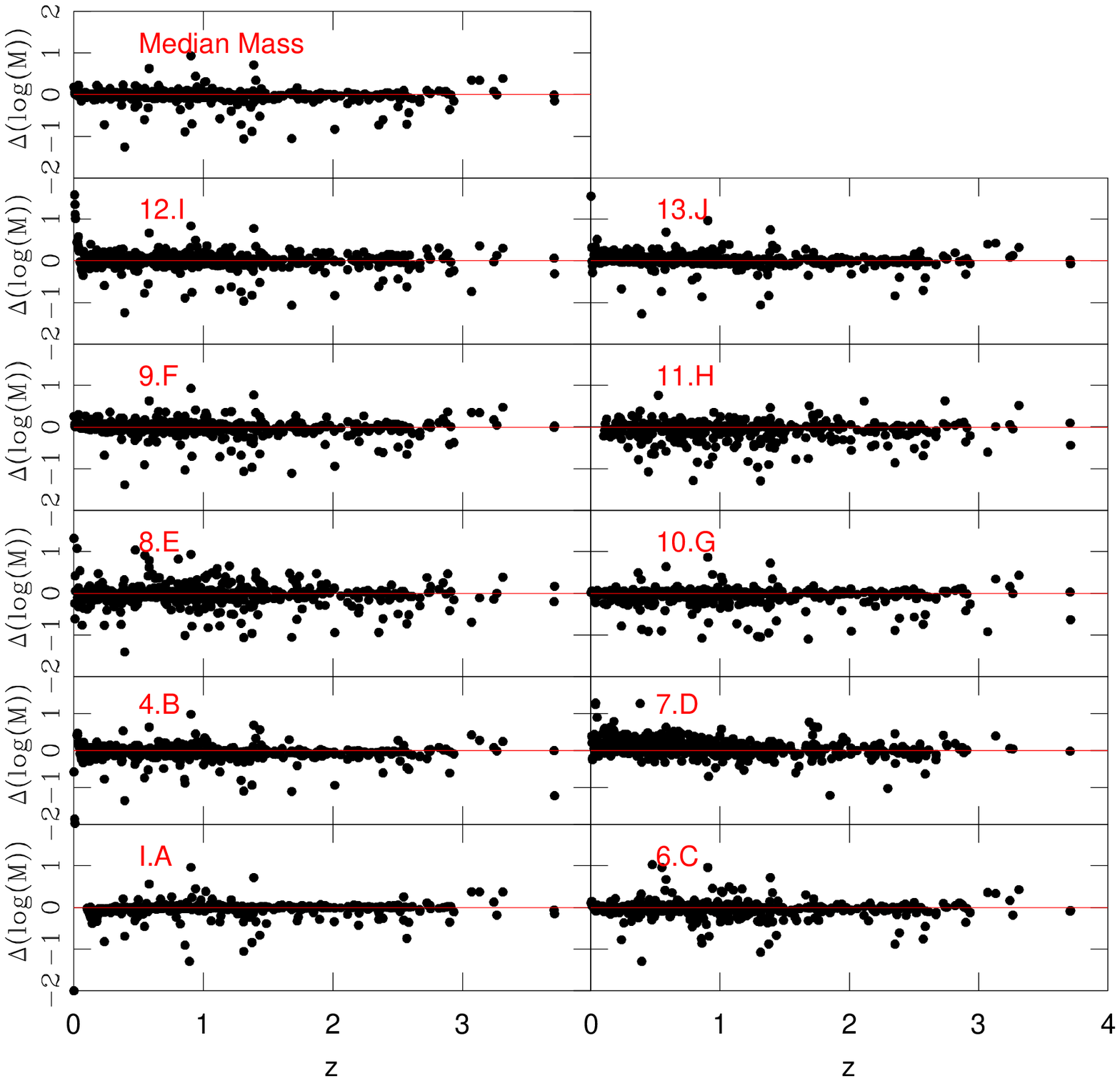}
\figcaption[TEST4_Stell_mass_comp_1n.eps]{2a)- Top Panel: Dependence of $\Delta log(M)$ on
H-band ({\it F160W}) magnitudes in TEST-1, showing sensitivity of $\Delta log(M)$  on the photometric $S/N$. (2b)-Bottom Panel: The same as 2a but for redshift.All the input parameters in this test are fixed with the only variable being the codes/methods. These show the sensitivity of the stellar mass on the codes over the range of magnitudes and redshifts covered.
\label{fig1}}
\end{figure*}

For each method, we estimated the {\it rms}, outlier fraction and the bias in $\Delta log(M)$ values using galaxies in the mock catalogs (i.e the scatter in $\Delta log(M)$ from individual methods among all the galaxies).  Results are listed in Table 4. The outlier fraction is defined as the ratio of the number of galaxies with $|\Delta log(M)| > 0.5$ to the total number of galaxies while the bias factor is defined as $mean[\Delta log(M)]$.   Overall, there is good agreement between the estimated masses from different methods and the input mass. The {\it rms} values range from $0.141$\,dex (13.J) to $0.241$\,dex (11.H). 
The highest {\it rms} values and outlier fractions are for methods 7.D, 8.E and 11.H. Method 8.E uses the {\it MCMC} technique, which is different from what used in other methods (except for method 1.A).  Both 7.D and 11.H also show higher biases ($-0.059$ and $0.057$ respectively), contributing to the relatively higher {\it rms} scatter. All these codes have relatively low resolution $E(B-V)$ and age grids.
The lowest {\it rms} scatter
is associated with codes: 6.C, 10.G and 13.G which have a relatively higher resolution in $E(B-V)$ and age grids.

\begin{table*}
\caption{The {\it rms} scatter, bias and outlier fraction (OLF) in $\Delta log(M)$ for mock sample in TEST-1.}
\centering
\begin{tabular}{rllcll}
Code & $rms$ & $rms$            & bias$^1$ & outlier$^1$  \\
     &       & no outliers &      & fraction \\
1.A &   0.134 & 0.100 & 0.026 & 0.014\\
4.B &0.175 & 0.127 & 0.047 & 0.030\\
6.C &  0.167 & 0.124 & 0.023 & 0.021\\
7.D &  0.214 & 0.177 & -0.059 & 0.040\\
8.E & 0.228 & 0.164 & 0.010 & 0.049\\
9.F & 0.180 & 0.110 & 0.024 & 0.034 \\
10.G & 0.172 & 0.123 & 0.000 & 0.026\\
11.H & 0.241 & 0.172 & 0.057 & 0.060\\
12.I & 0.181 & 0.129 & -0.014 & 0.032\\
13.J &0.141 & 0.106 & -0.018 & 0.015\\
      &       &       &        &\\
Median &  0.142 & 0.089 & 0.010 & 0.024\\
\tableline
\end{tabular}
\tablecomments{$^1$Outlier fraction is defined as the ratio of the number of galaxies with $\Delta(log(M)) > 0.5$ to the total number of galaxies where $\Delta log(M) = log(M_{input}) - log(M_{est})$. The bias is defined as $mean[\Delta log(M)]$}

\end{table*}

Figure 3 shows changes in the {\it rms}, outlier fraction and bias as a function of the $S/N$. There is a clear 
reduction in the {\it rms} and outlier fractions with increasing $S/N$ ratio. However, the estimated bias from all the methods is found to be independent of the $S/N$ ratio, with significant reduction in the bias when the outliers are removed. This supports our earlier conclusion that some of the differences in the stellar mass measurements from different methods could be attributed to low $S/N$ ratios in the photometric data.  
\begin{figure*}
\epsscale{0.5}
\plotone{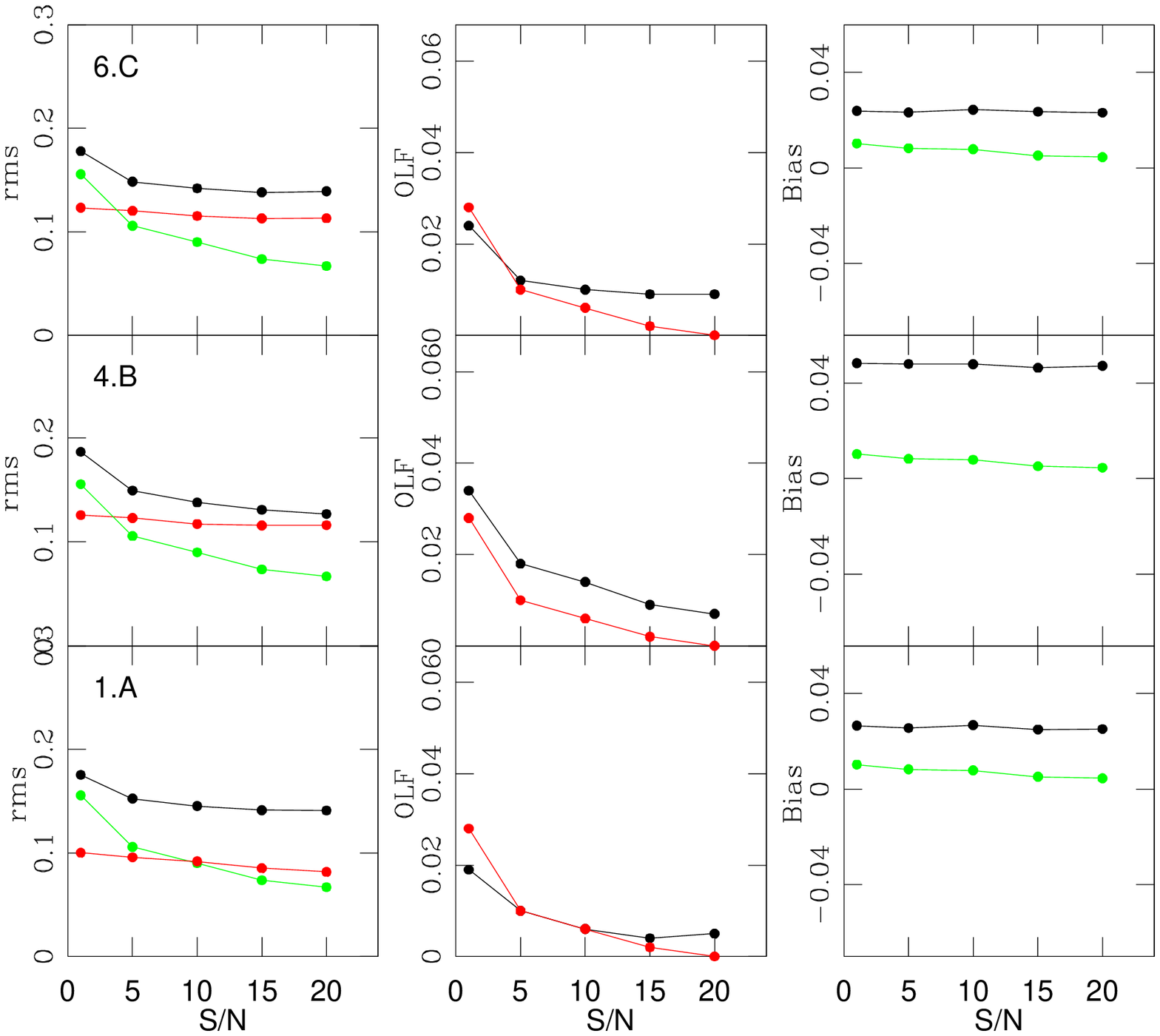}
\plotone{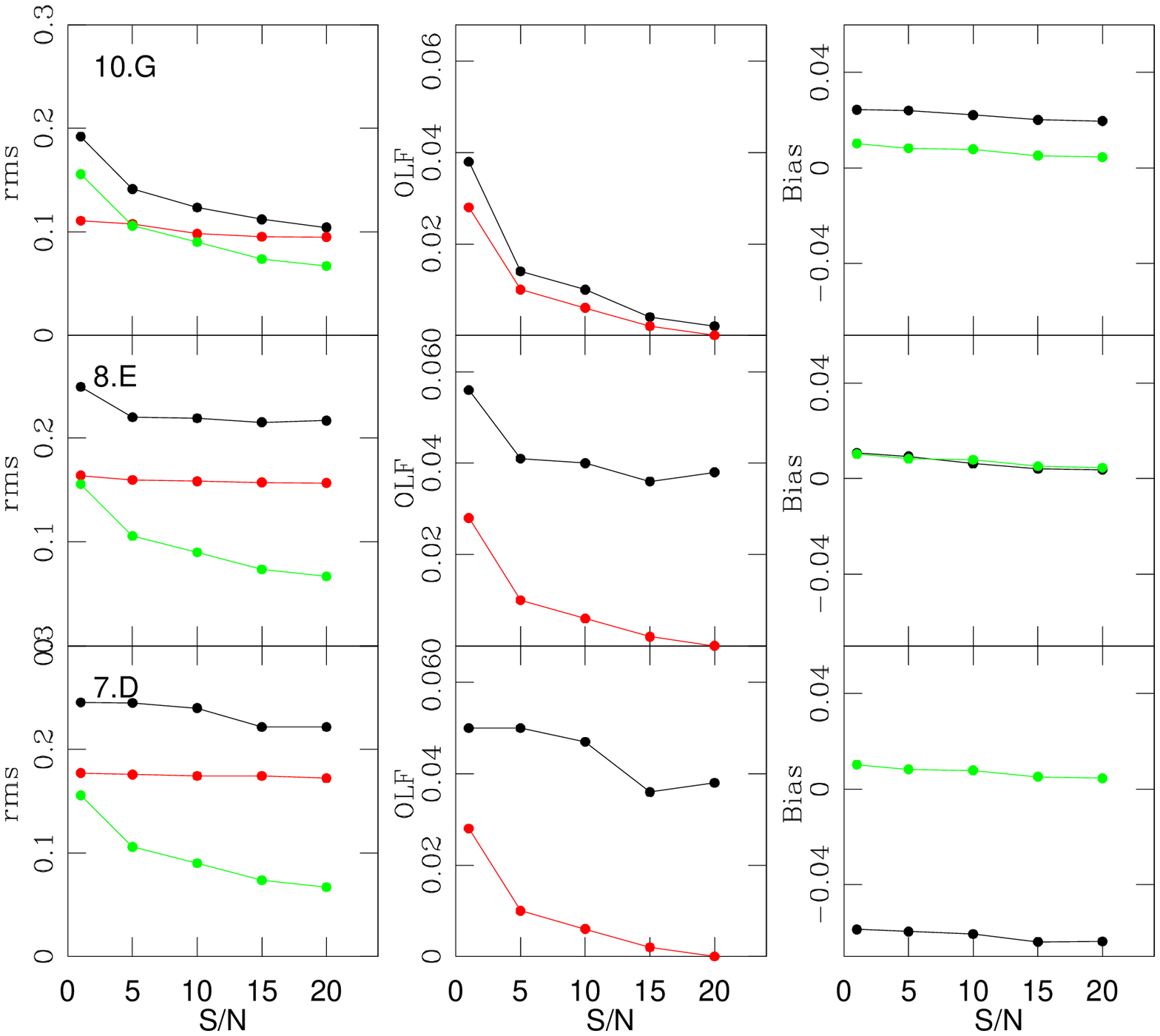}
\plotone{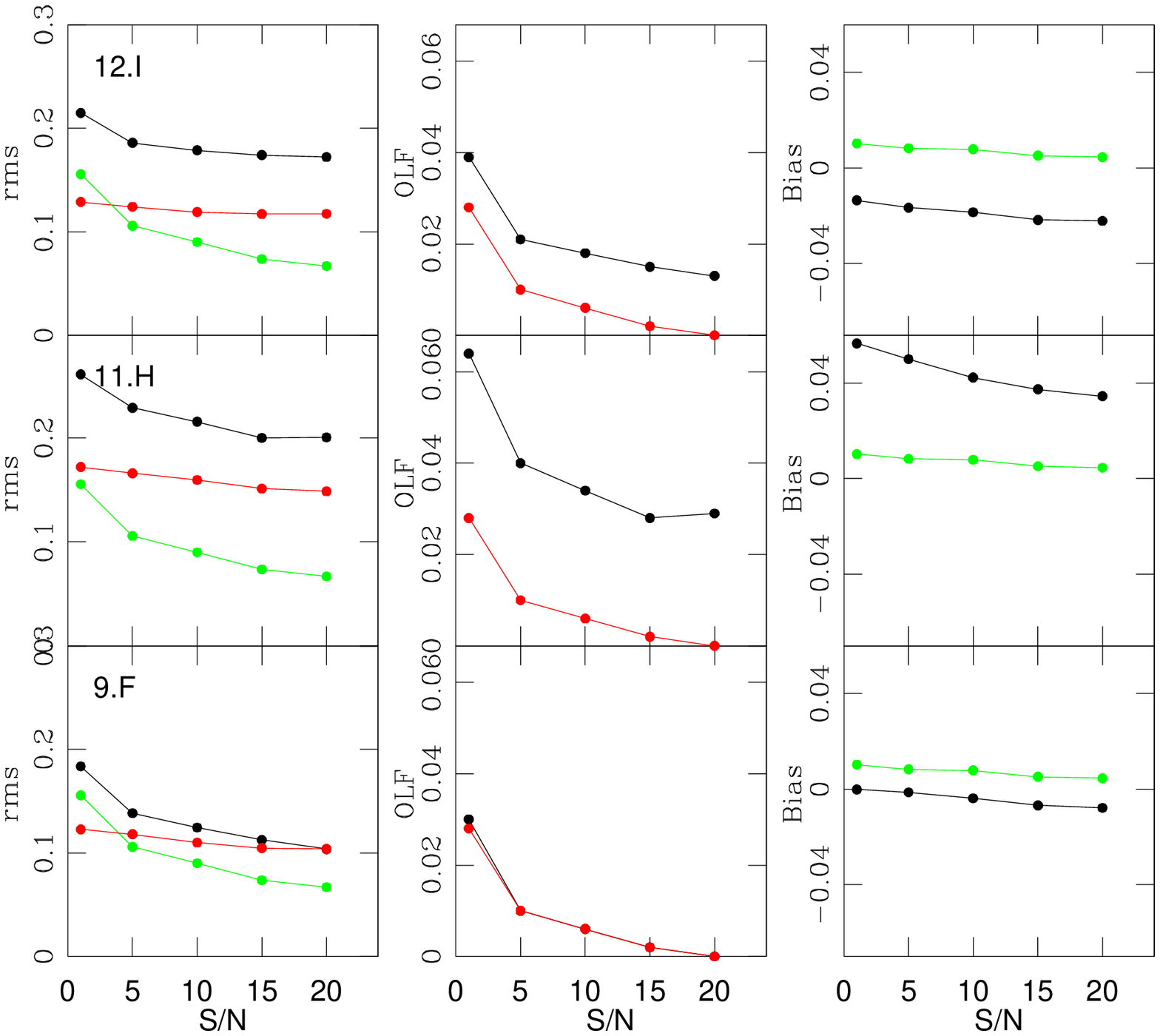}
\plotone{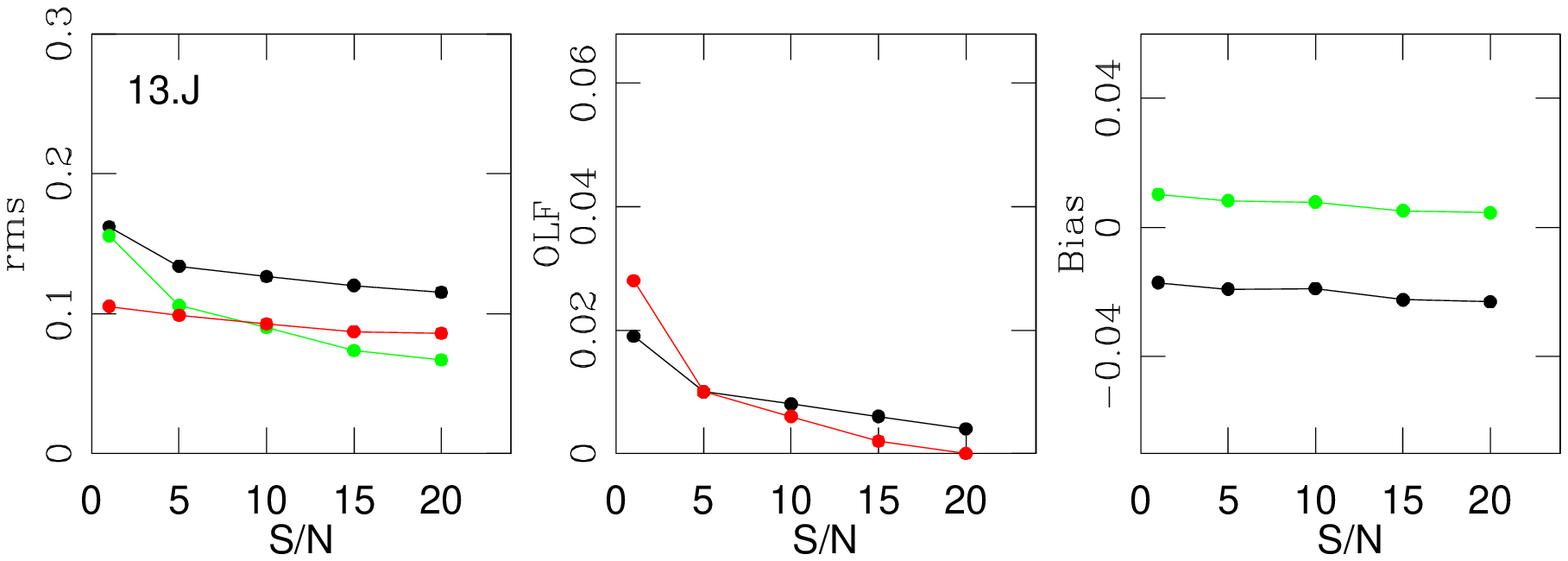}
\figcaption[TEST4_Stell_mass_comp_1n.eps]{Shows changes in the {\it rms} (left), outlier fractions (middle) and bias (right) for different methods in TEST-1, as a function of the photometric $S/N$ ratios. Different colors represent estimates for the whole sample (black line), with the outliers removed (green line) and those corresponding to the median mass (red lines.). The $S/N$ ratio is measured from the {\it F160W} band. }
%\label{fig2}}
\end{figure*}

We note that there is good agreement between the input and estimated stellar mass values when using the median of stellar masses measured for individual galaxies from different methods ($M_{med}$). The {\it rms} in $\Delta log(M_{med})$ is $0.142$\,dex where
$\Delta log(M_{med})=log(M_{input}) - log(M_{med})$. However, the median will be affected if some of the methods are biased. As shown in Table 4 and Figures 2a and 2b, for most of the methods, there is no indication of significant bias in the masses. Since the same input parameters are used for all the experiments in TEST-1, the median of the mass estimates for each galaxy measured from different methods is less affected by code-dependent uncertainties. Therefore,  
the smaller {\it rms} for the median suggests that the numerical noise (presumably due to different approximations and interpolations made in the fitting codes) is reduced by combining results from different methods. This numerical noise is small compared to other systematic uncertainties, so the gain from taking the median rather than using a single, well tested, code is likely to be useful only when values based on the same underlying set of assumptions are desired.

The {\it rms} scatter measured for the stellar masses in TEST-1 are based on galaxy samples which cover a range in luminosities and photometric $S/N$ ratios and also methods which handle these errors differently. This also contributes to the {\it rms} values in Table 4.  To quantify this, we measure the {\it rms} in $\Delta log(M)$ for {\it individual} galaxies in the mock catalog from each method separately. In this case, 
 the {\it rms} in $log \Delta(M)$, estimated for each galaxy from different methods, represents the genuine scatter {\it among} different codes/methods, only depending on the way each code/method treats the photometric error. 
Changes in the {\it rms} scatter as a function of the $S/N$ ratios for galaxies in TEST-1 is presented in Figure 4. Given that for a single galaxy the photometric errors are fixed, the relation between the {\it rms} values in $\Delta log(M)$ (corresponding to individual galaxies as measured from different codes) and the $S/N$ ratios reveals the extent to which handling the photometric errors by each code affects the resulting stellar mass estimates.  
The {\it rms} reduces with increasing the $S/N$ and asymptotes 
around $rms =0.05dex $
(at $S/N > 40$), where the photometric uncertainties become very small. This 
gives a measure of the systematic effects in the stellar mass measurement 
entirely due to the methods used (when all the rest of the parameters are 
fixed and photometric errors are negligible). 

The {\it rms} in $\Delta log(M)$, estimated for method 1.A (Table 4), is mainly due to contribution from photometric errors in stellar mass measurement and not the method or the SED templates used (because the template SEDs in TEST-1 were generated by this code and were used again to estimate the observable parameters after introducing photometric noise to the SEDs). Therefore, we estimate the intrinsic uncertainty in $\Delta log(M)$ associated with each method (in Table 4), $\sigma_{method, i}$, as $\sqrt {\sigma_i^2 - \sigma_{1.A}^2}$ where, $\sigma_i$ and $\sigma_{1.A}$ are the {\it rms} values for individual methods and for method 1.A (corresponding to photometric uncertainties) respectively. Here we assume that differences in $\sigma_i$ due to treatment of age and extinction among different codes is negligible (however, see the next section). The total uncertainty in $\Delta log (M)$ due to differences in codes/methods used is therefore, $\sigma_m ={1\over \sqrt{n}}\sqrt{\Sigma_{i=1}^n \sigma_{method,i}^2}=0.136$\,dex, where $n$ is the number of methods/codes used. Using the median {\it rms} value from all the methods (Table 4), we estimate $\sqrt{\sigma_{median}^2 - \sigma_{1.A}^2} = 0.047dex  $. This is close to $rms=0.050$\,dex we 
estimated for systematic errors from Figure 4 and is significantly smaller than the {\it rms} scatter of $0.136$\,dex due to different methods/codes used.  This confirms that the median mass (among all the methods/codes) provides the closest estimate to the ``real'' stellar mass.

\begin{figure*}
\epsscale{0.7}
\plotone{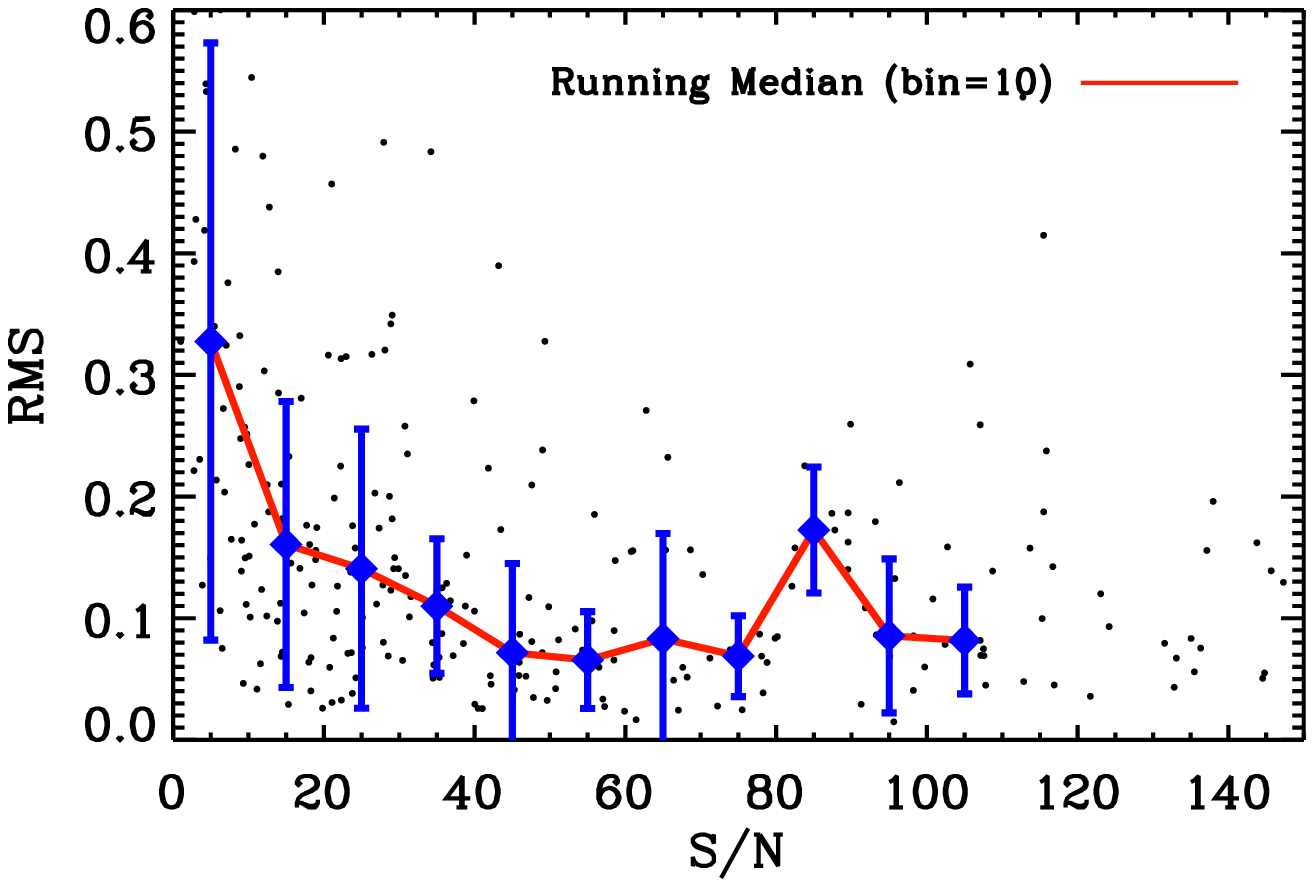}
\figcaption[mass_rms_sn.eps]{Changes in the {\it rms} with $S/N$ values. The {\it rms} is 
the scatter in stellar mass values ($\Delta log(M)$) for individual galaxies in TEST-1, based on different methods. The filled circles are the median values in $S/N$ bins with the errorbars corresponding to Poisson statistics. The scatter at a given $S/N$ represents the dispersion in the stellar mass values among different methods.
\label{fig1}}
\end{figure*}

%\begin{figure*}
%\epsscale{1.}
%\plotone{mass.rms.galaxy.mag.eps}
%\figcaption[TEST4.Stell.mass.comp.1n.eps]{Changes in rms as a function of H-bandmagnitudes for galaxies in TEST-1, showing an increased scatter towards fainter magnitudes. This is due to divergence of the stellar mass estimates at faint magnitudes between different stellar mass codes.  
%\label{fig1}}
%\end{figure*}

A Spearman Ranking Test was performed between the input, $M_{input}$, and estimated mass, $M_{est}$, for each galaxy as measured by applying different codes on the mock sample in TEST-1. Combined with the Pearson correlation coefficients from this test, as listed in Table 5, this confirms very close ranking of the stellar masses measured from different codes (i.e. the codes consistently produce the mass sequence for galaxies in the catalog). 

{\it We conclude that the uncertainties in the estimated stellar mass are dependent on the resolution of color excess ($E_{B-V}$) and age grids as well as the photometric $S/N$ ratios. We find an rms scatter of 0.135dex in $\Delta log(M)$ due to code-dependent effects. The estimated uncertainty in $log(M)$ due to photometric errors is $0.134$\,dex while using the median mass, it reduces to $0.05$\,dex. No evidence is found for bias in any of the methods in Table 4. For each galaxy, the median stellar mass between different methods gives the most accurate stellar mass with the errors mainly dominated by systematic effects.}

\begin{table*}
\caption{Estimated Spearman rank coefficients (column 1) and Pearson correlation coefficients (column 2) for TEST-1, using $M_{input}/M_\odot$ as the reference mass}
\centering
\begin{tabular}{lcc}
Code ID & $1$ & $2$\\
    &      &\\
1.A  & 0.91 & 0.92\\
4.B  & 0.98 & 0.96 \\
6.C  & 0.98 & 0.98\\
7.D  & 0.96 & 0.82\\
8.E  & 0.96 & 0.96\\
9.F  & 0.98 & 0.98\\
10.G & 0.98 & 0.98\\
11.H & 0.96 & 0.95\\
12.I & 0.96 & 0.96\\
13.J & 0.99 & 0.98\\
\end{tabular}
\tablecomments{$^1$ Spearman rank correlation coefficient
$^2$ Pearson correlation coefficient}
\end{table*}

\subsection {\bf TEST-1: The Effect of Age and Extinction on Stellar Mass Estimates}

A serious problem in stellar mass measurement for galaxies 
through SED fitting
is the interplay between the mass, age and extinction, leading to 
correlated errors among these parameters. The problem is compounded by the fact that there is no direct and 
model-independent measure for these parameters, although there is independent 
constraint on extinction with mid to far-IR dust measurements 
(eg. \citealt{reddy}), which narrows the range of allowed age 
and extinction values.  Therefore, the only way to 
constrain them is through simulations, where we know {\it apriori} the input 
values for each galaxy. The mock catalog in TEST-1 
also provides the input age and extinction for each simulated galaxy, 
providing a reference with which to compare their predicted values.   
In this section we study the uncertainty introduced to the estimated 
stellar mass values due to the interplay between age and extinction. Here age 
is defined as the time since the on-set of star formation and an exponentially declining SFH is assumed. 

Figure 5 shows the dependence of $\Delta log(M)$ on the input (expected) 
age and extinction ($E_{(B-V)}$) 
for different methods. The sample is divided into three different age and 
$E_{B-V}$ bins (corresponding to their input  
values from the simulation). On the 
$\Delta log(M)$-$E_{B-V}$ and $\Delta log(M)$-$log(age)$ plots, 
these respectively correspond to:
$7 < log(age) < 8$ (blue); $8 < log(age) < 9$ (black); 
$9 < log(age) < 10$ (red) and $0 < E_{B-V} < 0.3$
(blue); $0.3 < E_{B-V} < 0.6$ (black); $E_{B-V} > 0.6$ (red). There is 
significant scatter in $\Delta log(M)$ at a given age or extinction
interval. As expected, the number 
of old galaxies (age $ > 10^9$ yrs) with high extinction is small. In particular, the
scatter is higher for younger galaxies, independent of the 
 extinction. 

\begin{figure*}
\epsscale{0.6}
\plotone{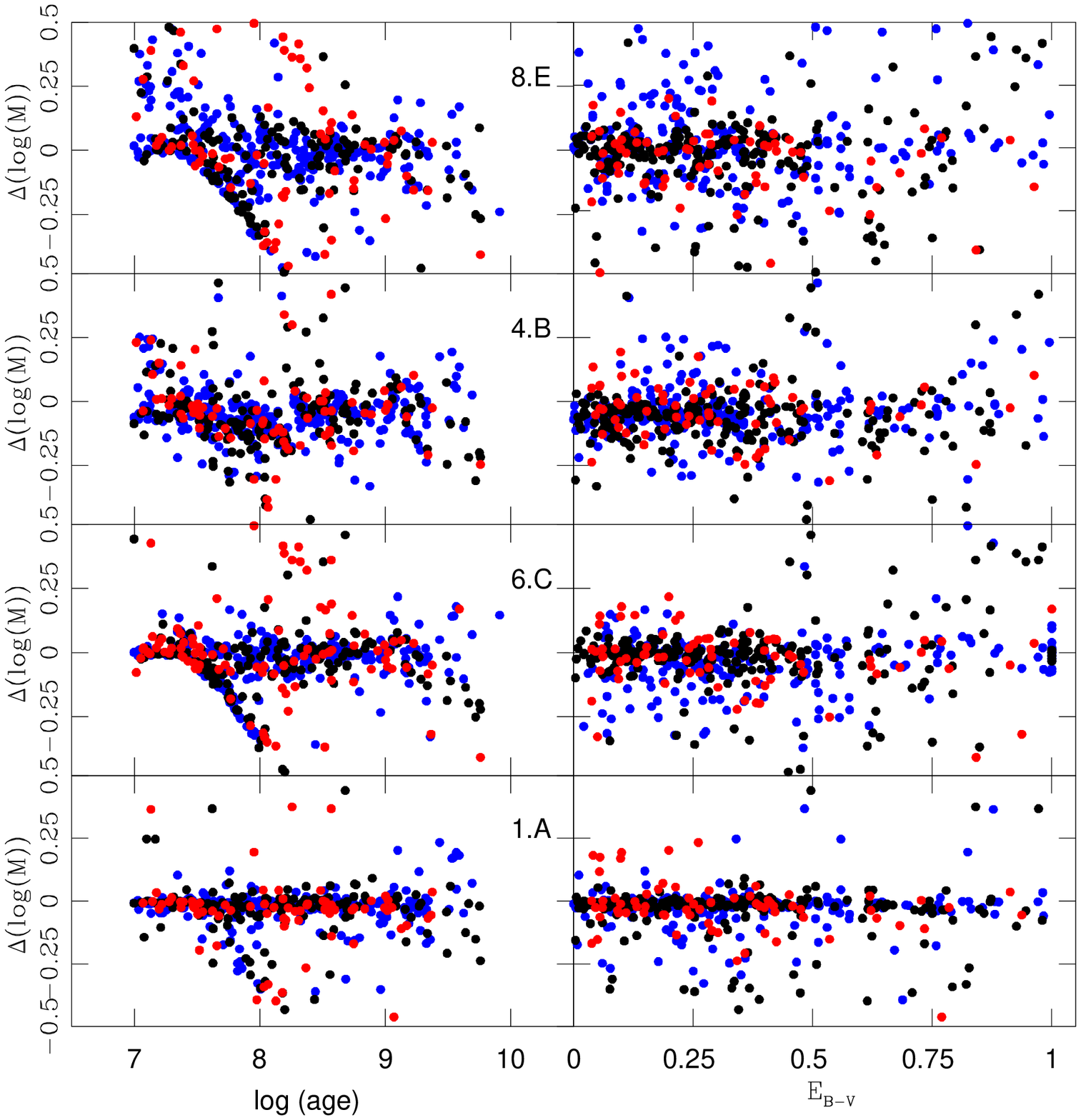}
\plotone{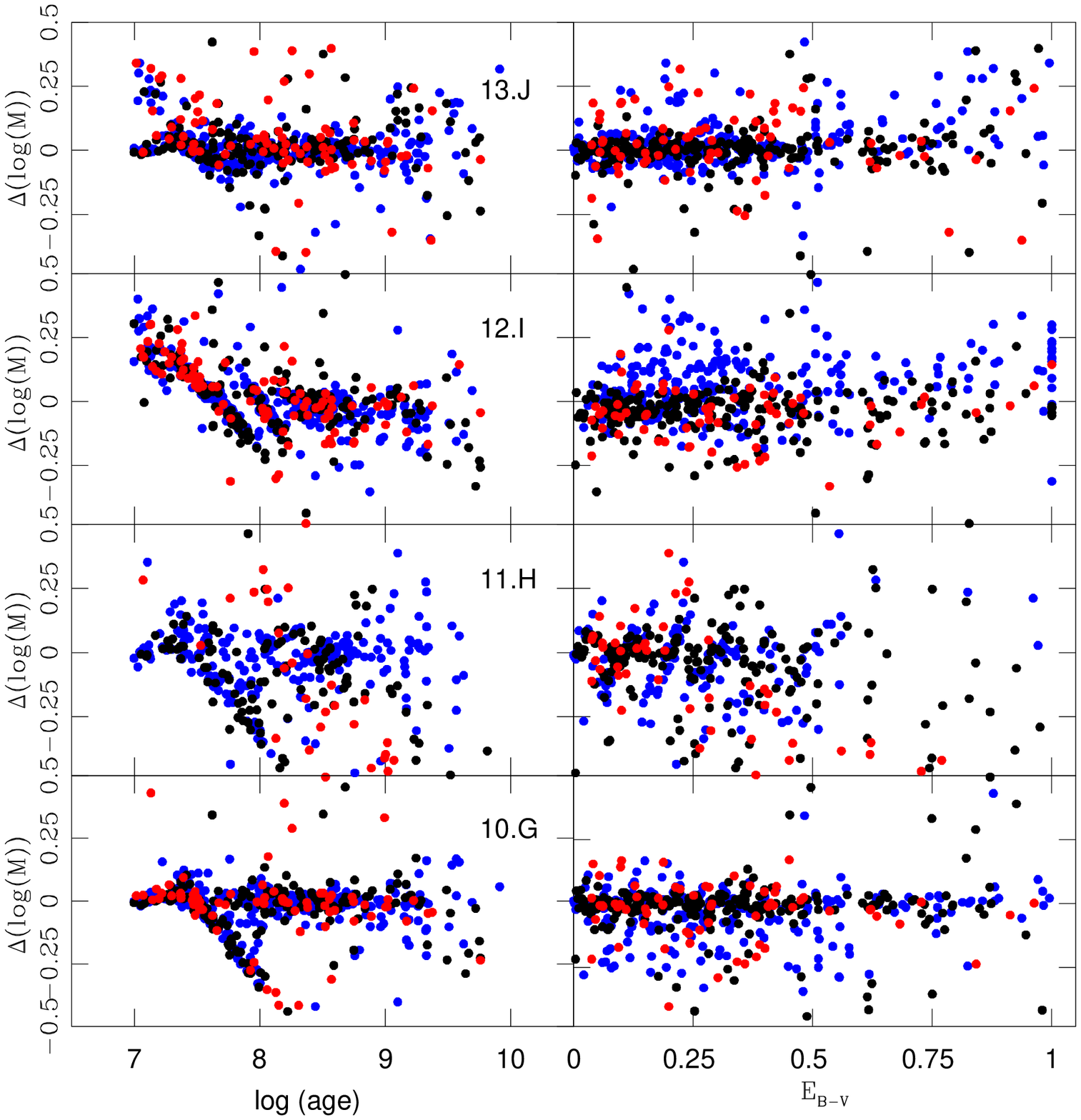}
\figcaption[mass_age_ext_1-3.eps]{The deviations in the stellar mass estimates from their input values ($\Delta log(M) = log (M_{input}) - log (M_{est})$) from TEST-1 are plotted against the input age and extinction.
Left panels: objects are divided into three different extinction intervals- 
$0 < E_{B-V} < 0.3$ (blue), $0.3 < E_{B-V} < 0.6$ (black); $E_{B-V} > 0.6$ (red).
Right panels: objects are divided into three different age intervals- 
$7 < log(age) < 8 $ (blue); $8 < log(age) < 9 $ (black); $9 < log(age) < 10 $ (red). This separates the contributions due to age, extinction and code/method to errors in the stellar mass estimates.  
\label{fig1}}
\end{figure*}

For some of the models (6.C, 8.E, 10.G, 11.H, 12.I) in Figure 5, 
we find a sequence of galaxies with ages $< 10^8 $ yrs clearly separated from 
the $\Delta log(M)=0$ line. These galaxies all have wrong stellar 
masses (i.e. large $\Delta log(M)$ values). Furthermore, this does not
depend on a particular code and SED fitting method as many of the methods show the same sequence. To find about sources of 
uncertainty in the stellar mass measurement,  
we need to understand the cause of such deviations. A large fraction of
the deviant galaxies have intermediate to high extinctions ($E(B-V) > 0.3$)
indicating they are likely dusty starburst systems. The degeneracy between 
the SED fitting parameters for these galaxies is higher as their SEDs mimic 
both the dusty starbursts and quiescent systems.   

We now explore the 
extent to which age and extinction are responsible for the sequences seen
in Figure 5 and for uncertainties in stellar mass measurement. 
Using the input age and extinction values for simulated 
galaxies in TEST-1, we compare
$\Delta log(M) $  
with both $\Delta log(age)$ and $\Delta(E_{B-V})$ (respectively defined as
$\Delta log(age)= 
log (age_{est}) - log(age_{input})$ and 
$\Delta (E_{B-V}) =  E_{(B-V),est} - E_{(B-V),input}$) for each method, with results presented in Figure 6. 
 All the experiments show a strong correlation 
between deviations in the stellar mass and age. This indicates that galaxies
with uncertain stellar mass estimates also have uncertain ages (ie. large 
$\Delta log(M)$ and $\Delta(age)$ values). In other
words, the errors in the stellar mass and age for mock galaxies, when 
constraining other parameters (as in TEST-1), are correlated.  
The observed divergence between the age estimates for younger
galaxies ($< 10^8$ years) is partly due to the varying $M/L$ ratios among these 
systems. The observed trend in Figure 6 is somewhat 
weaker on the $\Delta log(M)$ vs. $\Delta E(B-V)$ plane, 
indicating that for the range in $-0.5 < \Delta E(B-V) < 0.5$, there is a wide
range in  $\Delta log(M)$- ($-1 < \Delta log(M) < 1$), caused by differences in the estimated age 
values. By constraining galaxies only to those
with ages $ > 10^8$ years, the observed trend in Figure 6 in both extinction 
and stellar mass is reduced.

%The main results from the analysis in Figure 6 are: (a). the uncertainties in the stellar mass measurements are coupled with those in age and extinction; (b). these uncertainties are more tightly coupled with the errors in the age than with the extinction; (c). the observed correlations in Figure 6 imply that the {\it same} galaxies are outliers on both the stellar mass and age, regardless of the method used  (i.e. a galaxy with large error in stellar mass, also has large error in its age estimate). 

\begin{figure*}
\epsscale{0.5}
\plotone{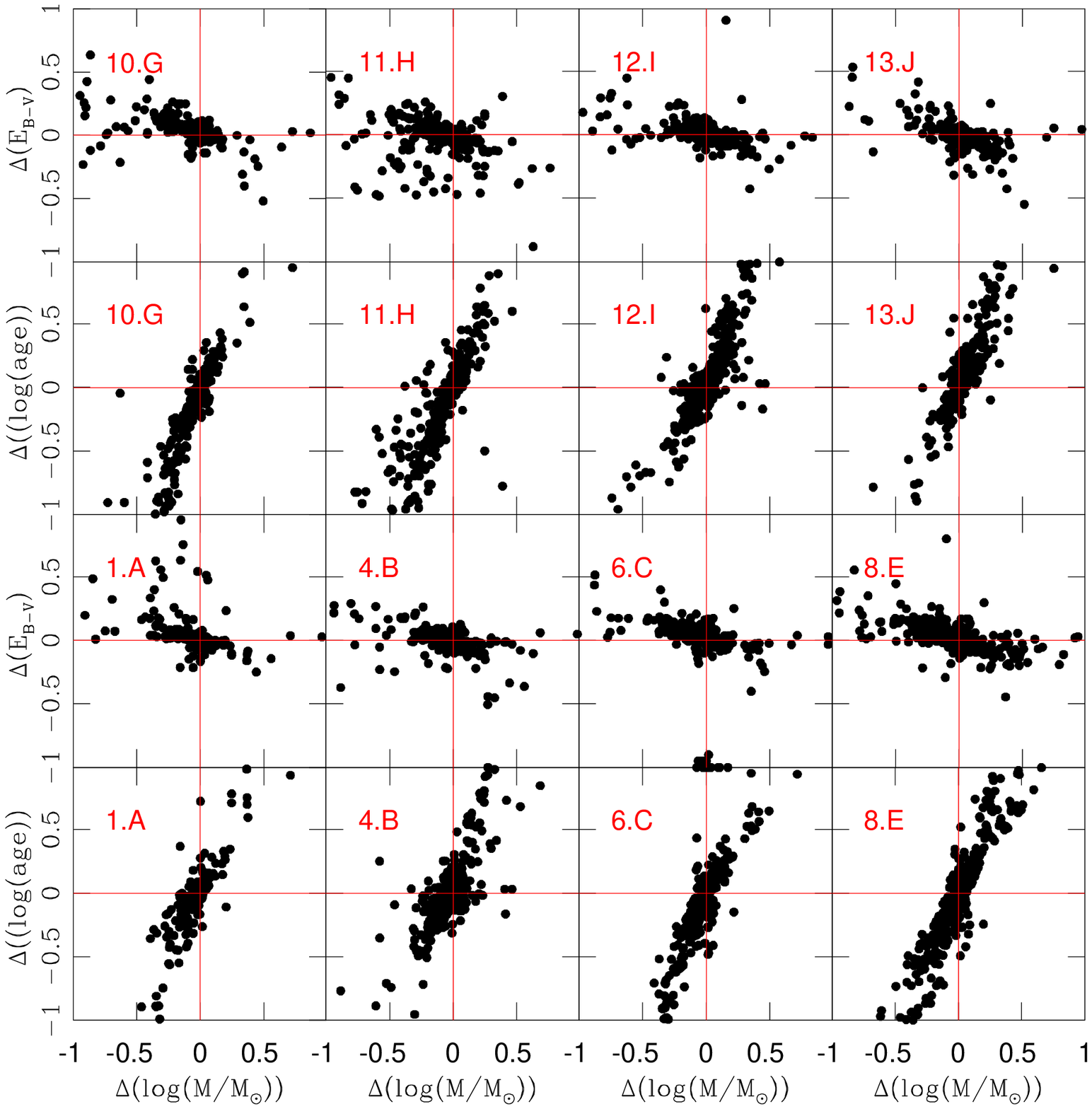}
\figcaption[dage_dext_dmass.eps]{Differences between the estimated and input 
stellar mass values ($\Delta log(M)$) are compared with deviations in age ($\Delta log(age)$) and extinction ($\Delta E(B-V)$) from TEST-1. The trend between the mass and age residuals indicates that galaxies which have uncertain mass estimates also have uncertain ages. The residuals in age range from $-1$\,dex to $1$\,dex while for the extinction they span the range $-0.5$\,dex to $0.5$\,dex.
\label{fig1}}
\end{figure*}

Given the degeneracy between age and extinction and to understand the error 
budget in the stellar mass estimates, 
we now disentangle contributions from these parameters by 
dividing the sample into three different age and extinction intervals
and estimating the {\it rms} in $\Delta log(M)$ values for each 
interval. The result is a covariance 
 matrix representing the error budget where the rows and columns are 
the age and extinction respectively,  
with the matrix elements being the {\it rms} values in $ \Delta log(M)$ ie.   
the stellar mass within 
a given age-extinction grid. As in Figure 5, the sample is divided into age and extinction intervals: 
$10^7 < age < 10^8$; $10^8 < age < 10^9$; $10^9 < age < 10^{10}$ 
years and  $E_{B-V} < 0.3$; $0.3 < E_{B-V} < 0.6$; 
$E_{B-V} > 0.6$. The error budget matrices corresponding to each of the methods 
are presented in Table 6.  
For any given method, the elements of the matrix correspond to 
the {\it rms} scatter in $\Delta log(M)$ for a given age and 
extinction.  
Using these error budget matrices, we separate relative contributions due to  
method, age and extinction to observed uncertainties in the stellar mass.  

Overall, the methods agree well per age-extinction
grid. Also, 
for any given method, the {\it rms} scatter in $\Delta log(M)$ (Table 6)
increases for redder ($E(B-V) > 0.6$) and older (age $> 10^9$ years) galaxies. 
The total error budget matrix (the overall uncertainty in $\Delta log(M)$ for different age and $E(B-V)$) from all methods combined, is estimated as 
$\sigma_{ij} = {1\over\sqrt{ n}}\sqrt{\Sigma_{k=1}^{k=n}\sigma_{ij,k}^2}  $, 
where $\sigma_{ij,k}$ are the matrix elements at any given age, $i$, and extinction, $j$, grid corresponding to the method, $k$.  The total error budget matrix is also given in Table 6. 

The {\it rms} scatter in $\Delta log (M)$ from method 1.A is likely  
dominated by photometric errors.  
Therefore, the error matrix associated with this method in Table 6 
provides a lower
limit to uncertainties in the stellar mass measurement (for any  
age/extinction combination) caused by photometric errors.  

{\it In conclusion, we find that uncertainties in stellar mass measurement are coupled with those in age and extinction, being more tightly coupled with errors in age. The same galaxies are outliers in both stellar mass and age regardless of the code used. We find serious degeneracy for galaxies with ages $ < 10^8$ years, with the rms scatter in stellar mass increasing for redder and older systems. Relative contributions due to age and extinction are disentangled by forming a covariance matrix.}

\begin{table*}
\caption{Error budget matrices for the methods in Table 3, when applied to TEST-1. The ``total'' error budget matrix represents the uncertainties for any given age-extinction grid regardless of the code/method. The uncertainties in stellar mass estimates due to photometric errors correspond to the error budget matrix associated with method 1.A}
\begin{tabular}{rllll}
 Method& $E_{B-V}$ & $< 0.3$ & $0.3-0.6$ & $>0.6$ \\
&&&&\\
& log(Age) & & & \\
 1.A&& && \\
& 7.5&	0.074&	0.100&	0.251\\	
& 8.5&	0.076&	0.169&	0.250\\	
& 9.5&	0.167&	0.267&	0.165\\
 4.B&& &&\\
& 7.5&	0.112&	0.155&	0.242\\	
& 8.5&	0.128&	0.168&	0.310\\
& 9.5&	0.232&	0.242&	0.394\\	
6.C&& &&\\
& 7.5&	0.109&	0.120&	0.265\\	
& 8.5&	0.072&	0.168&	0.321\\	
& 9.5&	0.221&	0.229&	0.164\\	
8.E&& &&\\
& 7.5&	0.176&	0.197&	0.344\\	
& 8.5&	0.156&	0.220&	0.374\\	
& 9.5&	0.322&	0.157&	0.346\\	
10.G&& &&\\
& 7.5&	0.116&	0.122&	0.230\\	
& 8.5&	0.109&	0.173&	0.314\\	
& 9.5&	0.292&	0.234&	0.504\\	
11.H&& &&\\
& 7.5&	0.130&	0.158&	0.501\\	
& 8.5&	0.136&	0.236&	0.313\\	
& 9.5&	0.385&	0.452&	0.585\\
12.I&& &&\\
& 7.5&	0.138&	0.144&	0.250\\	
& 8.5&	0.115&	0.147&	0.237\\	
& 9.5&	0.278&	0.246&	0.440\\	
13.J&& &&\\
& 7.5&	0.087&	0.104&	0.275\\
& 8.5&	0.073&	0.133&	0.228\\
& 9.5&	0.223&	0.244&	0.281\\	
   &&&&\\
Total Error Budget&& &&\\
& 7.5&   0.121&  0.141&  0.307\\
& 8.5&   0.112&  0.180&  0.297\\
& 9.5&   0.273&  0.271&  0.387\\

\end{tabular}
\end{table*}

\section {\bf TEST-2: Effect of Free Parameters on Stellar Mass Measurement}

The tests performed in the last section were used to quantify the deviation in the estimated mass of galaxies (from their expected values) due to different methods and to disentangle the effects of age and extinction in stellar mass measurement.  Here, we explore the effect of free parameters (i.e. degeneracies in the SED fits) on the stellar mass estimates. First, we perform SED fits to the mock data, allowing all the parameters to be free (except for the IMF which is chosen to be Chabrier and the redshift, which is fixed to its input value)-(TEST-2A). Second, we 
fix all the parameters in the SED fits and repeat the analysis (TEST-2B). The participating teams estimated the stellar masses following the above prescriptions. By comparing results between TEST2-A and TEST-2B for each method, we eliminate  the code-dependent effects. The difference then reveals the 
effect of free parameters on the stellar mass estimate.   

Figures 7a and 7b compare the input and estimated stellar mass values from different methods for TEST-2A and TEST-2B respectively. The {\it rms} scatter, bias and outlier fractions are estimated and presented in Table 7. For some of the methods in TEST-2A, there is a clear bi-modality between the expected and estimated stellar mass values (eg. 1.A, 4.B and 6.C). All the methods underpredict the stellar masses at $M < 10^8 M_\odot$, with the {\it rms} values changing among the methods from $0.172$\,dex to $0.394$\,dex. Also, some of the methods show a systematic offset in the estimated stellar mass from their ``true'' values.   
In Figure 7a, we also examine the distribution of galaxies as a function of extinction, measured for individual galaxies- $E_{(B-V)}=0$ (green); $0 < E_{B-V} < 0.3 $ (blue); $ 0.3 < E_{B-V} < 0.6 $ (black); $ E_{B-V} > 0.6 $ (red). 
There are two clear sequence of galaxies on the mass comparison plots in Figure 7a (TEST2-A), separated depending on their extinction values. The sequence is particularly evident for mrthods 1.A, 4.B, 6.C and 10.G. For 1.A, there is a clear separation of galaxies depending on their extinction, with redder galaxies ($ E_{B-V} > 0.3$) having a smaller (estimated) mass. Similar effects are found for experiments 4.B and 6.C where there is a complete absence of sources with high extinction 
($ E_{B-V} > 0.6$). Also, sources with medium extinction ($0.3 < E_{B-V} < 0.6 $) are mostly associated with galaxies with higher stellar masses. This indicates a possible interplay between stellar mass and extinction when both parameters are estimated simultaneously through the SED fits.  

The observed bi-modality disappears in TEST-2B (Figure 7b) when the free parameters are fixed. However, there is a mass-dependent effect in TEST-2B where most of the methods underestimate the stellar mass for low ($M < 10^8 M_\odot $) and high ($M > 3\times 10^9 M_\odot $) mass systems.  TEST-2B confirms that 
the observed bi-modality detected in TEST-2A is likely caused by the interplay between the free parameters. The {\it rms} in $\Delta log(M)$ values between the two tests are comparable, with TEST-2B having slightly higher {\it rms} (Table 7).     
Both 1.A and 4.B have higher {\it rms} values. They use templates generated from CB07 population synthesis models (for TEST-2A), which is different from the BC03 model templates used to generate the mock catalog. Furthermore, 1.A and 4.B use constant and hybrid star formation histories (for TEST-2A) respectively, which is different from the exponentially declining model assumed for the majority of the methods here.   Once the population synthesis model used to generate the template SEDs are adopted consistently with those for the mock data (BC03), as in TEST-2B, the observed bi-modalities disappear (Figure 7b- also see section 6.2). However, for almost all the methods there is a relatively higher bias in TEST-2B compared to TEST-2A. 
    
There is a significant offset in the result for the experiment 10.G in TEST-2A, corresponding to a bias of $0.183$\,dex. This method uses templates generated from \cite{maraston2005} with a hybrid SFH (consisting of exponentially declining, constant at 0.1, 0.3 and 1 Gyrs and zero afterwards)-(Figure 7a). The offset is completely removed in TEST-2B where BC03 was adopted. The observed offset in 10.G shows the sensitivity of the results to template SEDs generated from the two population synthesis codes (BC03 {\it vs.} M05). The templates resulting from 
\cite{maraston2005} include contributions from pulsating Asymptotic Giant Branch (AGB) stars, making them 
different from the templates based on the BC03 code, which include less
contribution from these stars.  This leads to an underestimation of the
stellar mass of galaxies when including the AGB contribution in the SEDs.  The scatter in method 11.H and 13.J, based on TEST-2A, are small with no offsets observed. These methods both use a SFH and synthetic population models similar to those adopted in TEST-2A. It is clear from Figures 7a and 7b that using the median of all measured stellar masses, gives smaller {\it rms} errors when compared to the expected stellar mass. However, we note that the median stellar mass measured for TEST-2A is not meaningful since the masses from this test are based on different input parameters (i.e. population synthesis models).

\begin{figure*}
\epsscale{0.8}
\plotone{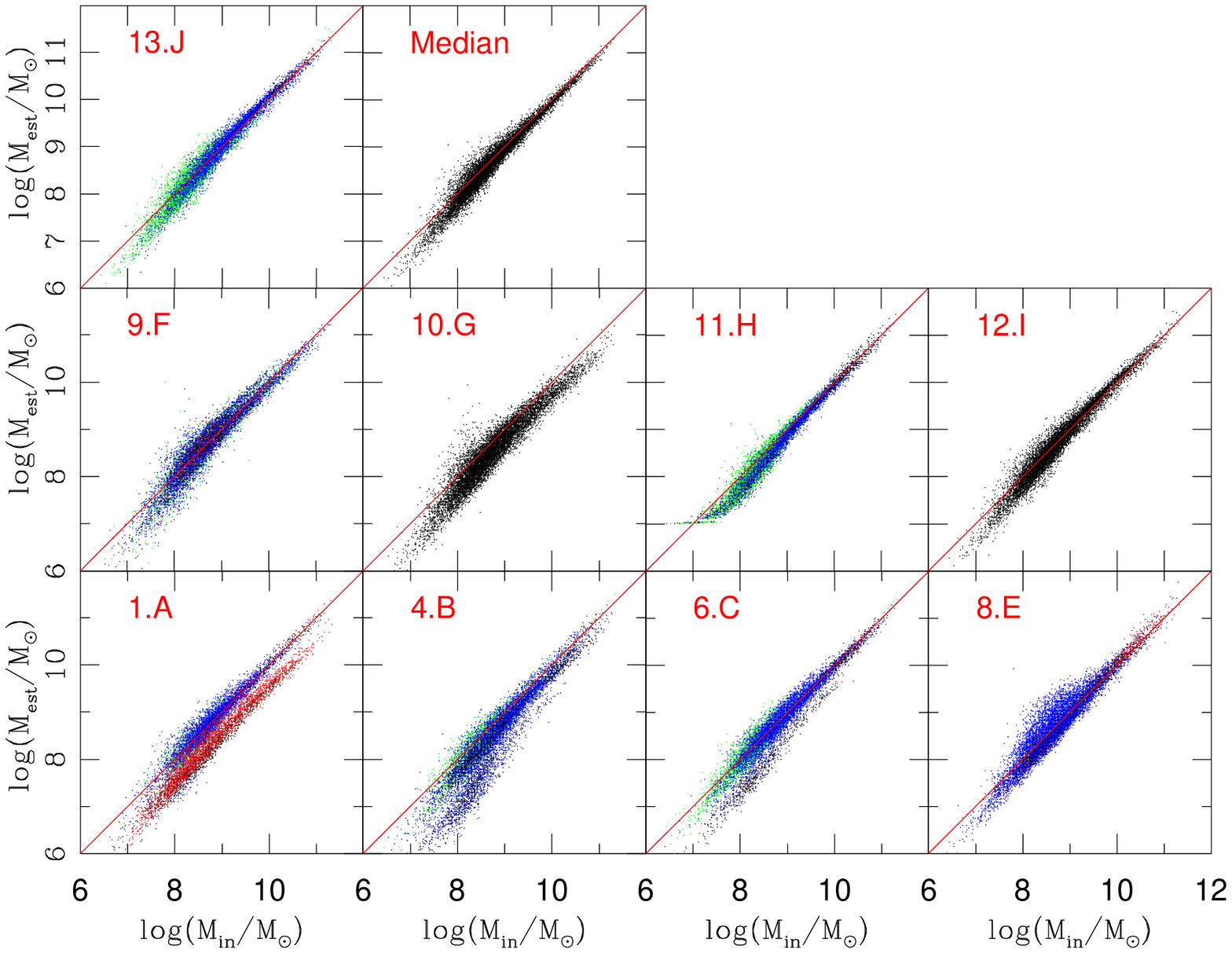}
\vspace {-3cm}
\plotone{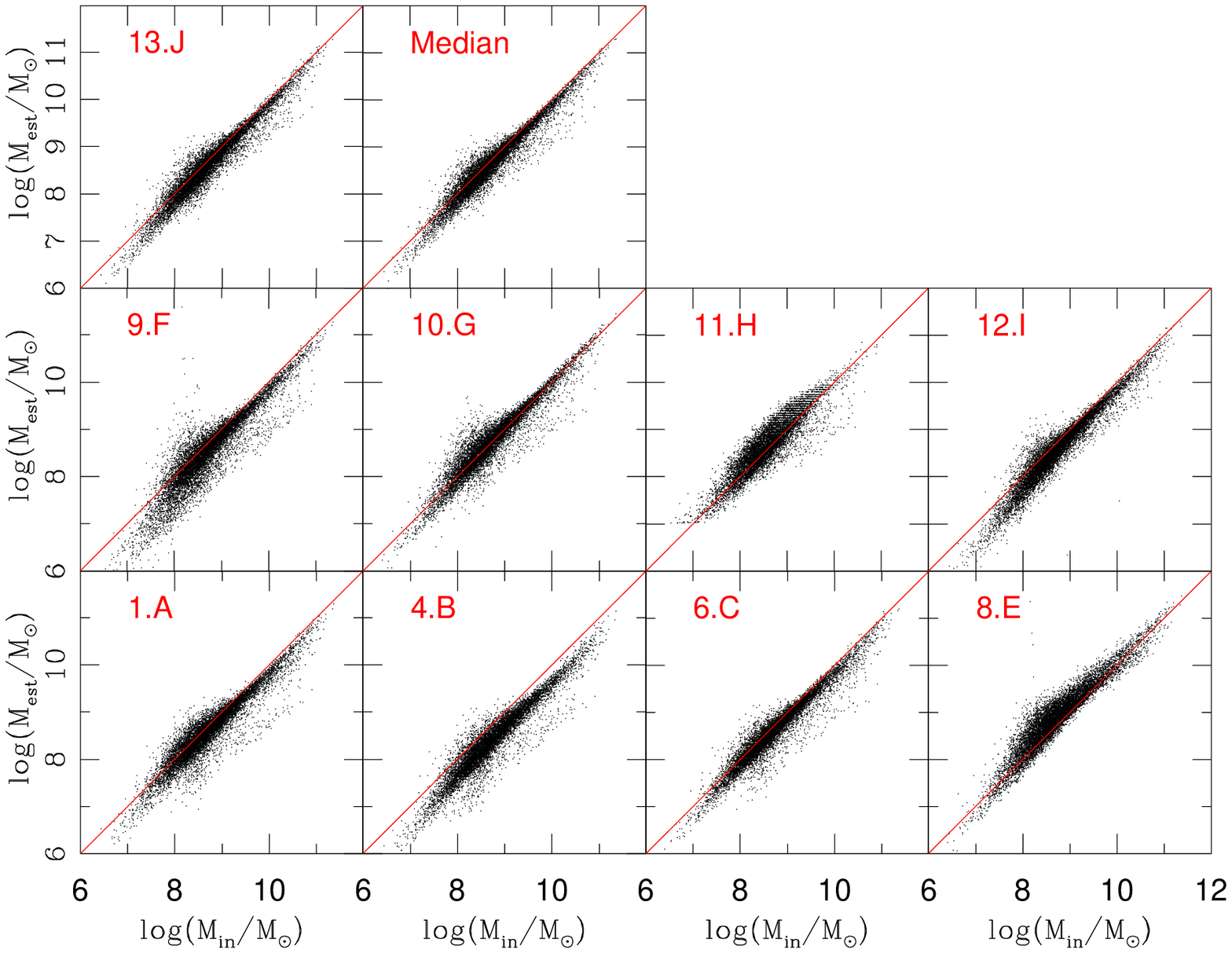}
\figcaption[TEST4_Stell_mass_comp_1n.eps]{(a)-Top: Comparison between the input and estimated stellar masses for TEST-2A. The colors correspond to extinction associated with each galaxy, as estimated from the SED fits- 
$E_{(B-V)}=0$ (green); $0 < E_{B-V} < 0.3 $ (blue); $ 0.3 < E_{B-V} < 0.6 $ (black); $ E_{B-V} > 0.6 $ (red). For the methods 10.G and 12.I no E(B-V) values are available.  There is a clear bi-modality in some cases. The red line corresponds to slope 1. Most methods underestimate the stellar masses for galaxies with $M < 10^8$ M$_\odot$. 
(b)-Bottom: Comparison between the expected and estimated stellar masses for TEST-2B. The observed bi-modality in TEST-2A largely disappears when parameters are constrained. This test is designed to study the effects of free parameters on the estimated stellar mass by leaving all the parameters free (TEST-2A) and by constraining them (TEST-2B).
}
%\label{fig2}}
\end{figure*}

\begin{table*}
\caption{The {\it rms} scatter, bias and outlier fraction (OLF) in $\Delta log(M)$ 
for TEST-2A (first line) and TEST-2B (second line). $rms[2A-2B]$ column gives
the difference (in quadrature) between $\sigma$ values for TEST-2A and TEST-2B, defined as 
$rms[2A-2B]=\sqrt{\sigma_{2A}^2 - \sigma_{2B}^2}$. This quantifies contribution from free parameters to uncertainties in the stellar mass}.
\centering
\begin{tabular}{rllclll}
Code & $rms$ & $rms$       &$rms$        & bias & outlier  \\
     &       & no outliers &[2A-2B]             &    & fraction \\
1.A &   0.328 & 0.234 &0.191&    0.087 & 0.164\\
   &   0.267 & 0.201 & &    0.096&  0.056\\
4.B & 0.394 & 0.235 &  0.085&    0.157 & 0.157\\
   & 0.403 & 0.314 &   &    0.275 & 0.161\\
6.C &  0.228 & 0.166 & 0.037&    0.030 & 0.057\\
   &   0.225& 0.177 &  &    0.098 &  0.038\\
7.D & 0.343  & 0.245 & 0.133&    0.065 & 0.133\\
   & 0.368  & 0.224 &  &    0.005 & 0.153\\
8.E & 0.230 & 0.194 &  0.165&    0.005 & 0.038\\
   & 0.283 & 0.223 &   &    -0.131 & 0.079\\
9.F & 0.219 & 0.189 &  0.264&    0.012 & 0.029\\
    &  0.343 & 0.220&  &    0.128 & 0.132\\
10.G & 0.311 & 0.261 & 0.215&    0.183 & 0.096 \\
   &  0.225& 0.170 &   &    -0.009&  0.045\\
11.H & 0.202 & 0.192 & 0.167&    0.132 & 0.014\\
    &  0.279& 0.200 &  &    0.119 & 0.082\\
12.I & 0.203 & 0.186 & 0.161&     0.066 & 0.020\\
    & 0.259&   0.222 & &     0.152 & 0.053\\
13.J &0.172 & 0.163 &  0.129&    0.026 & 0.009\\
    &  0.215&0.187 &        &    0.095 & 0.030\\
      &       &       &     &   &\\
Median &  0.175 & 0.168 & 0.110&    0.069 & 0.008\\
       &  0.203 & 0.174 &   & 0.068 & 0.028\\
\tableline
\end{tabular}
\end{table*}

For each method, we estimate the difference in quadrature between the {\it rms} 
values for TEST-2A and TEST-2B 
($rms[2A-2B]=\sqrt{\sigma_{2A}^2 - \sigma_{2B}^2}$) and present it in 
Table 7. This gives the 
contribution to the error budget in the stellar mass due to 
degeneracy in the SED fits and changes from 0.037dex (for
6.C) to 0.264dex (for 9.F). In Figure 8 we compare results between different methods, expressed by
their {\it rms} and bias (in stellar mass) estimates, as listed in Table 7. 
The smallest {\it rms} value is associated with methods 11.H, 12.I and 13.J
as well as the smallest outlier fractions. Method 13.J also has the least
bias, indicating that this method provides 
the closest mass estimates to the ``real'' values.

\begin{figure*}
\epsscale{0.8}
\plotone{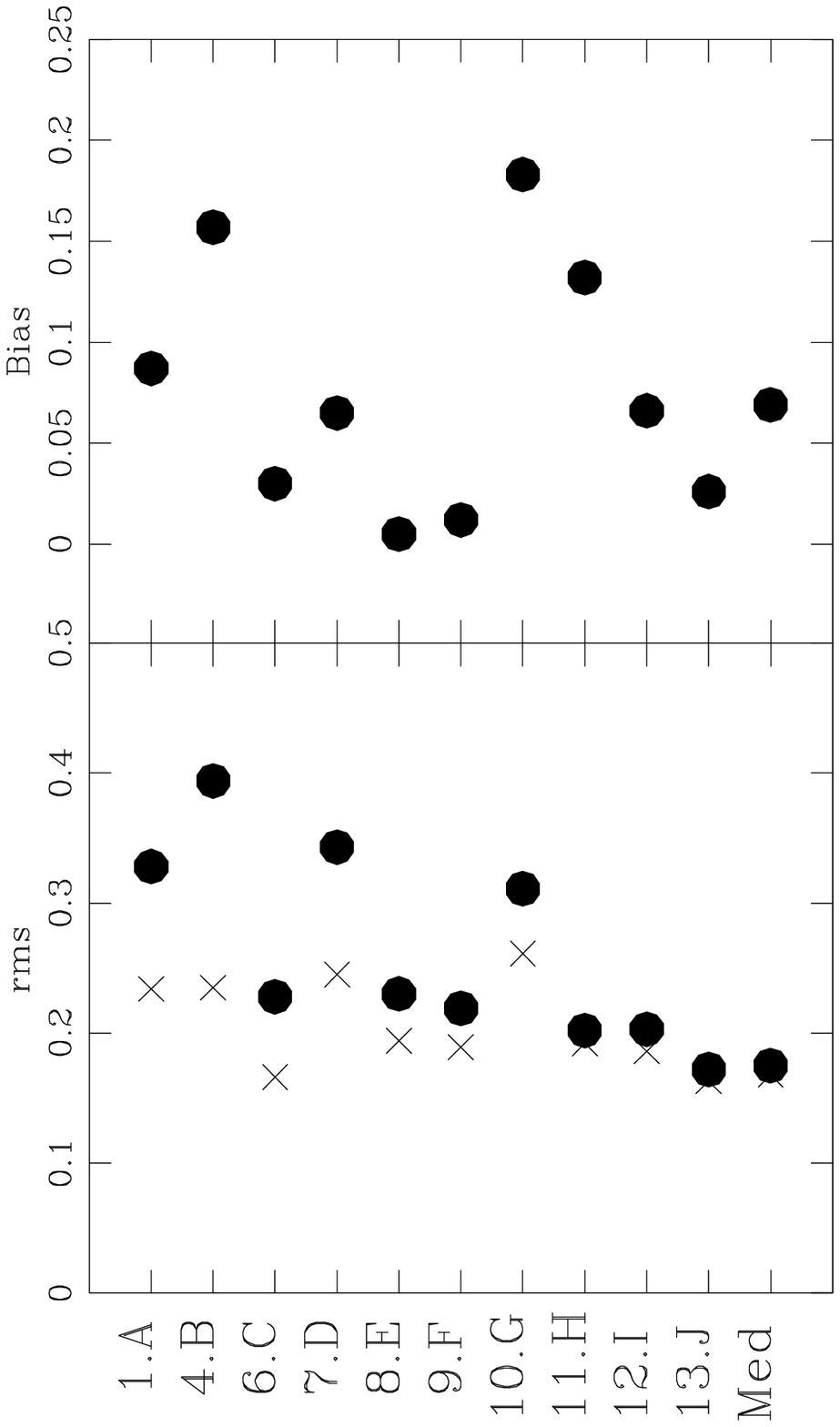}
\figcaption[TEST4_Stell_mass_comp_1n.eps]{The {\it rms} (bottom) and bias (top) in stellar masses measured from different codes are compared for TEST-2A. In the lower panel, filled
circles represent rms estimates based on all the data while crosses are
{\it rms} values with outliers excluded.
}
%\label{fig2}}
\end{figure*}

The simulations in TEST-2A are the most realistic.  
Therefore, it is instructive to further  
investigate the main sources of scatter in $\Delta log(M)$ values based on this test. In Figure 9 we show  
$\Delta log(M) $ distributions as measured from TEST-2A, plotted in H-band ({\it F160W}) 
magnitude intervals for each method separately.  It is clear that for any given
method, there is an increase in the width of the distributions from bright 
to faint magnitudes, indicating the effect of photometric $S/N$ ratios on the
stellar mass measurement. For some methods, there is
an offset from $\Delta log(M)=0 $, likely caused by systematic effects in stellar mass measurement. There are also differences in the 
distributions among different methods even over the same luminosity range. 
Figure 9 shows the median $\Delta log(M) $  has  
a narrow distribution at all luminosities, and is strongly peaked 
at $\Delta log(M)\sim 0$. This indicates that
the median of stellar masses for each galaxy, measured from all the methods in 
Table 3, successfully reproduces the input stellar mass. However, although
this is the closest simulation to real data, the results here
should be interpreted with caution as the simulations in TEST-2A are based on 
``free'' input parameters in the fit (i.e. the SFH, population synthesis templates, metallicities, age and extinction were not fixed), the effect of which could be reflected on the median stellar mass (ie. the input parameters are not the same among different methods, which could affect the estimated median values). 
Considering other independent results where the majority of the input parameters are fixed, as listed in Table 4 (and 2{\it nd} line in Table 7), one could assert that the median of the independently estimated stellar masses gives the closest agreement with the expected (input) mass.
 
%\label{fig2}}

\begin{figure*}
\epsscale{0.55}
\plotone{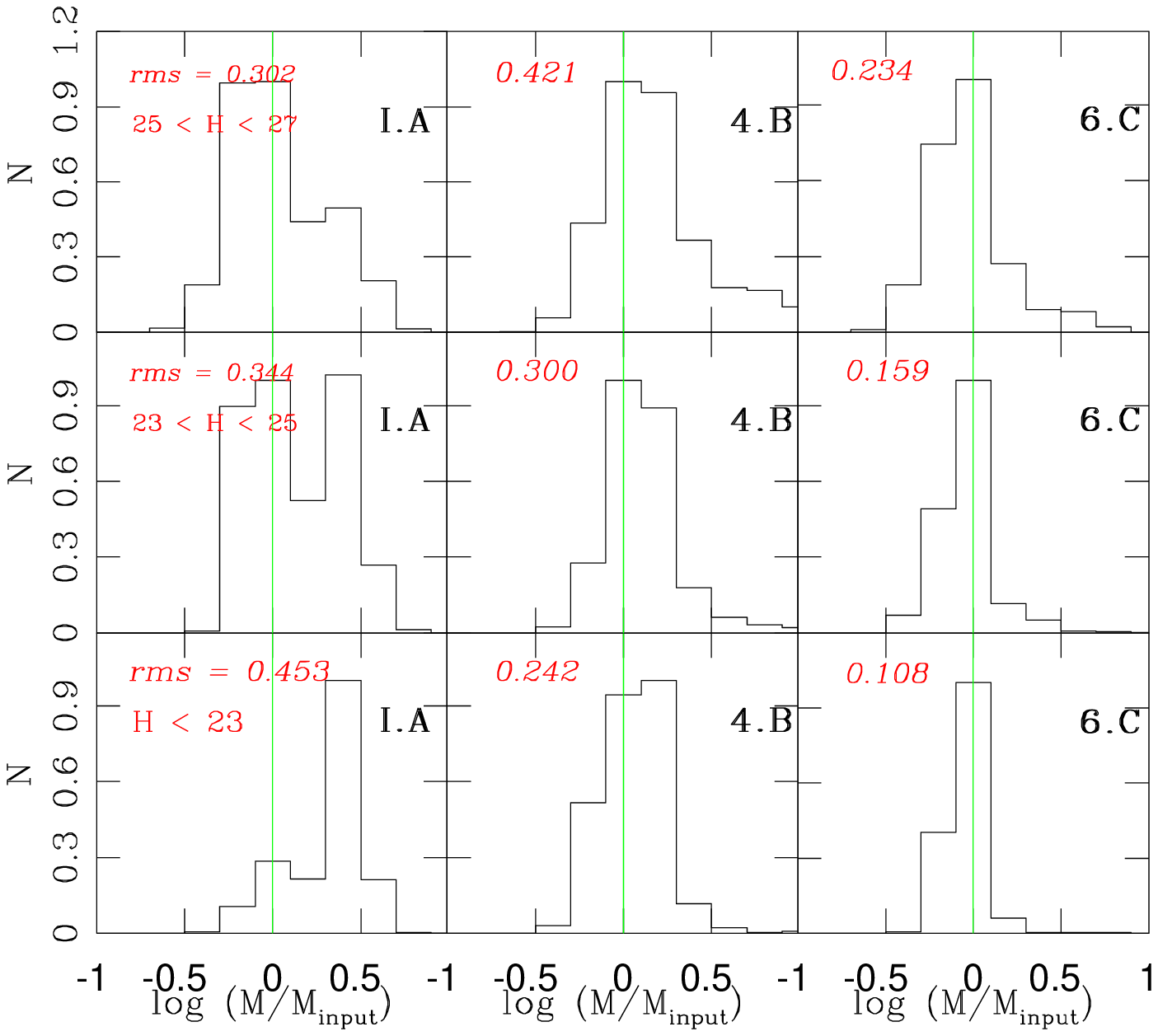}
\plotone{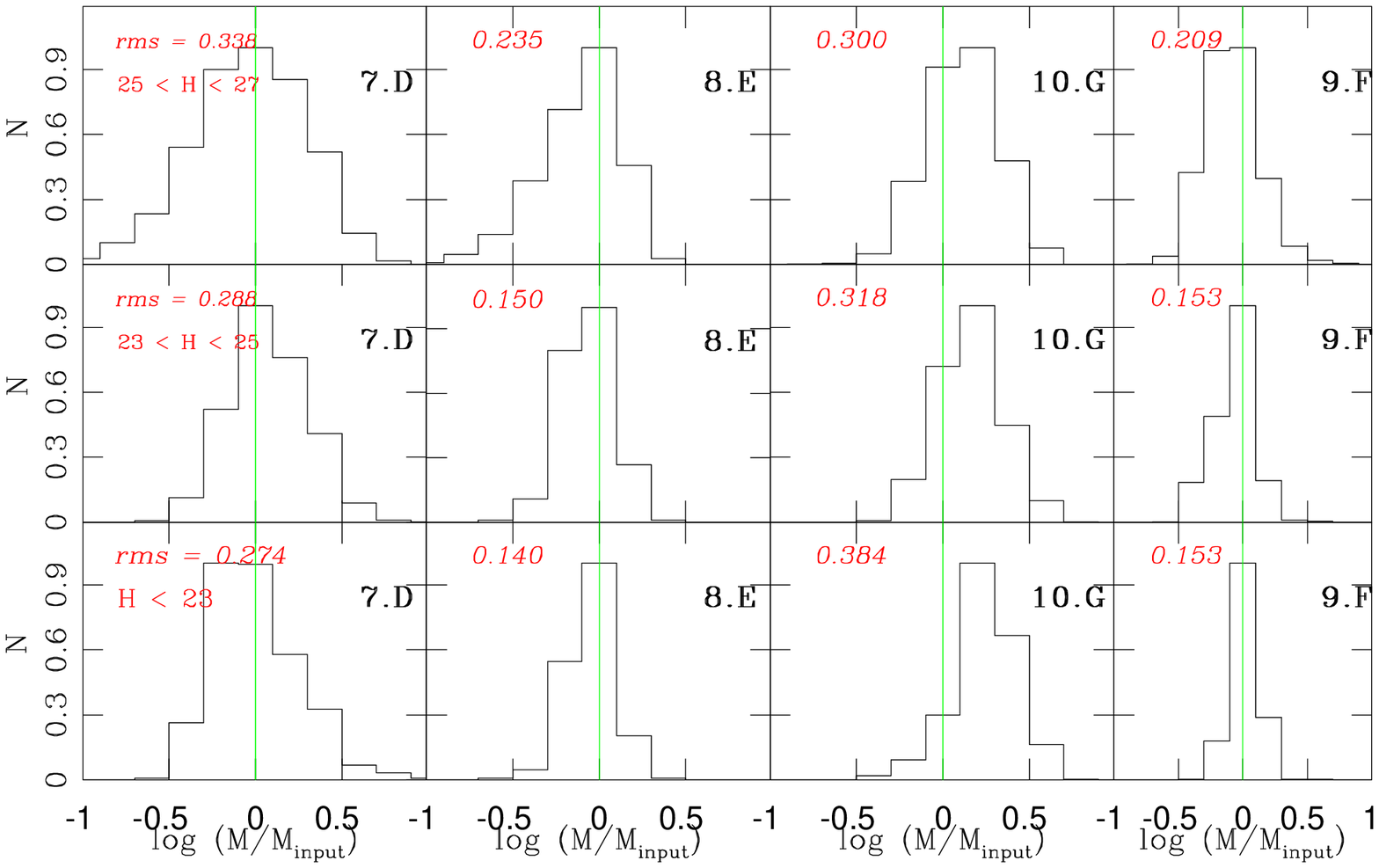}
\plotone{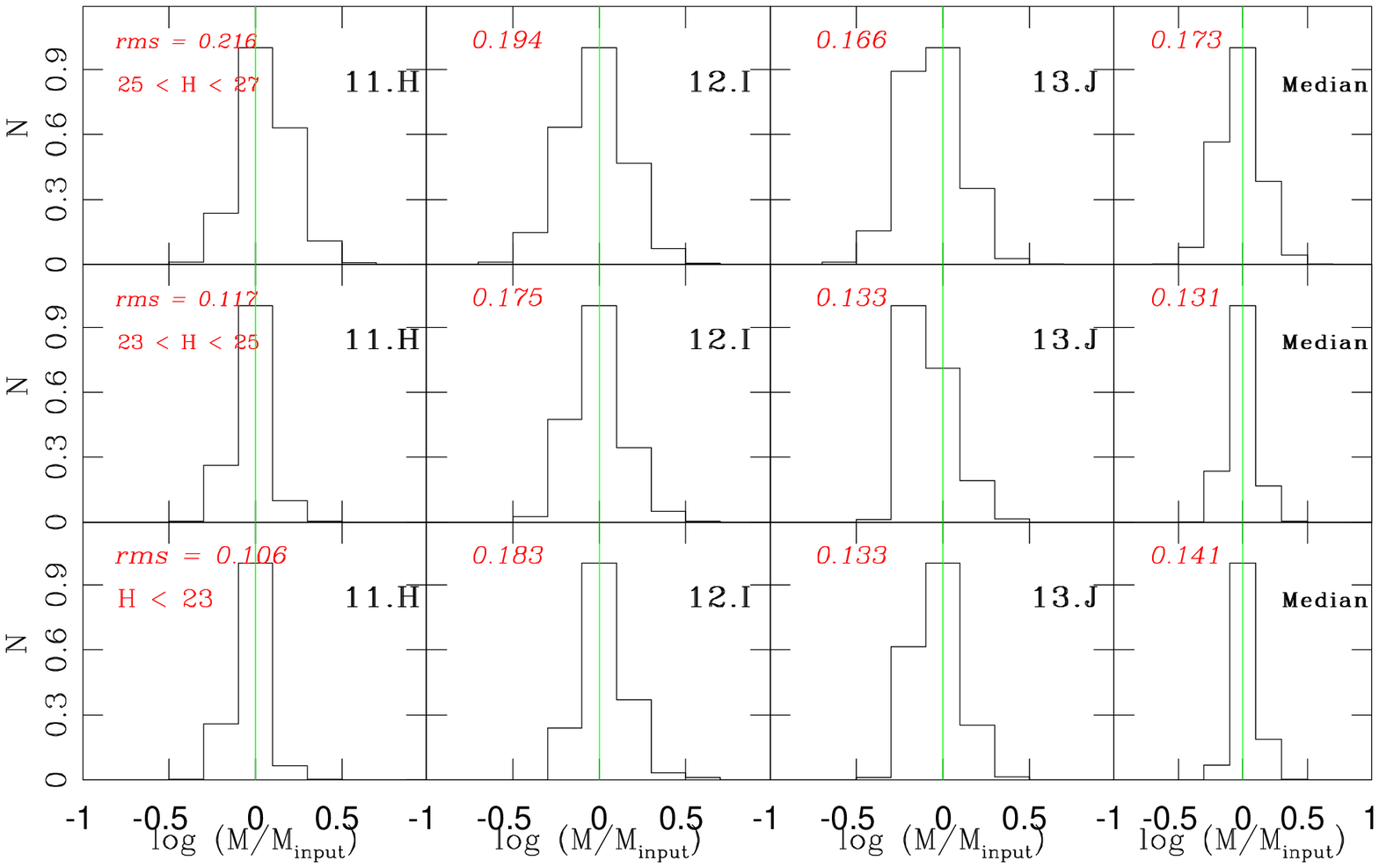}
\figcaption[TEST4_Stell_mass_comp_1n.eps]{Histogram of 
$\Delta log(M/M_{input}) $ values (from TEST-2A) in H-band magnitude intervals, 
estimated for each method separately. The {\it rms} values corresponding to
each distribution are also
shown. There is a clear increase in the width of the histogram towards 
fainter magnitudes. Also, there is a difference between the methods in
terms of the spread in $\Delta log(M/M_{input})$. The median of the stellar masses provides narrow distribtions indicating that the median of different independent methods is a stable measure of the stellar mass.
}
%\label{fig2}}
\end{figure*}

Figures 10a and 10b present the relation between $\Delta log(M)$ (from TEST-2A) 
and photometric $S/N$ ratios and redshifts respectively. For most of the 
methods, an offset is present in $\Delta log(M)$ for high $S/N$ ratios, 
indicating that the errors in stellar masses are not necessarily caused 
by photometric uncertainties. There is an increase in the scatter 
at lower $S/N$ values 
(i.e fainter galaxies). Furthermore, we find a clear trend in $\Delta log(M)$ 
as a function of redshift (Figure 10b), with some methods showing 
significantly larger scatter in $\Delta log(M)$ at a given redshift. 
At higher redshifts, all the methods overestimate the stellar masses 
while the same methods underestimate the stellar mass for lower redshift 
galaxies. This is similar to result from Figure 7, where the stellar masses 
were underestimated at $M < 10^8 M_\odot$. The observed trend in Figure 10b is likely caused by a variety of different reasons. This is likely due to changes in the functional forms assumed for SFHs at different redshifts and the diversity of this parameter within the SAMs. For example, at high-z almost all galaxies have rising SFHs while at low-z there is a mix of quenched and star-forming galaxies. Furthermore, changes in extinction among galaxies, lower photometric $S/N$ ratios for some or 
re-cycling and mass loss could contribute to the observed trend. 

\begin{figure*}
\epsscale{0.6}
\plotone{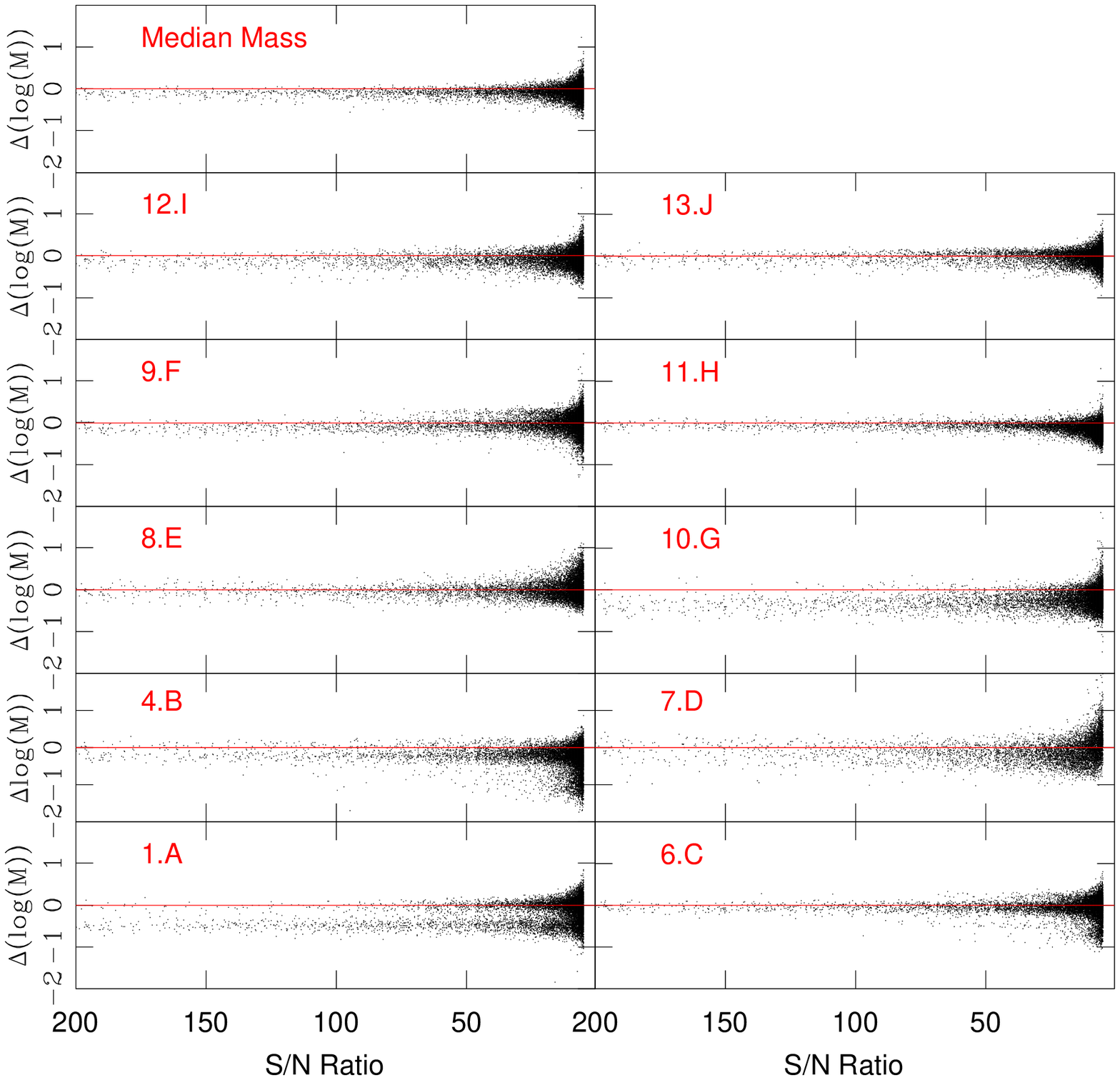}
\plotone{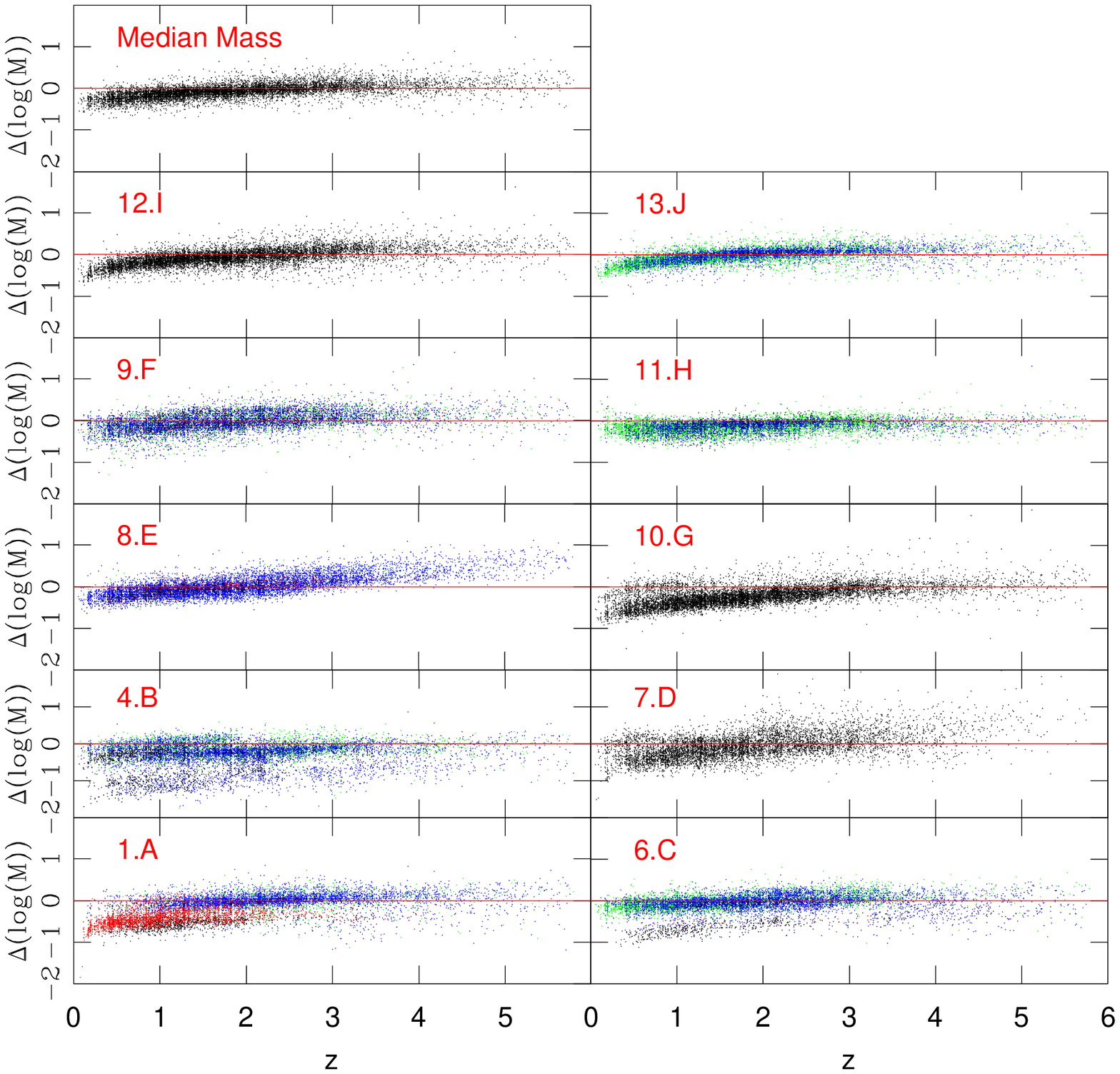}
\figcaption[TEST4_Stell_mass_comp_1n.eps]{(a)-Top Panel: $\Delta log(M)$ from TEST-2A as a function of the photometric $S/N$ ratios. 
(b)-Bottom Panel: $\Delta (log((M))$ from TEST-2A as a function of redshift. Colors correspond to different extinction values:
$E_{(B-V)}=0$ (green); $0 < E_{B-V} < 0.3 $ (blue); $ 0.3 < E_{B-V} < 0.6 $ (black); $ E_{B-V} > 0.6 $ (red).
}
%\label{fig2}}
\end{figure*}

The simulated 
templates based on the SAMs are generated from a diversity of SFHs 
(declining, increasing and constant) while the methods use simple 
prescriptions for the SFHs, causing an inconsistency in the mass
estimation process. 
To explore if extinction is responsible for the observed trend and bimodality, 
we identify galaxies in Figure 10b by their input E(B-V) values. 
For method 1.A, high extinction ($E(B-V) > 0.6$) appears to be responsible 
for some of the observed bimodality but this is not the case for other
methods.  Methods that show bimodality in Figure 10b (1.A, 4.B and
10.G) use different population synthesis models (CB07 and M05) than the
one used in the SAMs (BC03) from which the mock catalog is constructed. 
This introduces bias or additional errors 
in the mass estimate and hence, is responsible for the observed bimodality
and the trend with redshift. This is particularly the case as the difference 
in the stellar mass estimates due to differences in the population synthesis
codes (CB07, M05 and BC03) is dependent on redshift (see section 6.2).  
However, method 6.C shows serious bimodality while
using the same stellar synthesis model as the SAMs. 
Furthermore, since there is a change in the photometric 
$S/N$ ratios with redshift, it is
probable that photometric uncertainties is partly responsible for the 
observed trend in Fig 10b.This is explored by restricting the 
sample in TEST-2A to galaxies with $S/N > 10$. This does not remove the 
observed bimodality or the trend in the $\Delta(log(M))-z$ relation, 
indicating that photometric errors are not responsible for the
observed distribution of galaxies.  

The observed filters refer to different rest-frame wavelengths and different redshift intervals. Therefore, the observed 
redshift dependence could be due to the fact that more of the light from shorter wavelengths (i.e. 
UV/optical light sensitive to 
SFR, reddening and age) is contributing to the observed light from 
high-z galaxies while, the longer wavelengths (i.e. optical/infrared light sensitive to stellar mass) are dominating the light for low-z galaxies. This inherently introduces a redshift-dependent bias by weighting the fit towards different galaxy types. \cite{pforr2012} showed that high-z galaxies are easier to fit because the parameter space for degeneracies (specially age and dust) is more limited due to the small age of the Universe at those redshifts. Using rest-frame $U-V$ colors, we divided the mock catalog into the red and blue galaxies and measured their respective stellar masses in redshift intervals. No significant difference was found between $\Delta(M)$ values from these two populations. 

It is also important for the observed SEDs to cover the spectral breaks 
at any given redshift, as these breaks are essential for estimating physical parameters of galaxies. To quantify this, we identified the
redshift interval where a certain break moves in or out of the observed
wavelength range. We then measured and compared the median $\Delta (M)$ values
for the two sets, separating galaxies in redshift bins to those with/without 
the spectral features lying in that bin. 
If the observed redshift-dependence was due to this effect, we
would expect to see a difference between the median $\Delta(M)$ values in
redshift intervals. We find an average difference of only 
 $0.03$\, dex
between the ($\Delta log(M)$) values from the two samples, too small to 
be responsible for the observed trend by itself.

The effects of free parameters and in particular the population synthesis
models are examined by studying the same relations using the data in TEST-2B, 
where all the teams used templates from BC03 (similar to the ones from which 
the mock catalogs are generated), zero extinction was assumed 
and the free parameters were all fixed (Table 1). 
The results are presented in Figures 11a and 11b. 
 The bi-modality observed for TEST-2A disappears however, the trend with redshift is still present. 

As mentioned above, a possible cause of the observed trend in Figures 10b and 11b is different treatment of recycling and mass loss in the SAMs compared to the fitted models. The mock catalog here is generated using SAMs, which predict the multi-band photometry based on the BC03 model in the same way as the SED fitting codes, but predict stellar mass using the instantaneous recycling approximation, which does not accurately take into account the stellar mass loss as a function of time.  
In this scenario, the stellar mass is underpredicted at an early epoch after the stellar mass is formed, and overpredicted at a later epoch when the real stellar mass loss exceeds the adopted return fraction. 
 We estimate that the change in the stellar mass due to instantaneous recycling is around $0.04$\,dex in $\Delta log(M) $, with a clear trend with redshift. The 
expected trend due to recycling and mass loss is shown in Figure 11b (green boxes), indicating that it only plays a minor role in explaining the observed trend.  The conclusion is that although none of the effects, described above, could individually explain the observed trends in Figs 10b and 11b, the combined contribution from the individual effects, could fully explain it.

\begin{figure*}
\epsscale{0.6}
\plotone{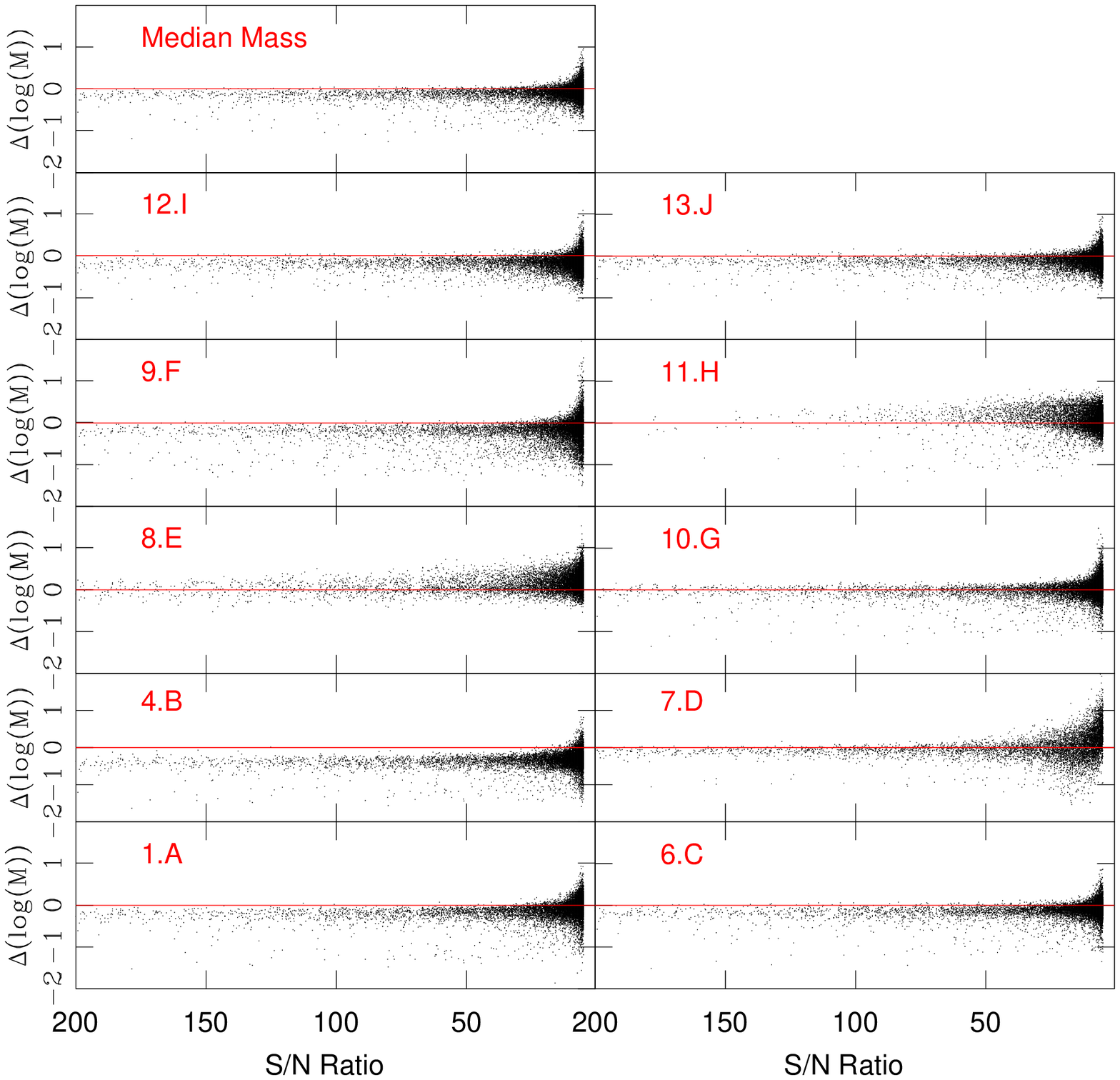}
\plotone{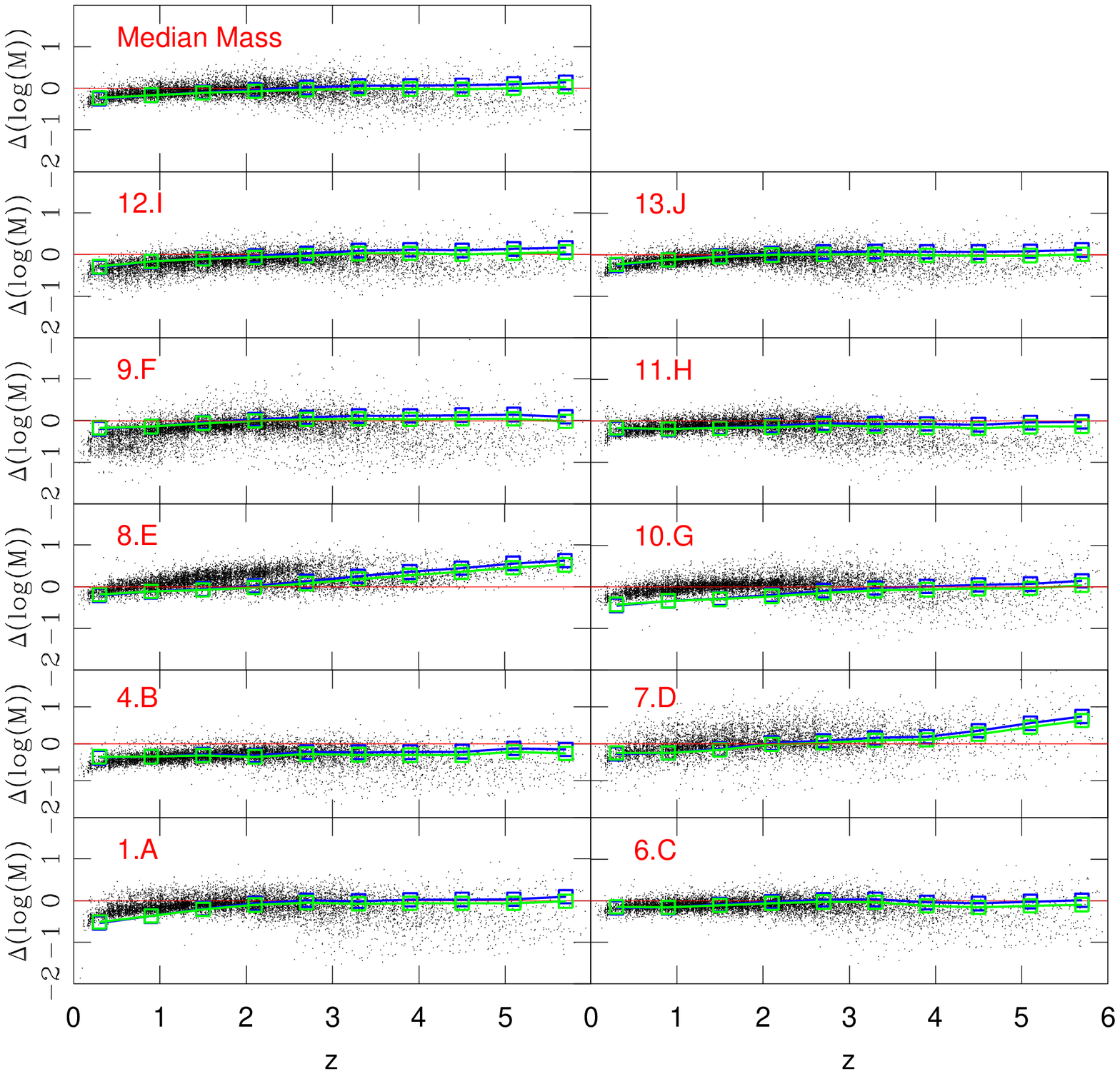}
\figcaption[TEST4_Stell_mass_comp_1n.eps]{(a)-Top Panel: $\Delta(log(M)$ from TEST-2B as a function of the photometric $S/N$ ratios. 
(b)-Bottom Panel: $\Delta log(M)$ from TEST-2B is plotted as a function of redshift for TEST-2B. The blue boxes indicate the mean $\Delta log(M)$ values in redshift bins based on TEST2-A (Fig 10b). This shows changes in $\Delta log (M)$  per redshift interval between TEST-2A and TEST-2B. The green boxes show the expected trend due to re-cycling and mass loss (see the text for details). The green line connects the green boxes.
}
%\label{fig2}}
\end{figure*}

In Figure 12 we compare the {\it rms}, bias and outlier fractions in $\Delta log(M)$ between TEST-2A and TEST-2B. The {\it rms} values, even after removing the outliers, are still high ($\sim 0.2$\,dex). The green point in Fig 12 corresponds to the median mass. The observed scatter in the bias and outlier fractions between the two tests are identical, with methods that have higher {\it rms} scatter
in TEST-2A also have high values in TEST-2B.  

\begin{figure*}
\epsscale{0.5}
\plotone{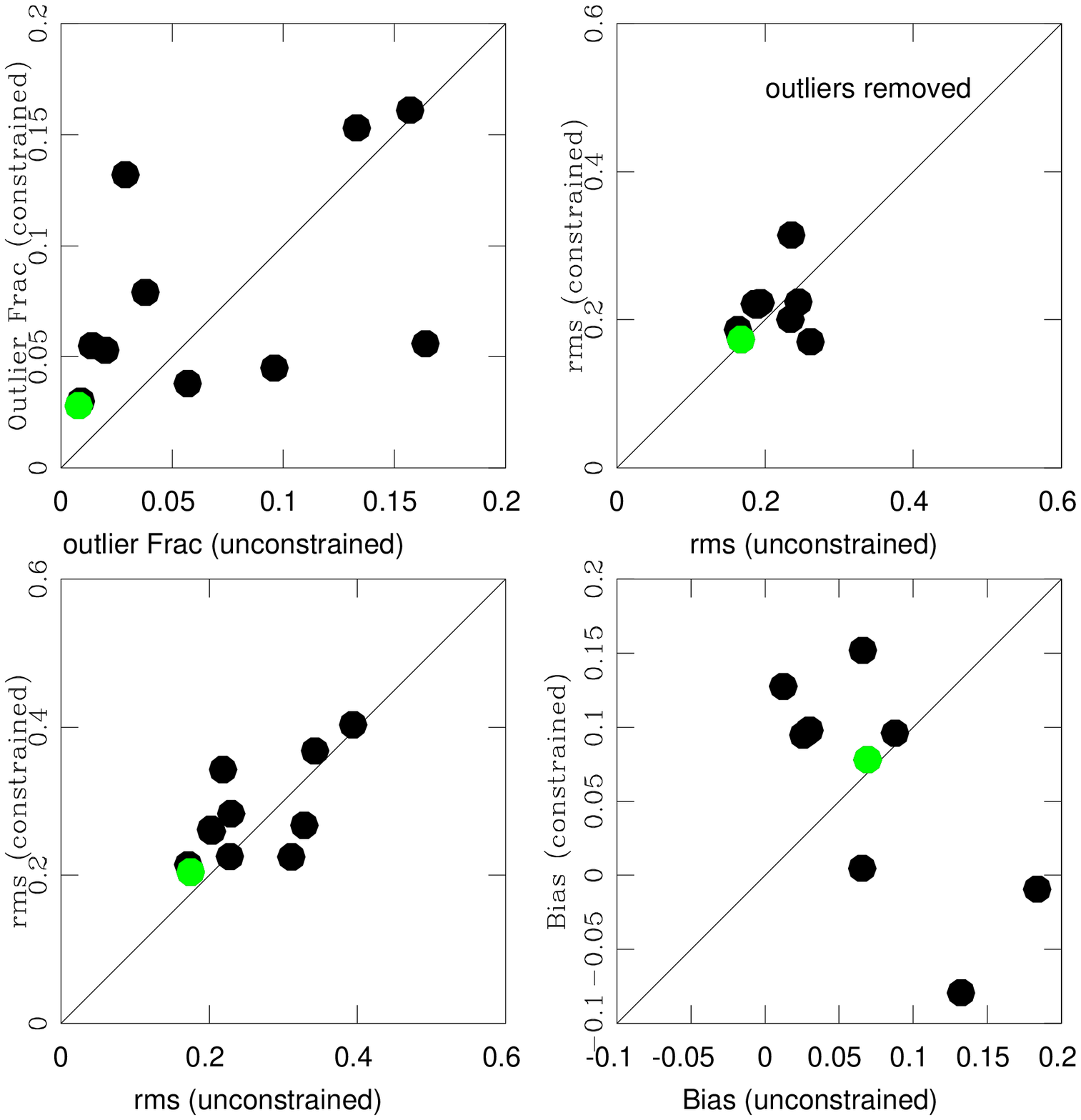}
\figcaption[TEST4_Stell_mass_comp_1n.eps]{Comparison between the {\it rms}, bias and outlier fractions for TEST-2A (horizontal) and TEST-2B (vertical). The green point corresponds to the median mass from all measurements.}
%\label{fig2}}
\end{figure*}

To explore the effect of photometric errors, for each galaxy in TEST-2A 
simulation we estimate the {\it rms} in 
$\Delta log(M)$ between the values measured from different methods and 
plot it aganist the photometric $S/N$ ratios in Figure 13. The scatter at
any given $S/N$ indicates the {\it rms} in $\Delta log(M)$ among 
different methods.  
As expected, there is significant scatter at lower $S/N$ ratios,
with that decreasing towards higher values. The median rms in
$S/N$ intervals are also shown in Figure 13. For TEST-2A, the {\it rms}
distribution asymptotes around $rms\sim 0.2 $\, dex at $S/N > 20$. At these high 
$S/N$ values, the effect of photometric uncertainties on stellar mass measurement is 
negligible and all the scatter is due to systematic and code-dependent effects. 
From TEST-1 we estimated that the contribution to the total rms due to 
method/code is $0.136$\,dex. Subtracting this, in quadrature, from the total 
rms for TEST-2A gives $rms=0.146$\, dex, which is the {\it rms} scatter in stellar mass
estimate, due to the effect of the free parameters.   
We carried out a linear fit to the median values in Figure 13 and find: 
$rms = (-0.013\pm 0.023) S/N + (0.409 \pm 0.230)$. 
Using this 
fit, one could estimate the {\it rms} values in the stellar mass for any given photometric
$S/N$ ratio.

\begin{figure*}
\epsscale{0.5}
\plotone{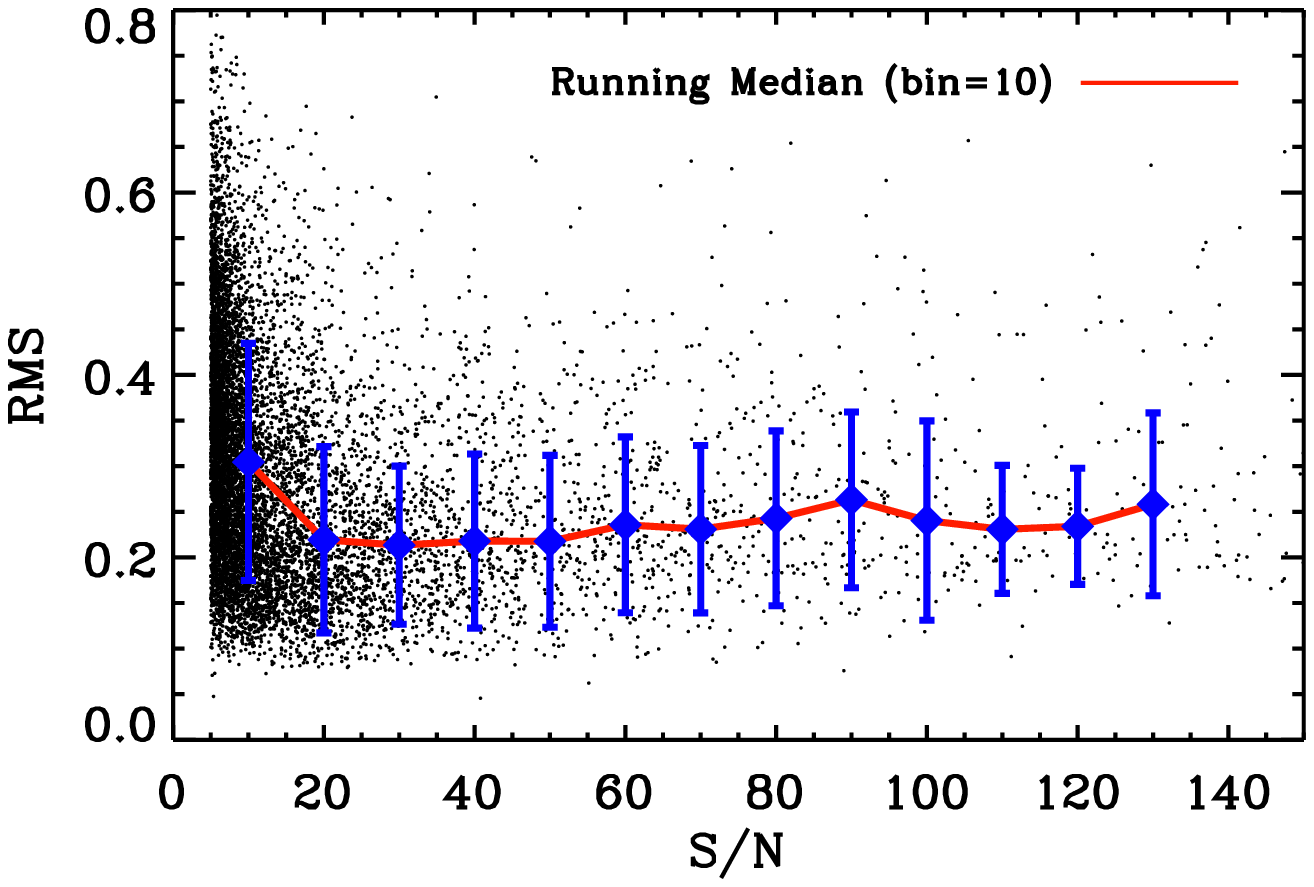}
\plotone{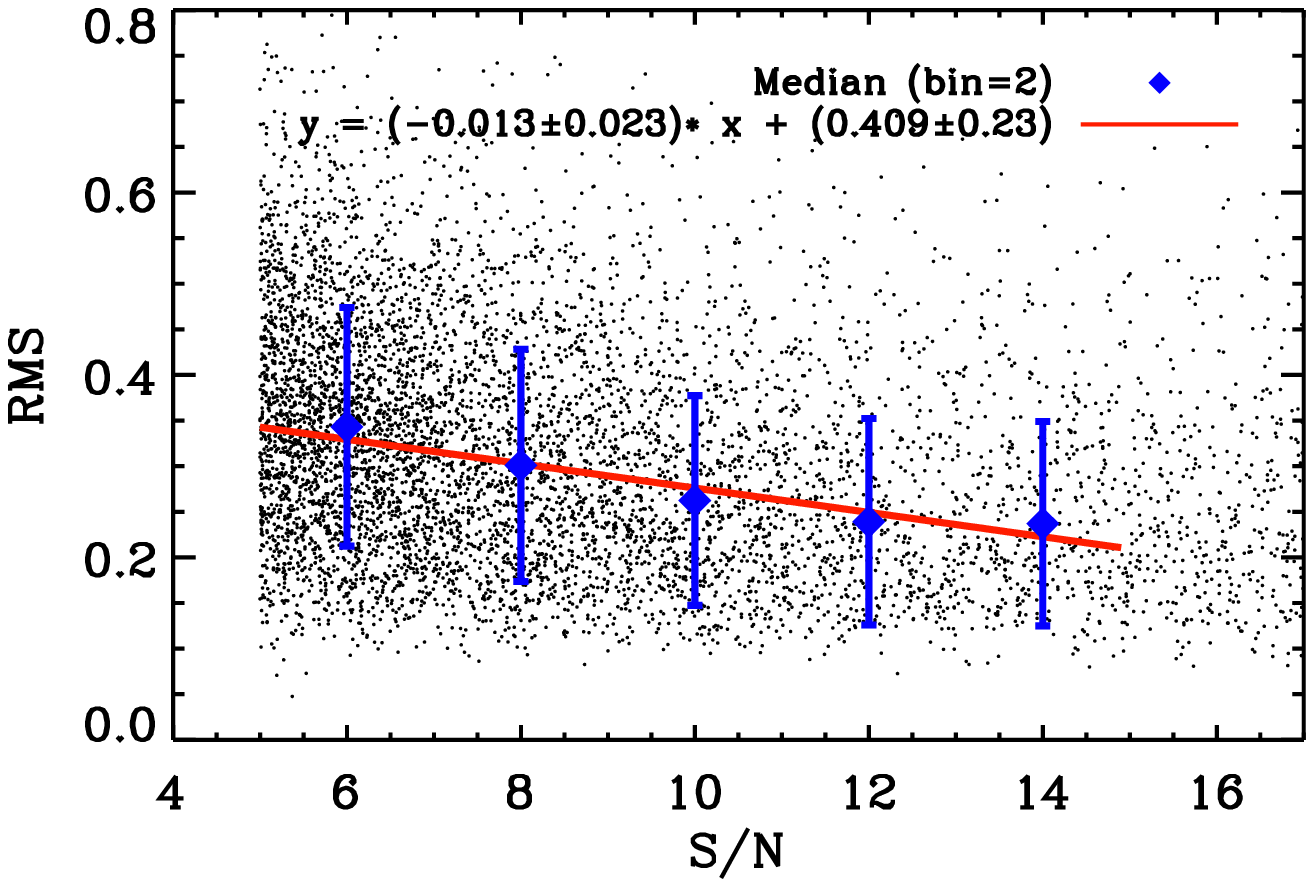}
\figcaption[TEST4_Stell_mass_comp_1n.eps]{Left panel- The photometric $S/N$ ratios are plotted {\it vs.} the {\it rms} in $\Delta log(M)$ for each galaxy, measured from different methods. Right Panel-  the same as the left panel but plotted over a limited range of
$S/N$ values. The boxes are the median $\Delta log(M)$ values 
measured in $S/N$ intervals. The line
is the least squares fit to the median points. The equation of the line can be used to estimate the {\it rms} values in stellar mass as a function of the $S/N$ ratios. the $S/N$ ratios correspond to the H-band ({\it F160W}) photometry.}

%\label{fig2}}
\end{figure*}

{\it In conclusion, using realistic simulations from TEST-2A, we find the difference between the input (expected) and estimated masses ($\Delta (M)$) to follow a distribution that broadens from bright to faint magnitudes. At a given magnitude interval, while some methods show a relatively larger scatter in $\Delta (M)$, some show a systematic offset from the $\Delta(M) = 0$ line. The observed offset in stellar mass is likely due to degenaracy between the free parameters (i.e. age and extinction). The offset is significantly reduced in TEST-2B where the input parameters are fixed. A trend was found between $\Delta(M)$ and redshift for both TEST-2A and TEST-2B. The most likely cause is the diversity of the SFHs used in the SAMs, from which the mock catalogs were constructed  (and the fact that the methods mostly use simplified SFHs).}

\section{\bf TEST-3: Uncertainties in Stellar Mass Measurement from Observed Data}

\subsection {\bf Internal Tests of Stellar Mass Measurement Methods}

In this section we study the {\it internal consistency} in stellar mass estimates between
different methods, using observed data (TEST-3). Unlike the mock catalogs, in case of the observational data we do not have prior knowledge of the expected 
stellar masses and therefore, this only provides an internal test of the consistency of mass measurements.    
Given this, we need to define a ``reference'' mass as a base to compare 
all the other masses with. Since the median mass is shown to be relatively 
unbiased (eg. Figure 9), we adopt that as the ``reference'' mass. 
We note that the median of the stellar masses based on methods
with different input parameters (TEST-3A) is not meaningful. Also, any bias
in individual mass estimates would be reflected on to the median. However, 
this is only aimed to provide a relative test between different methods
and the choice of the ``reference'' mass will not affect the results in this 
section. Furthermore, while 
a mass estimated from any other method here is equally acceptable as 
the ``reference'', it would still be susceptible to the above problems. 

In Figure 14 we compare stellar masses predicted from different methods 
using TEST-3A and TEST-3B with the median mass, $M_{med}/M_{\odot}$, for each 
method.  The {\it rms}, bias and outlier fractions in 
$\Delta log(M_{med})= log (M_{est}) - log (M_{med})$ is estimated and listed 
in Table 8. These should only be considered as relative measures, providing 
estimates of the overall agreement between masses from different methods and
for individual methods between TEST-3A and TEST-3B. In case of TEST-3A, some methods show large scatter (eg. 1.A and 4.B) and large outlier fraction (8.E) while others
closely agree (11.H and 12.I). The behavior of these methods is   
consistent with results from the mock catalogs (TEST-2A). Also, 
the scatter between the estimated stellar masses from individual methods and the
``reference'' values are significantly reduced for most models when using TEST-3B (Figure 14 and Table 8).  The sum of the {\it rms} values (in quadrature) in Table 8 gives the dispersion between different methods, corresponding to $0.39$\,dex 
(for TEST-3A) and $0.22$\,dex (for TEST-3B). 
 The scatter in TEST-3A constitutes all the observational errors
including photometric uncertainties and errors in the SED fitting process and
hence, provides an estimate of the observed error associated with mass 
measurement
from any given technique. The reduction in the {\it rms} scatter in TEST-3B is 
a result of the absence of constraints on the free parameters and the way different methods handle the
SED fits.

\begin{figure*}
\epsscale{0.4}
\plotone{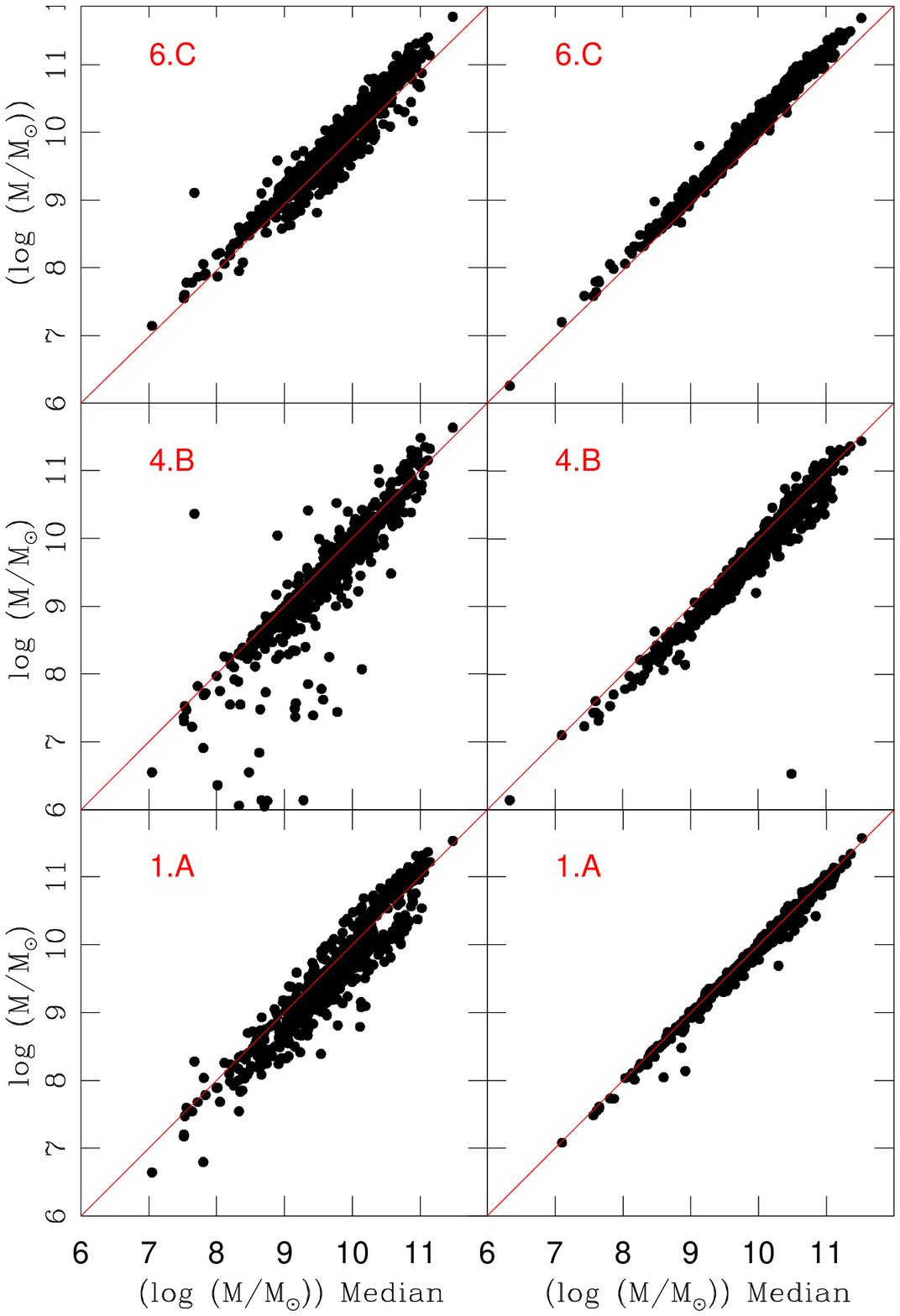}
\plotone{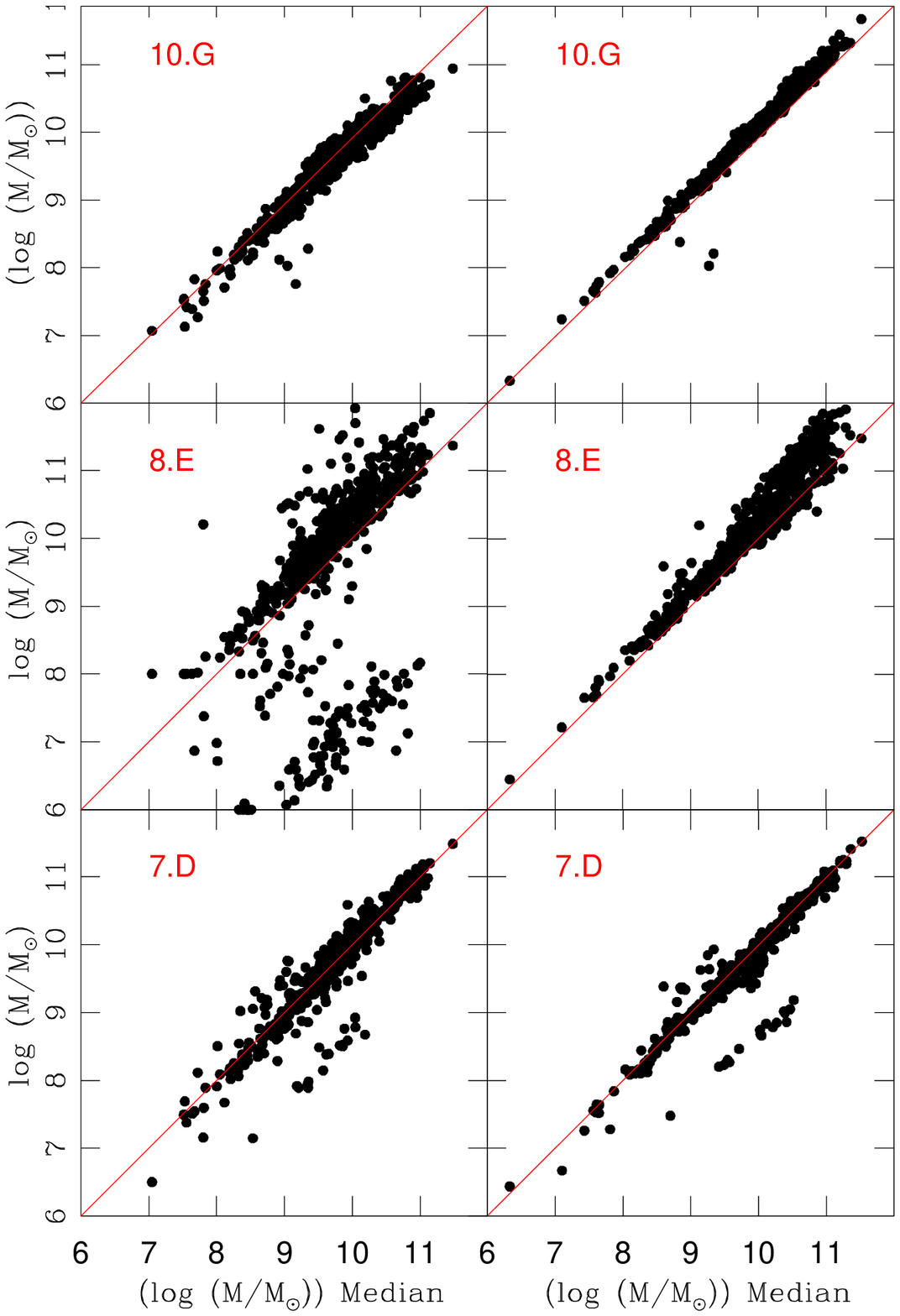}
\plotone{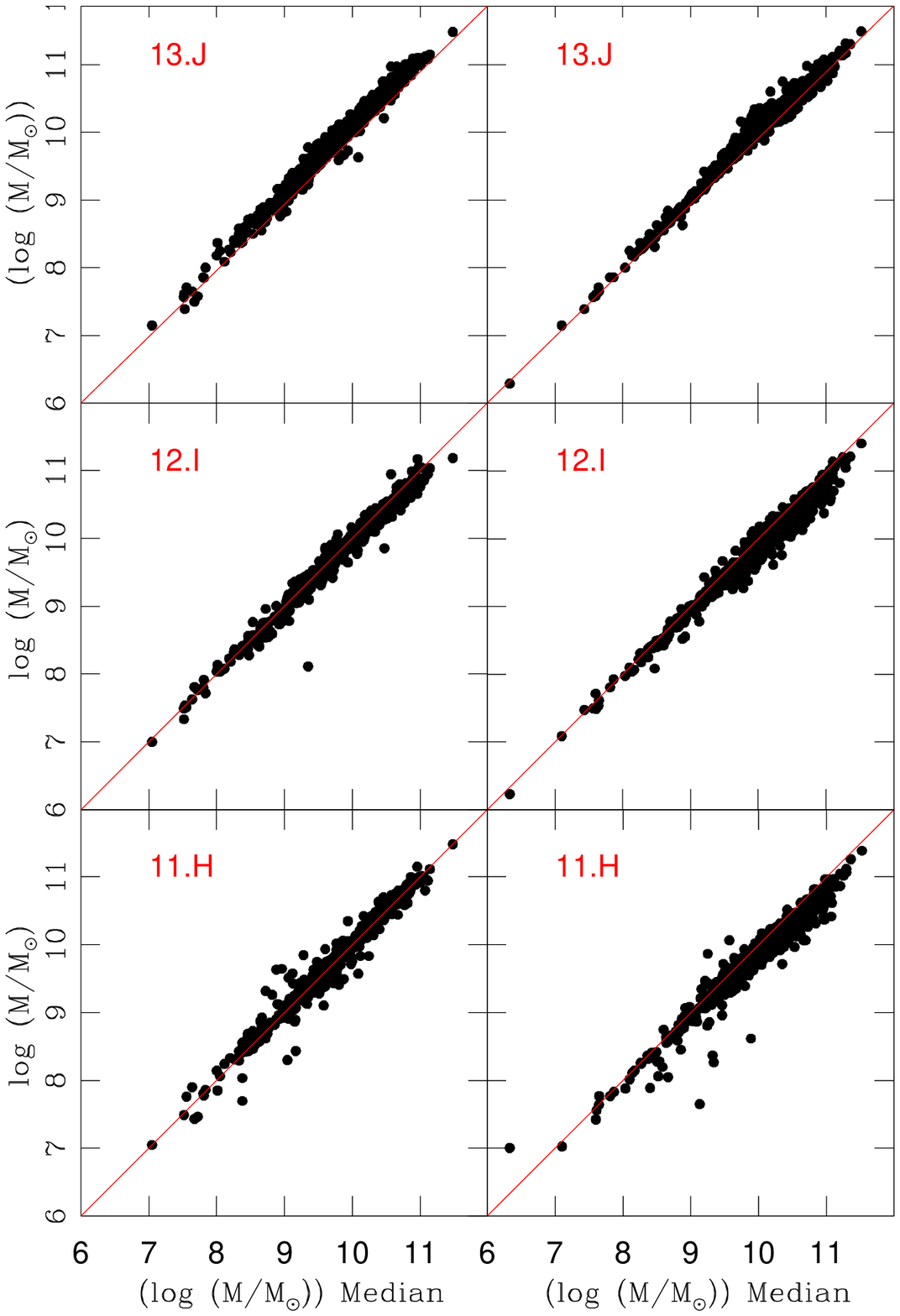}
\figcaption[TEST4_Stell_mass_comp_1n.eps]{Relations between the median of stellar masses between all the methods and the estimated stellar mass from individual methods for both TEST-3A (left panels) and TEST-3B (right panels). There is a significant reduction in the scatter and outlier fraction in the case of TEST-3B.}
%\label{fig2}}
\end{figure*}

\begin{table*}
\caption{The {\it rms} scatter, bias and outlier fraction (OLF) in
$\Delta log(M_{med})= log (M_{est}) - log (M_{med}) $ for different methods applied to TEST-3A (first line) and TEST-3B (second line). $M_{med}$ is the median of the stellar masses for a given galaxy, measured by different methods.}
\centering
\begin{tabular}{rllcll}
Code & $rms$ & $rms$            & bias & outlier  \\
     &       & outliers removed &      & fraction \\
  1.A & 0.327&   0.256&   0.111&   0.108\\
      & 0.093&   0.080&   0.046&     0.005\\
  4.B & 0.378&   0.203&   0.100&   0.074\\
      & 0.247 &   0.230&   0.171&     0.028\\
  6.C & 0.223&   0.206&  -0.040&   0.014\\
      & 0.167&    0.164&   -0.130&    0.003\\
  7.D & 0.294&   0.149&  -0.007&   0.055\\
      & 0.268&    0.114&   0.043 &    0.044\\
  8.E & 0.938&   0.297&  -0.237&   0.345\\
      & 0.365&    0.242&  -0.194 &    0.196\\
  10.G & 0.235&   0.216&   0.147&   0.014\\
       & 0.143 &   0.125&   -0.101&    0.003\\
 11.H & 0.132&   0.106&  -0.003&   0.014\\
      & 0.225&   0.184&   0.137&     0.029\\
 12.I & 0.114&   0.098&   0.030&   0.003\\
      &  0.183&   0.175&   0.128&     0.010\\
 13.J & 0.146&    0.146&   -0.110&   0.00\\
      & 0.102&    0.102&     -0.012&  0.00\\
\tableline
\end{tabular}
\end{table*}

Figure 15 examines the consistency of the stellar mass and extinction estimates
between different methods. For any pair of methods, we find the difference 
between their estimated stellar mass ($\Delta (M)$) and extinction ($\Delta (ext)$)
values. Since these parameters are estimated simultaneously from the 
SED fits, this provides a direct and unbiased test of the consistency of
the stellar mass and extinction estimates between different methods. 
 It is clear that, in all cases, there is a shift on the $\Delta log(M)-\Delta (ext)$ plane from $\Delta log(M)=\Delta(ext)=0 $ point for any of the two methods
compared. Some methods agree on their estimated stellar mass and some on the extinction but none of the pair of methods agree in both.

\begin{figure*}
\epsscale{0.5}
\plotone{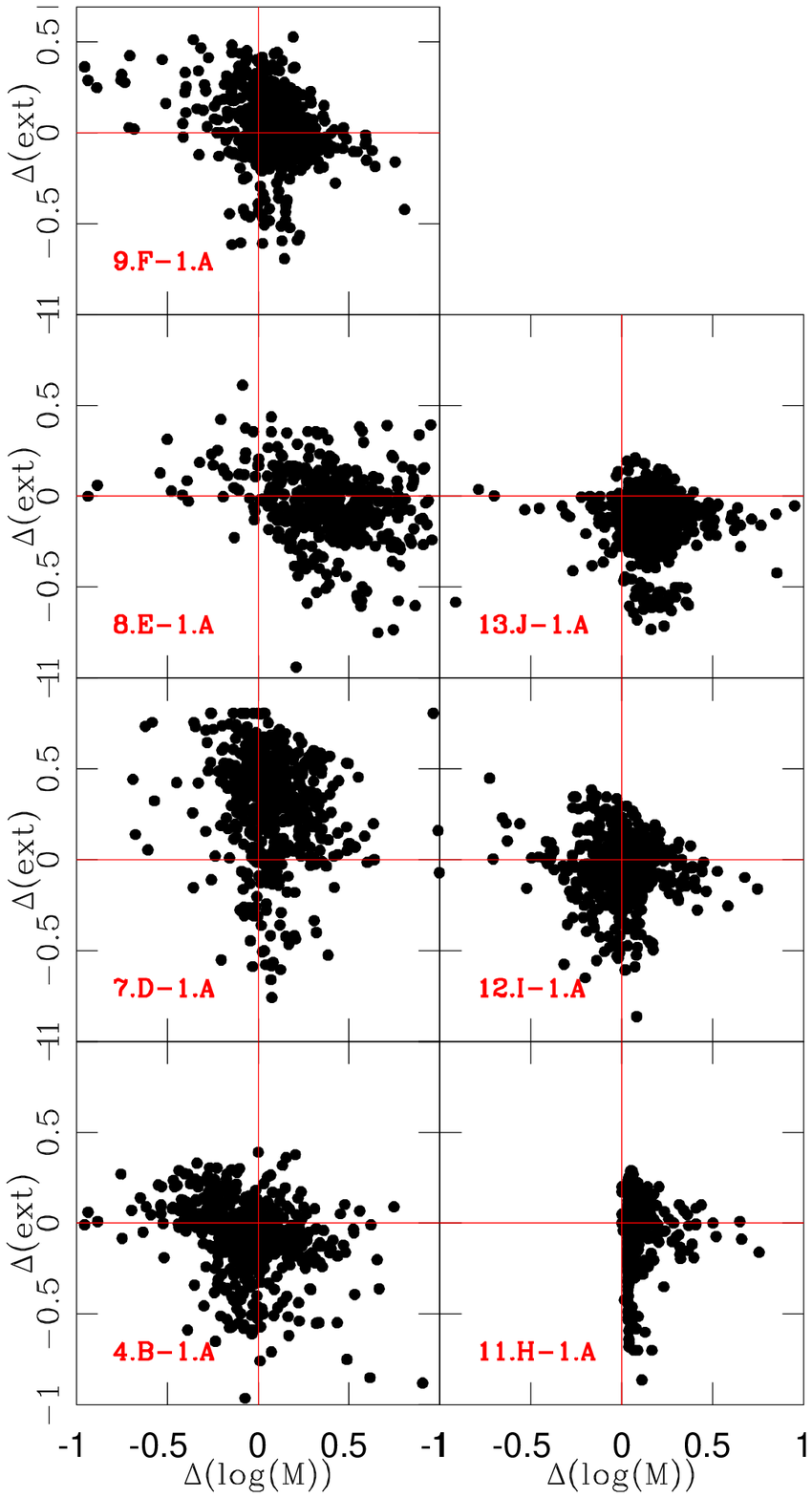}
\plotone{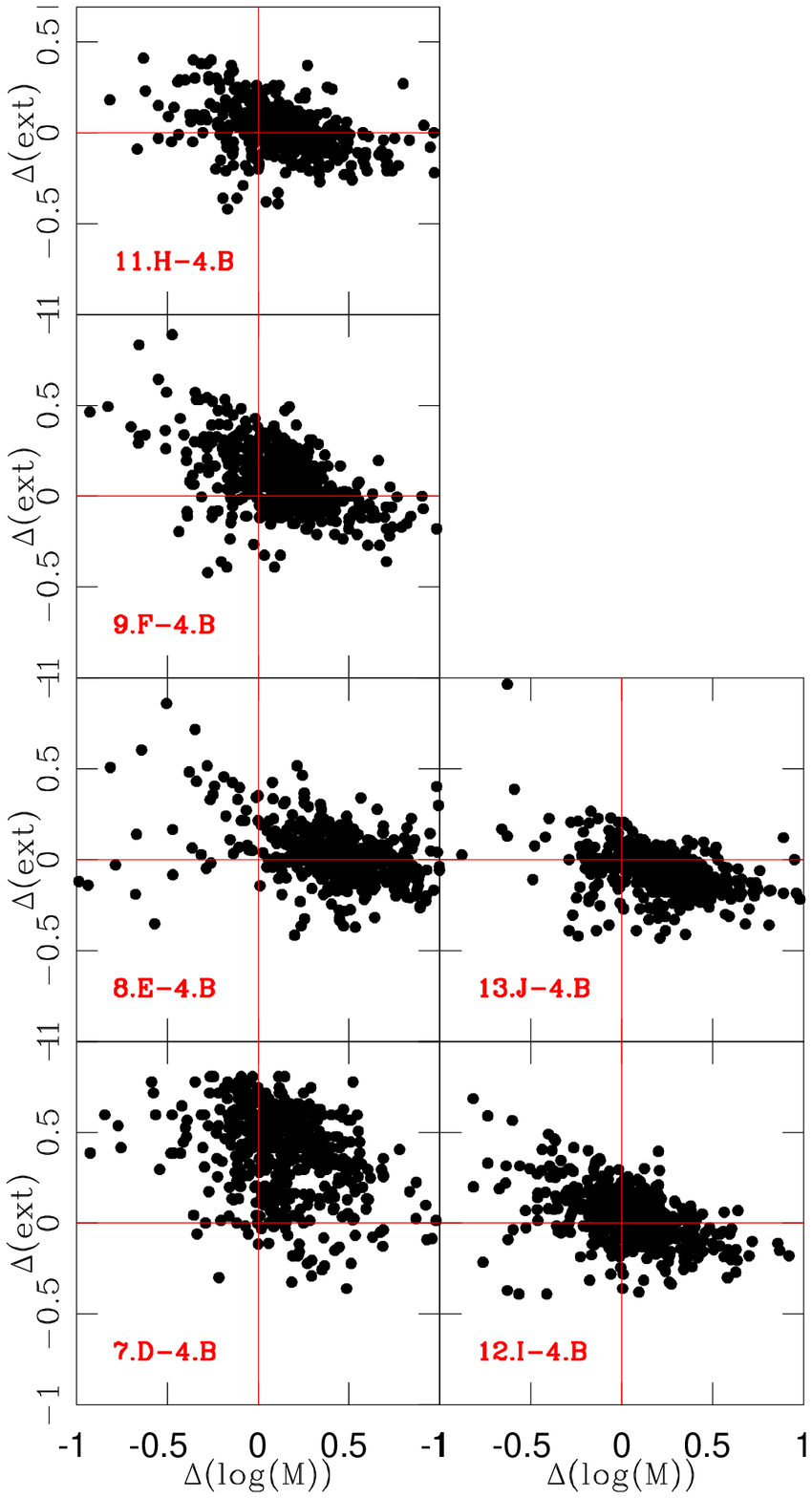}
\plotone{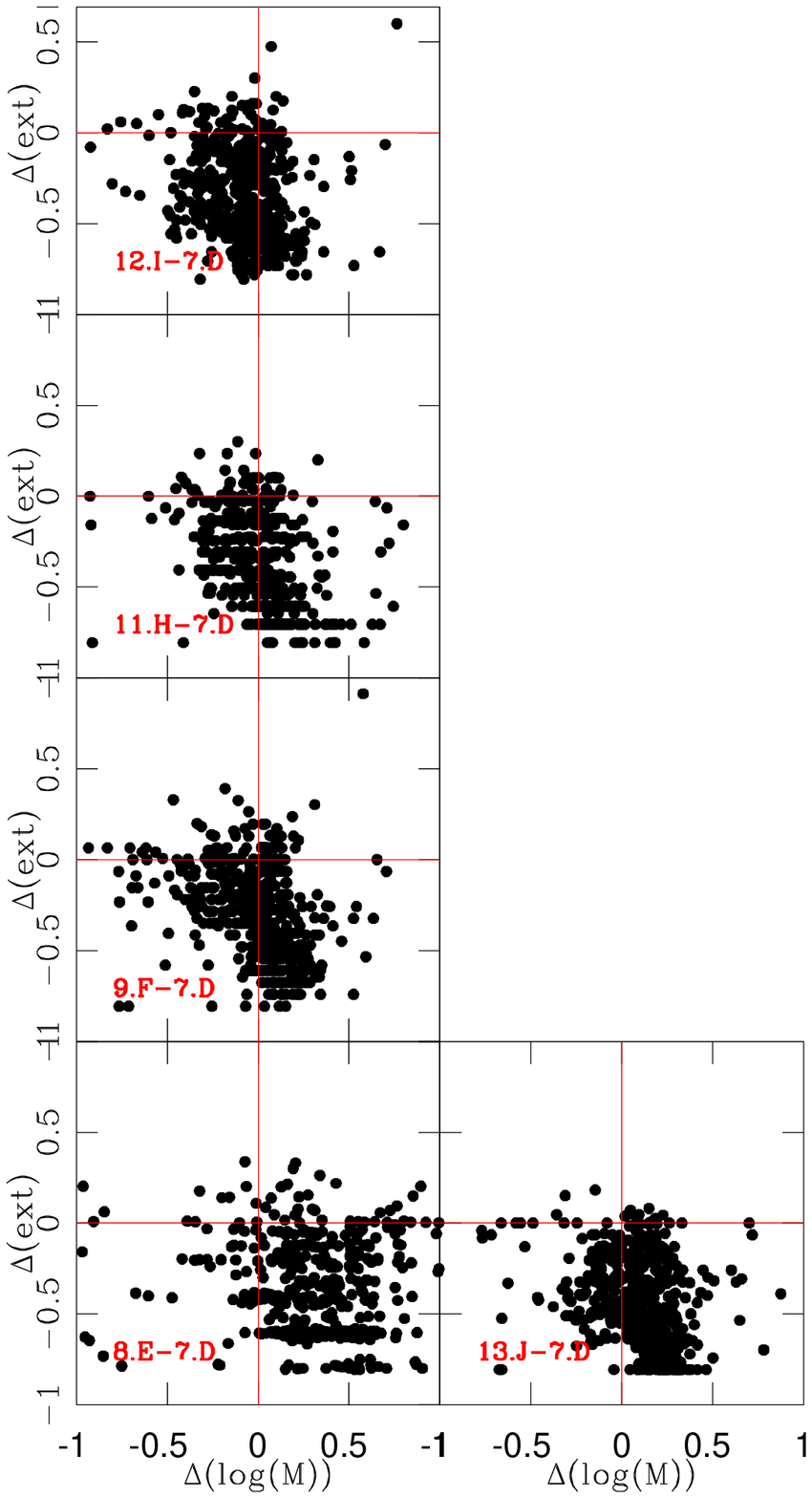}
\plotone{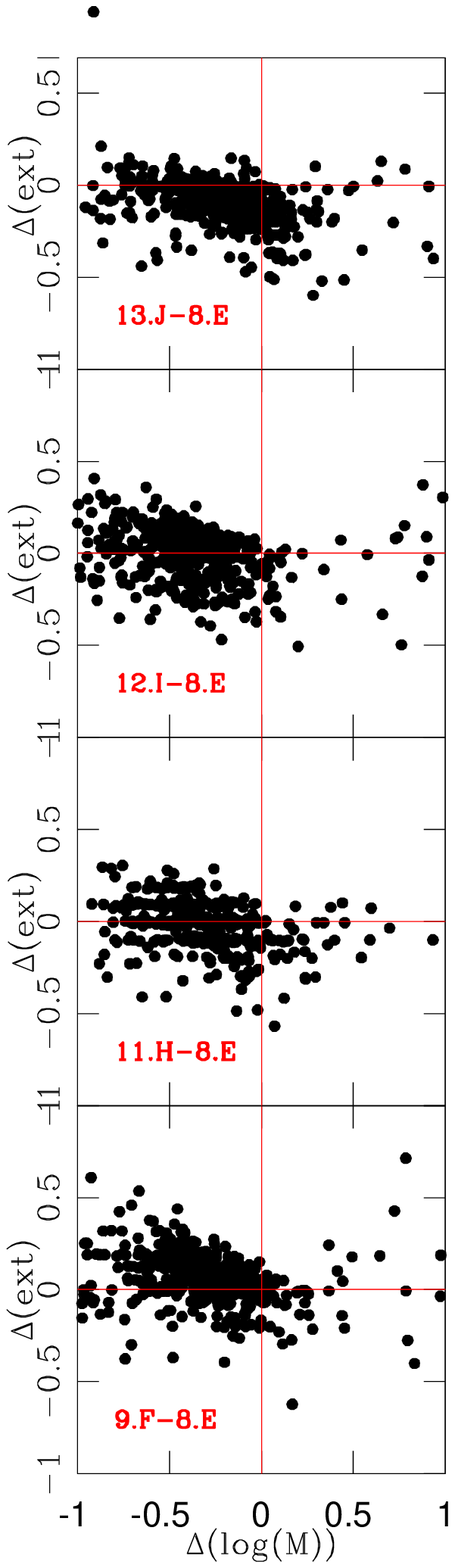}
\plotone{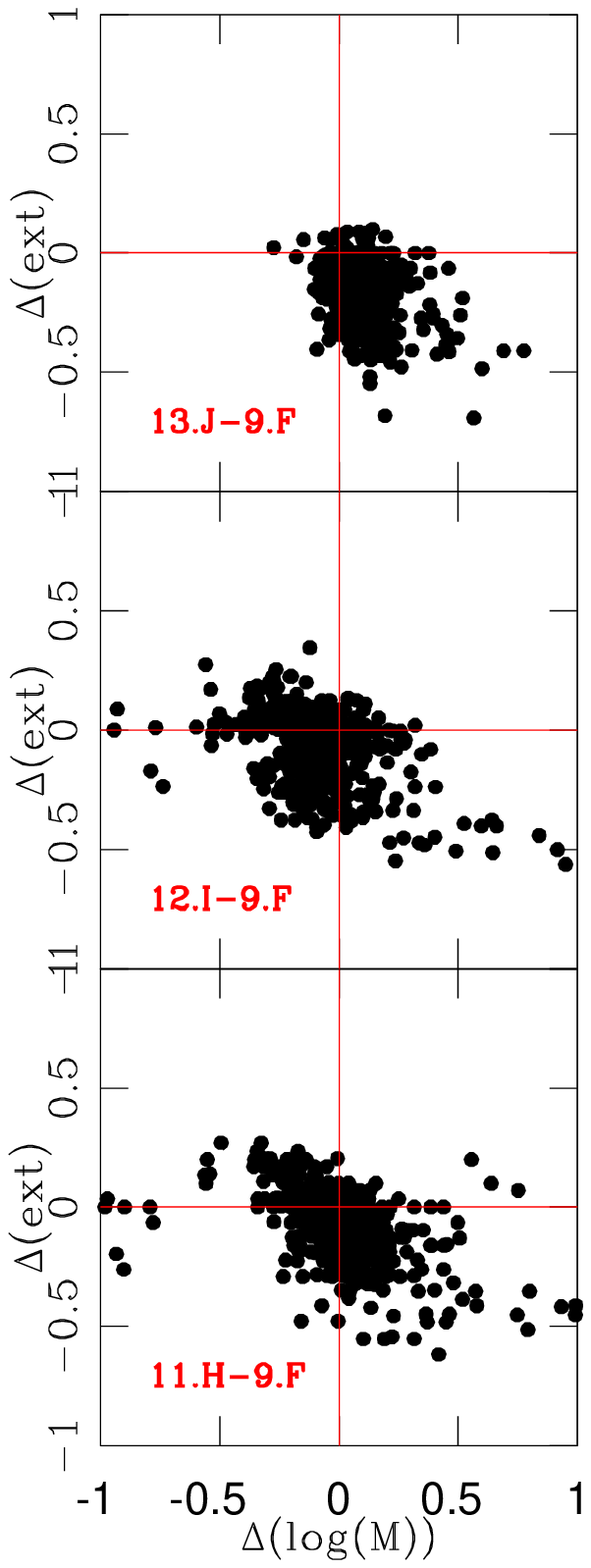}
\plotone{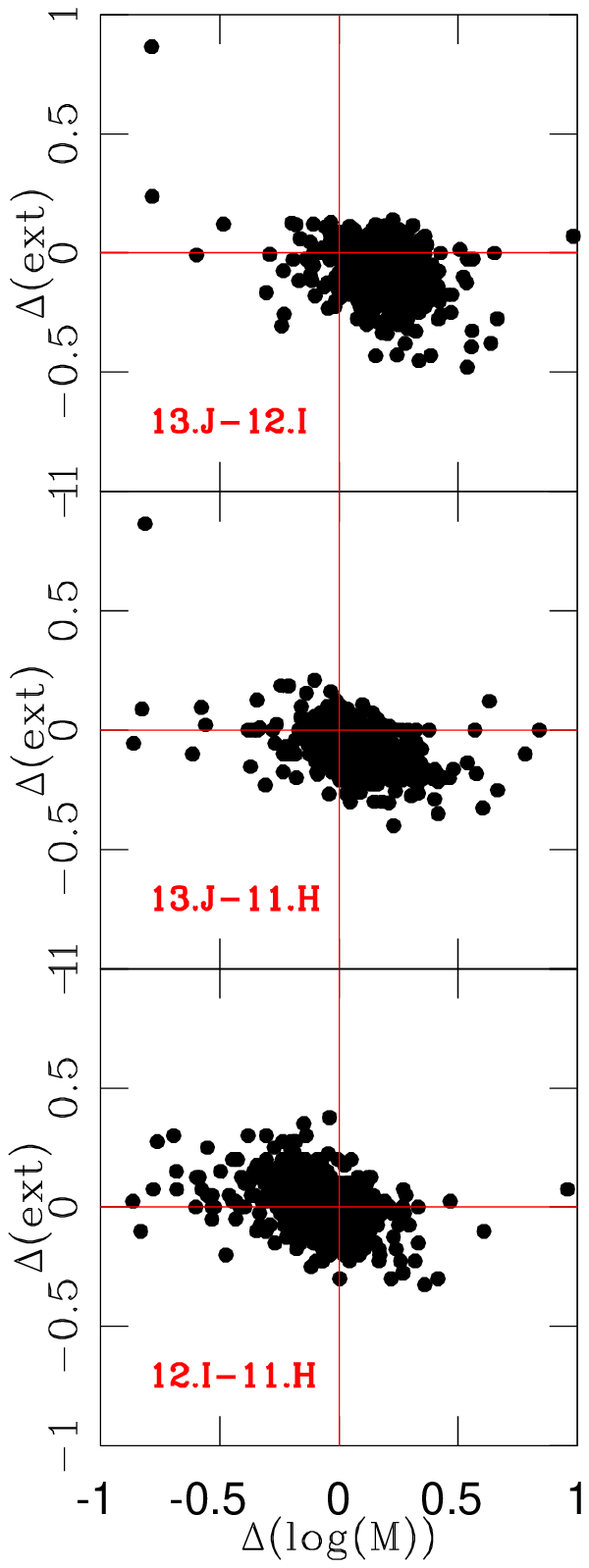}
\figcaption[TEST4_Stell_mass_comp_1n.eps]{Compares the difference in the stellar mass and extinction- $\Delta log(M)$ and $\Delta (ext)$ respectively- between any two methods using TEST-3A applied on 586 observed galaxies in GOODS-S. The scatter in these diagrams around the center ($\Delta log (M) = \Delta (ext) = 0$) indicates that stellar mass measurement methods cannot at the same time produce both the stellar mass and extinction.}
%\label{fig2}}
\end{figure*}

\subsection{\bf Dependence of the Stellar Mass on Population Synthesis Models}

The template SEDs generated by population synthesis models are the most 
fundamental components in measuring stellar mass of galaxies. It is therefore
instructive to quantify the effect of the 
population synthesis models on the 
estimated mass of galaxies, given differences in the composition and
data libraries used in these models. Here we estimate stellar masses using
templates generated from BC03 and CB07 models while keeping all the rest
of the parameters the same. The main difference between these two models is 
the addition of pulsating Asymptotic Giant Branch stars to the CB07 
model. For this experiment we use observational data from TEST-3, 
consisting of a sample of 586 galaxies. 
All the galaxies in this sample have spectroscopic data, used to 
fix redshifts of the galaxies when performing the SED fits. Since this is
an internal comparison between results from the two models, 
there is no dependence on the ``true'' stellar mass values. The 
 difference between the stellar masses using templates 
from BC03 and CB07 is plotted aganist redshift and mass in Figure 16, showing 
an offset of $\sim 0.2$\,dex in $log(M_{BC03}/M_{CB07})$, with higher masses from
the BC03. The difference between the stellar masses 
reduces at higher redshifts ($z > 3$) while it is constant over the
entire stellar mass range studied here. 

Given that the observational sample here is confined to brighter galaxies 
(for which spectroscopic data are available), it is possible that the above 
result is
biased. To examine this, we apply the same procedure on the simulated
data in TEST-2A (which is the most realistic). We find a similar shift
$\sim 0.2$\,dex in $log(M_{BC03}/M_{CB07})$ as for the observational data. 
The simulated galaxies 
also show closer agreement between the stellar masses at higher 
redshifts, in agreement with the result from TEST-3. The observed offset
here is mainly due to the addition of the pulsating AGB stars to the 
CB07 model, affecting near-infrared part of the SEDs generated from it. 
At higher redshifts ($z\sim 3$) the near-infrared light shifts outside 
the wavelength 
range spanned by the SEDs here and hence, the results (stellar masses) 
become unaffected by the AGB contribution, leading to
a better agreement in the estimated stellar mass between BC03 and CB07. 

\begin{figure*}
\epsscale{1}
\plotone{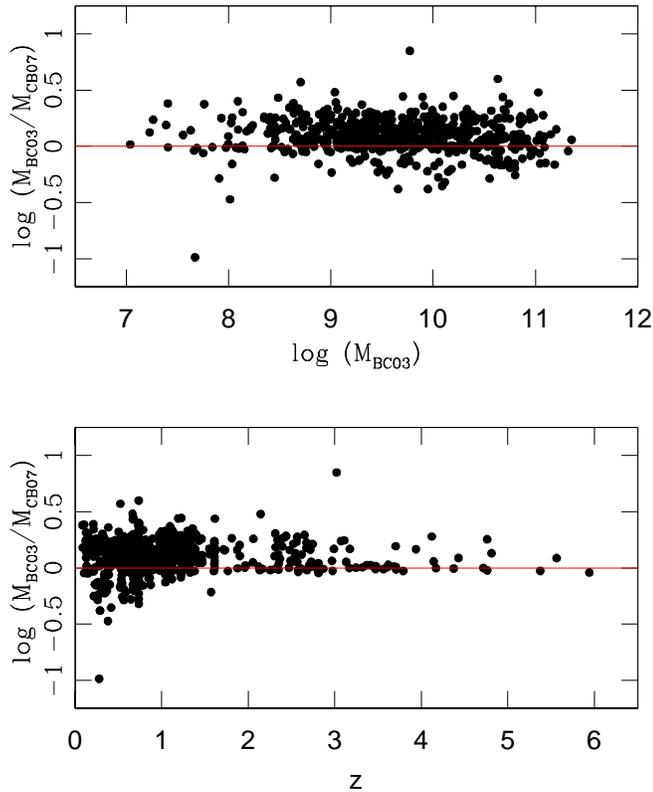}
\figcaption[TEST4_Stell_mass_comp_1n.eps]{Comparison between the stellar mass estimates based on the BC03 and CB07 population synthesis models using observational data from TEST-3. All the parameters in the SED fits are fixed, with the only difference being the population synthesis models which generate the template SEDs.}
%\label{fig2}}
\end{figure*}

\subsection{\bf Uncertainties in Stellar Mass Measurement due to Contributions from Nebular Emission}

It is well known that contribution from nebular emission lines is a non-negligible component of the observed flux from a galaxy at
certain redshifts,   
leading to appreciable differences in the parameters estimated from their SEDs 
(\citealt{debarros}; \citealt{schenker13b}).  In the absence of correction for nebular emission, one overestimates both the stellar mass and age, as the nebular emission mimics an increase in the observed flux at longer wavelengths,  
enhancement of Balmer Breaks and hence, increased mass and age. 

However, it is difficult to accurately quantify the effect of nebular emission
in the estimated stellar mass in galaxies, as it depends on the redshift of the
galaxy, the filters used for the SED fitting process and the width of the 
filters. For example, the $H\alpha$ line shifts into the IRAC 3.6\, $\mu$m band at $z\sim 3.1$. 
Depending on the width of the filter, we get different fractional contributions
to the observed fluxes. Therefore, the contribution due to 
nebular emission lines needs to be taken into account depending on the redshift of the galaxy in question and the filters used.  

To quantify this, we estimated the stellar masses with and without correction for nebular emission lines, using the observed SEDs in TEST-3, keeping all the rest
of the parameters fixed. We find a difference of up to $\sim 0.3$\,dex 
in the estimated stellar mass, purely due to contribution from
nebular emission lines. 

{\it In conclusion, differences in the stellar mass and extinction between differet methods when using observational data, confirm that the majority of the methods do not converge on the estimates of BOTH the stellar mass and extinction. Dependence of the stellar mass on population synthesis models was investigated and found that inclusion of pulsating AGB stars would decrease the estimate of the stellar mass by 0.2dex. Finally, it was found that the contribution from nebular emission lines is to increase the stellar mass of galaxies by $\sim 0.3$\,dex, depending on the redshift of the galaxy in question. }

\section{\bf TEST-4: The Effect of Near-infrared Photometric Depth and Selection Wavelength on the Overall Stellar Mass}

To investigate the effect of near-infrared photometric depth on the 
estimated stellar mass, we designed TEST-4 which is similar to TEST-3A with 
the only difference being that it is based on a z-band selected sample 
(as compared to TEST-3A which was based on an H-band (F160W) selected sample)
and with shallower near-infrared (JHK) data. Since this test  
also depends on the {\it real} data, we do not know the expected stellar 
masses. 
Using the median of the stellar masses measured for each galaxy by different methods, we estimate $\Delta log(M_{med})=log (M_{est}) - log (M_{Median})$ for individual galaxies.   
The {\it rms} in $\Delta log (M_{med})$ is then calculated for each method 
using all the galaxies, and
for each galaxy using measurements from different methods. 
The results from TEST-3A and TEST-4 are compared in Figure 17, which 
also presents comparison between the {\it rms} values when outliers are removed and between the bias estimates. There is 
no significant difference in the average mass estimates between the 
optical (z-band)
and near-IR (H-band) selected samples. Also, no difference is found 
due to a relatively shallower near-IR photometry in TEST-4. Figure 18 presents the median {\it rms} values in $S/N$ bins for
TEST-3A and TEST-4. There is no significant difference between the masses estimated from these two methods in terms of $S/N$ ratios, both converging to {\it rms=0.2} at $S/N > 40$.  

\begin{figure*}
\epsscale{0.6}
\plotone{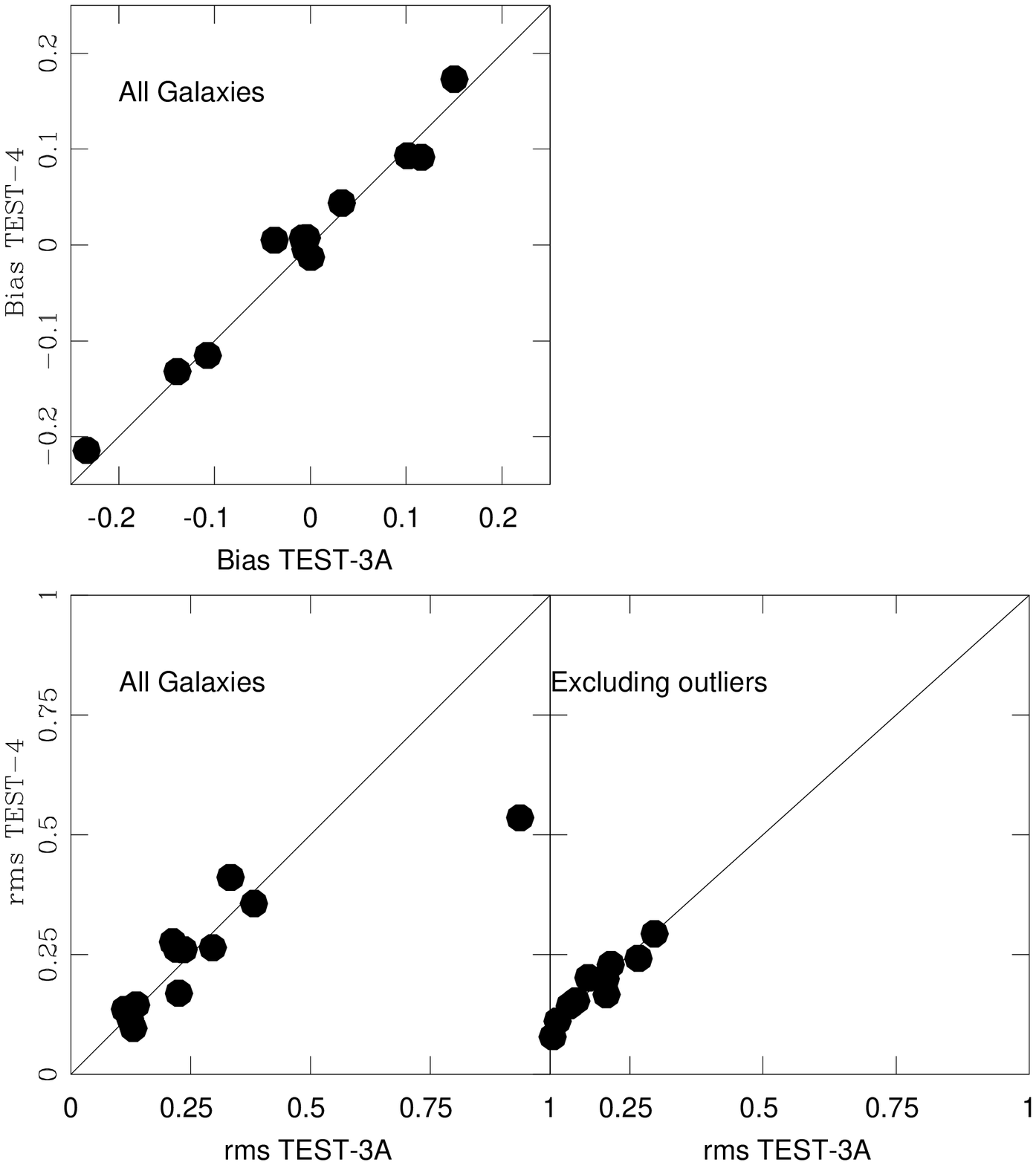}
\figcaption[TEST4_Stell_mass_comp_1n.eps]{Compares the {\it rms} values in $\Delta log(M_{Med})=
log (M_{est})- log(M_{Med})$, measured from different codes, between TEST-3A and TEST-4. $M_{med}$ is the median of the mass estimates for individual galaxies from different codes. 
Also, presented are the comparison between the {\it rms} values with the outliers removed and the bias resulted from different methods.}
%\label{fig2}}
\end{figure*}

\begin{figure*}
\epsscale{0.5}
\plotone{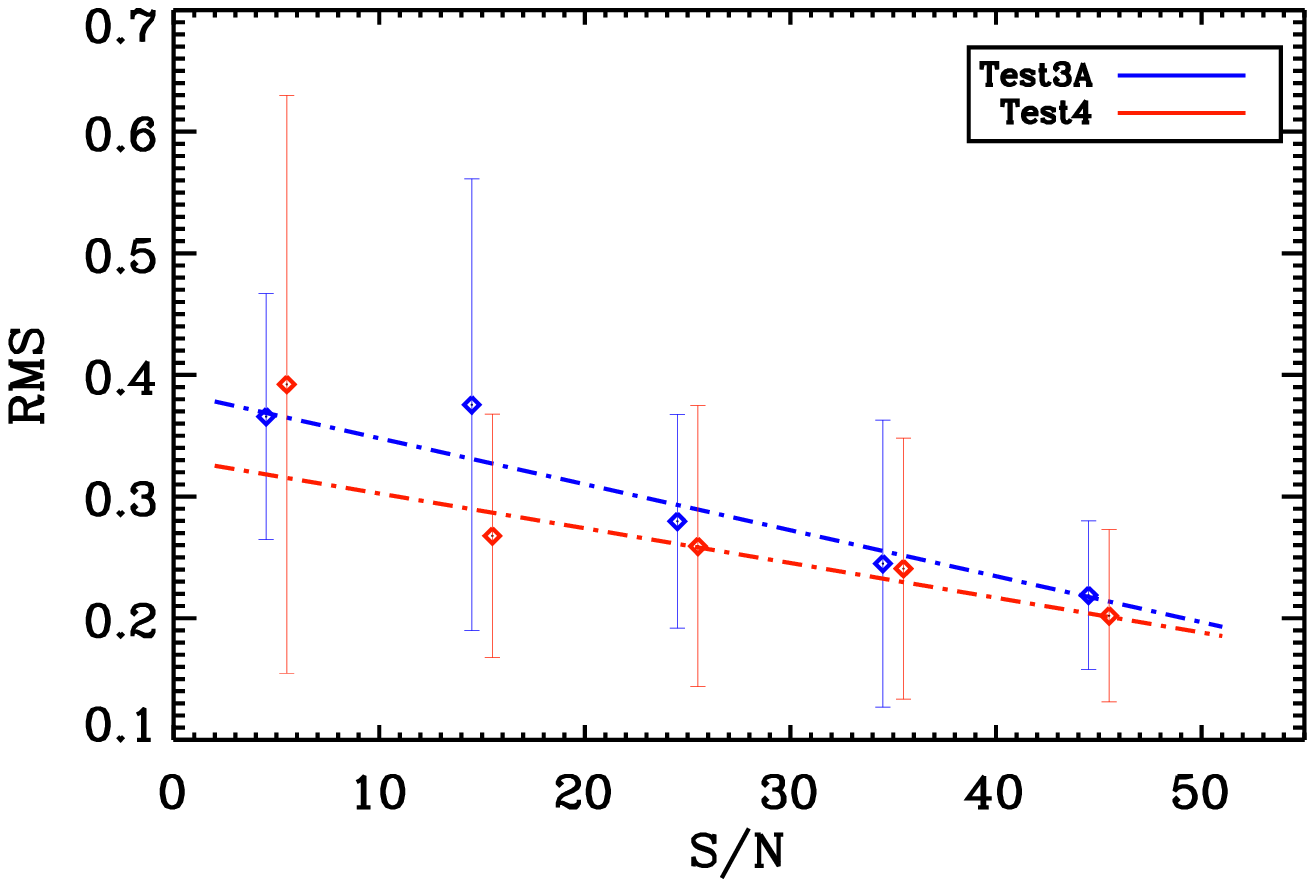}
\figcaption[TEST4_Stell_mass_comp_1n.eps]{Median of the {\it rms} values in $\Delta log (M_{med})=log (M_{est})- log (M_{Med})$ estimated in $S/N$ intervals. $M_{med}$ is the median of the mass estimates for individual galaxies from different codes. The plot shows results for both 
 TEST-3A and TEST-4. The lines are the best fits to the median values.}
%\label{fig2}}
\end{figure*}

{\it  In conclusion, no significant difference is found in the estimated stellar masses due to the selection wavelength of the survey or the depth of the near-IR data alone}. 

\section{\bf Comparison with other Studies}

In recent years several studies have addressed the dependence of the stellar mass on physical parameters using simulated catalogs with known input values (\citealt{wuyts2009}; \citealt{lee2009}; \citealt{pforr2012}). 
\cite {long2012} studied the
dependence of the estimated stellar masses  on age, metallicity, IMF and SFH for early-type galaxies at $1 <  z < 2$, using different stellar population
synthesis codes to model their SEDs. They found that, at a given IMF, the stellar masses cannot be recovered better than a factor of $2-3$.  

Using model templates based on BC03, assuming Calzetti extinction law with reddening in the range $0-4$, and three SFHs: SSP, constant SFR and a 
$\tau$ model with $\tau=0.3$ Gyr, \cite{wuyts2009} generated mock catalogs
in the redshift range $1.5 < z < 3$. When keeping redshift fixed, they 
underestimated the reddening, stellar mass and SFRs however, these estimates 
improved when redshift was used as a free parameter in the fit.  
While correctly predicting properties of spheroidal galaxies, they failed 
to reproduce input parameters for star-forming systems. 
Their results agree well with the independent study by 
\cite{pforr2012}. 

Concentrating only on a simulated sample of Lyman Break Galaxies at $z\sim 3.4$, 4 and 5, and using BC03, \cite{lee2012} found that both masses and SFRs are underestimated while the ages are overestimated. They attributed this to differences in the SFHs between the mock and $\tau$-model templates used in the fitting process. They further showed that data spanning over a  long wavelength range is essential to best recover the input parameters. 

\cite{pforr2012} performed a comprehensive study of uncertainties in the
estimated physical parameters in galaxies. Based on the SED fitting code 
of \cite{bolzonella} and population sythesis models from
\cite{maraston2005}, they found that the most important parameter in recovering the stellar mass is the SFH, 
in agreement with \cite{maraston2010}. This underlines the importance of
the physics of the model templates used in the SED fitting process.  
Using mock passive and star-forming galaxies in redshift range $0.5 < z < 3$, 
they examined the sensitivity of the stellar mass to redshift. 
When spectroscopic redshifts are known, they find best stellar mass estimates at low redshift when reddening is excluded and at high redshift using reddening and inverted tau models (Pforr et al. 2012). The inclusion of reddening at low redshift causes severe underestimation for the stellar masses. 
When redshift is a free parameter in the fit (e.g. when no spec-z are available), the additional degree of freedom allows for better mass estimates because redshift compensates for SFH and metallicity mismatch as well as the age-dust degeneracy (Pforr et al. 2013). This agrees well with results from the current study. At low redshifts, masses are still best determined excluding reddening from the fit.

In this paper we performed a critical study to quantify differences between the stellar masses estimated using 
different methods with  model templates from different population synthesis 
codes, considering the existing degenaracies between the physical parameters. By fixing the physical parameters (specifically redshifts), we find a
larger difference between the predicted and expected stellar masses. 
 For example, by allowing the redshift to vary in the fit, it compensates for the mismatch between SFH and metallicity and age-dust degeneracy and hence, improves the recovery. In agreement with previous studies, we find that our lack of knowledge of the correct SFH, combined with 
inherent degeneracy between age, dust and metallicity, are the main reasons
 for uncertainties in stellar masses. Moreover, the estimated uncertainty depends on the wavelength coverage at any given redshift. We also 
investigated the effect of photometric uncertainties on these parameters and confirm that their
effect is less serious than the above parameters.

\section{\bf The Error Budget}

In this section we quantify and compare relative contributions from 
the main sources dominating uncertainties in the stellar mass measurement.  
The situation becomes complicated by the fact that these parameters 
 are correlated. Therefore, one needs to disentangle their individual 
contributions, as investigated by simulations in previous sections.
In its general term, the uncertainty is defined as the {\it rms} scatter in
 $\Delta log(M) = log(M_{input}) - log (M_{est})$. The uncertainties in the stellar mass due to different parameters are listed in Table 9 and explained below:    

\begin{table*}
\caption{Error Budget for Stellar Mass Measurement}
\begin{tabular}{ll}
&\\
&\\
 & {\it rms} in $log(M/M_\odot)$\\
&\\
Method & $0.136$ \\
&\\
Systematic & $0.050$\\
&\\
Photometry & $0.135$ \\
&\\
Numerical & $0.045$\\
& \\
$log(Age)$ &$ E_{(B-V)}$  \\
& ($<0.3, 0.3-0.6, >0.6$)\\
$7-8$ & $(0.121, 0.141, 0.307)$\\	
$8-9$ &	$(0.112,0.180,0.297)$\\	
$9-10$&	$(0.273,0.271,0.387)$\\
&\\
Free Parameters& $0.110$  \\
&\\
Nebular lines & $0.300 $ \\
&\\
Combined Observational& $0.390$  \\
&\\
Population Synthesis Models& $0.200$\\
&\\
Depth (near-IR photometry) & $< 0.200 $\\

\end{tabular}
\end{table*}
 
\noindent {\bf Photometric errors:}  We examined this by estimating stellar masses for galaxies in mock catalogs (with known input mass) over a range of magnitudes.  
 By using the same parameters
to fit the SEDs as those used to generate the catalogs, we minimize the effect of other (free) parameters (TEST-1). Furthermore, by concentrating on 
individual codes, we avoid any code-dependent effect (Table 3).  Taking the above points into account, we estimate an uncertainty of $\sigma (\Delta log(M))=0.134$\,dex due to photometric errors. This dominates the error budget for galaxies with $m (160W) > 26$ (Figure 2a).
\\
\noindent {\bf Codes/Methods:} This was specifically tested by generating a simulated mock catalog (TEST-1) and constraining the input parameters, with the only free parameter being the code/method used.  
After subtracting the uncertainties due to photometric errors, we estimate the
scatter in $\sigma (\Delta log(M))$ among different codes in Table 4. We estimate an {\it rms} scatter of $\sigma (\Delta log (M))=0.136$\,dex due to differences in methods/codes used.
\\
\noindent {\bf Age and Extinction:} in order to disentangle the effects of age and extinction 
and estimate their individual contribution to the error budget, we 
constructed a covarience matrix with $\sigma (\Delta log(M))$ as matrix
elements measured in different age-extinction grids. All the other variables
were kept fixed. Results are listed in Table 6 and presented in Figure 5. 
The highest {\it rms} scatters were found for high extinction
($E_{B-V} > 0.6$) and age ($\sim 10^{9.5}$) yrs values.
\\
\noindent {\bf Numerical/Systematics:} even if all the above uncertainties are accounted for, we
still have an inherent ``base'' error, independent from photometric, 
code-dependent and the degeneracies mentioned above. We estimate this to be
$\sigma (\Delta log(M))=0.047$\,dex (Figures 3 \& 4). 
\\
\noindent {\bf Free Parameters:} this is estimated by performing a realistic simulation where
all the free parameters in the SED-fitting process were allowed to change
(TEST-2A) and compared the results with a similar test where the parameters
were kept fixed (TEST-2B). By subtracting (in quadrature) the {\it rms} 
estimates
from TEST-2A and TEST-2B, we find a scatter among different methods, 
(due to free parameters), in the range $\sigma (\Delta log(M))=0.037$\,dex to $0.264$\,dex (Table 7).  
The {\it rms} scatter associated with free parameters from the median stellar mass 
values (from TEST-3A) is $0.110$\,dex, which is taken as our estimate of the 
uncertainty in the stellar mass measurement caused by free parameters.
\\
\noindent {\bf Combined Observational Uncertainties:} using the observed data (TEST-3A), 
we measure the scatter in the estimated stellar mass values among different codes. This is estimated to be $\sigma (\Delta log(M))=0.390 $\,dex.
\\
\noindent {\bf Selection Wavelength and Photometric Depth:} TEST-4 was formulated to address
this and predicts a contribution $ < 0.2$\,dex in the total error budget due to
the selection of the wavelength and photometric depth of the sample.
\\
\noindent {\bf Nebular Line Correction:} In TEST-3 (based on the observational data), we compared the stellar mass estimates with and without correction for nebular emission (both line and continuum). We estimate an average error of 
$0.5 $\,dex in the stellar masses due to contribution from nebular line emission. 

\noindent {\bf Population Synthesis Models:} The templates used to measure stellar masses were generated by population synthesis models. We studied the effect of pulsating AGB stars on these templates and on the resulting stellar mass and find this to change the stellar mass by $\sim 0.2$\,dex.

\section{\bf Summary and Conclusions}

We performed a detailed study of the errors and main sources of uncertainty in stellar mass measurement in galaxies. Generating simulated galaxy catalogs with known input parameters (redshift, mass, SEDs), we investigated deviations in the estimated stellar mass from their input values ($\Delta log(M)$) and its dependence on the observable parameters. The stellar masses were measured by ten 
independent methods/codes with the results compared. Conclusions from this study are summarized below:

\begin{itemize}

\item{} When the same set of input assumptions are used, no significant bias is found between different methods. We find that the spread in the stellar mass of any given galaxy, using different methods is $\sigma (\Delta log(M))=0.136$\,dex. Fainter galaxies with lower photometric $S/N$ ratios ($H > 26$ mag) are responsible for most of this scatter. 

\item{} When the same population synthesis models and parameters are used, the median of the stellar masses from different methods provides the smallest {\it rms} scatter (with respect to the input stellar mass values) compared to individual methods.

\item{} We separated degeneracies in stellar mass measurements due to age and extinction and estimated their individual contribution to the total error budget. We find that the {\it rms} in stellar mass significantly increases for $E_{B-V} > 0.6$ for all ages. For any given method and extinction, there is an increase in the estimated stellar mass for ages $ > 10^{8.5}$ years.

\item{} From our simulations we found that errors in the stellar mass and age are strongly correlated (galaxies with large deviations in their stellar mass also show large deviations in age). A weaker trend is found with the extinction.

\item{} The effect of free parameters on stellar mass estimates was studied using mock photometric catalogs with known input stellar mass. 
We find $\sigma (\Delta log(M))=0.136$\,dex,  caused by degeneracy and interplay between parameters.

\item{} The effects of population synthesis models and corrction for nebular emission were investigated and found to change the stellar mass ($ \Delta (log(M))$) by $0.2$\,dex and $0.3$\,dex respectively. 

\end{itemize}

!\bibliographystyle{apj}
!\bibliography{reference2}

\begin{thebibliography}{}
\expandafter\ifx\csname natexlab\endcsname\relax\def\natexlab#1{#1}\fi

\bibitem[{{Acquaviva} {et~al.}(2012){Acquaviva}, {Gawiser}, {Bickerton},
  {Grogin}, {Guo}, \& {Lee}}]{Acquaviva12}
{Acquaviva}, V., {Gawiser}, E., {Bickerton}, S.~J., {et~al.} 2012, \apj, 749,
  72

\bibitem[{{Acquaviva} {et~al.}(2011){Acquaviva}, {Gawiser}, \&
  {Guaita}}]{acquaviva11}
{Acquaviva}, V., {Gawiser}, E., \& {Guaita}, L. 2011, \apj, 737, 47

\bibitem[{{Arnouts} \& {Ilbert}(2011)}]{arnauts}
{Arnouts}, S., \& {Ilbert}, O. 2011, {LePHARE: Photometric Analysis for
  Redshift Estimate}, astrophysics Source Code Library, ascl:1108.009

\bibitem[{{Ashby} {et~al.}(2013){Ashby}, {Willner}, {Fazio}, {Huang}, {Arendt},
  {Barmby}, {Barro}, {Bell}, {Bouwens}, {Cattaneo}, {Croton}, {Dav{\'e}},
  {Dunlop}, {Egami}, {Faber}, {Finlator}, {Grogin}, {Guhathakurta},
  {Hernquist}, {Hora}, {Illingworth}, {Kashlinsky}, {Koekemoer}, {Koo},
  {Labb{\'e}}, {Li}, {Lin}, {Moseley}, {Nandra}, {Newman}, {Noeske}, {Ouchi},
  {Peth}, {Rigopoulou}, {Robertson}, {Sarajedini}, {Simard}, {Smith}, {Wang},
  {Wechsler}, {Weiner}, {Wilson}, {Wuyts}, {Yamada}, \& {Yan}}]{ashby}
{Ashby}, M.~L.~N., {Willner}, S.~P., {Fazio}, G.~G., {et~al.} 2013, \apj, 769,
  80

\bibitem[{{Beckwith} {et~al.}(2006){Beckwith}, {Stiavelli}, {Koekemoer},
  {Caldwell}, {Ferguson}, {Hook}, {Lucas}, {Bergeron}, {Corbin}, {Jogee},
  {Panagia}, {Robberto}, {Royle}, {Somerville}, \& {Sosey}}]{beckwith6}
{Beckwith}, S.~V.~W., {Stiavelli}, M., {Koekemoer}, A.~M., {et~al.} 2006, \aj,
  132, 1729

\bibitem[{{Behroozi} {et~al.}(2013){Behroozi}, {Wechsler}, {Wu}, {Busha},
  {Klypin}, \& {Primack}}]{behroozi}
{Behroozi}, P.~S., {Wechsler}, R.~H., {Wu}, H.-Y., {et~al.} 2013, \apj, 763, 18

\bibitem[{{Bolzonella} {et~al.}(2000){Bolzonella}, {Miralles}, \&
  {Pell{\'o}}}]{bolzonella}
{Bolzonella}, M., {Miralles}, J.-M., \& {Pell{\'o}}, R. 2000, \aap, 363, 476

\bibitem[{{Bouwens} {et~al.}(2011){Bouwens}, {Illingworth}, {Oesch},
  {Labb{\'e}}, {Trenti}, {van Dokkum}, {Franx}, {Stiavelli}, {Carollo},
  {Magee}, \& {Gonzalez}}]{bouwens11}
{Bouwens}, R.~J., {Illingworth}, G.~D., {Oesch}, P.~A., {et~al.} 2011, \apj,
  737, 90

\bibitem[{{Brammer} {et~al.}(2008){Brammer}, {van Dokkum}, \&
  {Coppi}}]{brammer}
{Brammer}, G.~B., {van Dokkum}, P.~G., \& {Coppi}, P. 2008, \apj, 686, 1503

\bibitem[{{Bruzual} \& {Charlot}(2003)}]{bruzual03}
{Bruzual}, G., \& {Charlot}, S. 2003, \mnras, 344, 1000

\bibitem[{{Charlot} \& {Fall}(2000)}]{Charlot}
{Charlot}, S., \& {Fall}, S.~M. 2000, \apj, 539, 718

\bibitem[{{Dahlen} {et~al.}(2013){Dahlen}, {Mobasher}, {Faber}, {Ferguson},
  {Barro}, {Finkelstein}, {Finlator}, {Fontana}, {Gruetzbauch}, {Johnson},
  {Pforr}, {Salvato}, {Wiklind}, {Wuyts}, {Acquaviva}, {Dickinson}, {Guo},
  {Huang}, {Huang}, {Newman}, {Bell}, {Conselice}, {Galametz}, {Gawiser},
  {Giavalisco}, {Grogin}, {Hathi}, {Kocevski}, {Koekemoer}, {Koo}, {Lee},
  {McGrath}, {Papovich}, {Peth}, {Ryan}, {Somerville}, {Weiner}, \&
  {Wilson}}]{dahlen}
{Dahlen}, T., {Mobasher}, B., {Faber}, S.~M., {et~al.} 2013, \apj, 775, 93

\bibitem[{{de Barros} {et~al.}(2014){de Barros}, {Schaerer}, \&
  {Stark}}]{debarros}
{de Barros}, S., {Schaerer}, D., \& {Stark}, D.~P. 2014, \aap, 563, A81

\bibitem[{{Finkelstein} {et~al.}(2012){Finkelstein}, {Papovich}, {Ryan},
  {Pawlik}, {Dickinson}, {Ferguson}, {Finlator}, {Koekemoer}, {Giavalisco},
  {Cooray}, {Dunlop}, {Faber}, {Grogin}, {Kocevski}, \& {Newman}}]{finkelstein}
{Finkelstein}, S.~L., {Papovich}, C., {Ryan}, R.~E., {et~al.} 2012, \apj, 758,
  93

\bibitem[{{Giavalisco} {et~al.}(2004){Giavalisco}, {Ferguson}, {Koekemoer},
  {Dickinson}, {Alexander}, {Bauer}, {Bergeron}, {Biagetti}, {Brandt},
  {Casertano}, {Cesarsky}, {Chatzichristou}, {Conselice}, {Cristiani}, {Da
  Costa}, {Dahlen}, {de Mello}, {Eisenhardt}, {Erben}, {Fall}, {Fassnacht},
  {Fosbury}, {Fruchter}, {Gardner}, {Grogin}, {Hook}, {Hornschemeier}, {Idzi},
  {Jogee}, {Kretchmer}, {Laidler}, {Lee}, {Livio}, {Lucas}, {Madau},
  {Mobasher}, {Moustakas}, {Nonino}, {Padovani}, {Papovich}, {Park},
  {Ravindranath}, {Renzini}, {Richardson}, {Riess}, {Rosati}, {Schirmer},
  {Schreier}, {Somerville}, {Spinrad}, {Stern}, {Stiavelli}, {Strolger},
  {Urry}, {Vandame}, {Williams}, \& {Wolf}}]{giavalisco}
{Giavalisco}, M., {Ferguson}, H.~C., {Koekemoer}, A.~M., {et~al.} 2004, \apjl,
  600, L93

\bibitem[{{Grogin} {et~al.}(2011){Grogin}, {Kocevski}, {Faber}, {Ferguson},
  {Koekemoer}, {Riess}, {Acquaviva}, {Alexander}, {Almaini}, {Ashby}, {Barden},
  {Bell}, {Bournaud}, {Brown}, {Caputi}, {Casertano}, {Cassata}, {Castellano},
  {Challis}, {Chary}, {Cheung}, {Cirasuolo}, {Conselice}, {Roshan Cooray},
  {Croton}, {Daddi}, {Dahlen}, {Dav{\'e}}, {de Mello}, {Dekel}, {Dickinson},
  {Dolch}, {Donley}, {Dunlop}, {Dutton}, {Elbaz}, {Fazio}, {Filippenko},
  {Finkelstein}, {Fontana}, {Gardner}, {Garnavich}, {Gawiser}, {Giavalisco},
  {Grazian}, {Guo}, {Hathi}, {H{\"a}ussler}, {Hopkins}, {Huang}, {Huang},
  {Jha}, {Kartaltepe}, {Kirshner}, {Koo}, {Lai}, {Lee}, {Li}, {Lotz}, {Lucas},
  {Madau}, {McCarthy}, {McGrath}, {McIntosh}, {McLure}, {Mobasher},
  {Moustakas}, {Mozena}, {Nandra}, {Newman}, {Niemi}, {Noeske}, {Papovich},
  {Pentericci}, {Pope}, {Primack}, {Rajan}, {Ravindranath}, {Reddy}, {Renzini},
  {Rix}, {Robaina}, {Rodney}, {Rosario}, {Rosati}, {Salimbeni}, {Scarlata},
  {Siana}, {Simard}, {Smidt}, {Somerville}, {Spinrad}, {Straughn}, {Strolger},
  {Telford}, {Teplitz}, {Trump}, {van der Wel}, {Villforth}, {Wechsler},
  {Weiner}, {Wiklind}, {Wild}, {Wilson}, {Wuyts}, {Yan}, \& {Yun}}]{Grogin}
{Grogin}, N.~A., {Kocevski}, D.~D., {Faber}, S.~M., {et~al.} 2011, \apjs, 197,
  35

\bibitem[{{Guo} {et~al.}(2013{\natexlab{a}}){Guo}, {Ferguson}, {Giavalisco},
  {Barro}, {Willner}, {Ashby}, {Dahlen}, {Donley}, {Faber}, {Fontana},
  {Galametz}, {Grazian}, {Huang}, {Kocevski}, {Koekemoer}, {Koo}, {McGrath},
  {Peth}, {Salvato}, {Wuyts}, {Castellano}, {Cooray}, {Dickinson}, {Dunlop},
  {Fazio}, {Gardner}, {Gawiser}, {Grogin}, {Hathi}, {Hsu}, {Lee}, {Lucas},
  {Mobasher}, {Nandra}, {Newman}, \& {van der Wel}}]{Guo}
{Guo}, Y., {Ferguson}, H.~C., {Giavalisco}, M., {et~al.} 2013{\natexlab{a}},
  \apjs, 207, 24

\bibitem[{{Guo} {et~al.}(2013{\natexlab{b}}){Guo}, {Ferguson}, {Giavalisco},
  {Barro}, {Willner}, {Ashby}, {Dahlen}, {Donley}, {Faber}, {Fontana},
  {Galametz}, {Grazian}, {Huang}, {Kocevski}, {Koekemoer}, {Koo}, {McGrath},
  {Peth}, {Salvato}, {Wuyts}, {Castellano}, {Cooray}, {Dickinson}, {Dunlop},
  {Fazio}, {Gardner}, {Gawiser}, {Grogin}, {Hathi}, {Hsu}, {Lee}, {Lucas},
  {Mobasher}, {Nandra}, {Newman}, \& {van der Wel}}]{guo2013}
---. 2013{\natexlab{b}}, \apjs, 207, 24

\bibitem[{{Johnson}(2013)}]{Johnson}
{Johnson}, S. 2013, {SATMC: SED Analysis Through Monte Carlo}, astrophysics
  Source Code Library, ascl:1309.005

\bibitem[{{Klypin} {et~al.}(2011){Klypin}, {Trujillo-Gomez}, \&
  {Primack}}]{klypin}
{Klypin}, A.~A., {Trujillo-Gomez}, S., \& {Primack}, J. 2011, \apj, 740, 102

\bibitem[{{Koekemoer} {et~al.}(2011){Koekemoer}, {Faber}, {Ferguson}, {Grogin},
  {Kocevski}, {Koo}, {Lai}, {Lotz}, {Lucas}, {McGrath}, {Ogaz}, {Rajan},
  {Riess}, {Rodney}, {Strolger}, {Casertano}, {Castellano}, {Dahlen},
  {Dickinson}, {Dolch}, {Fontana}, {Giavalisco}, {Grazian}, {Guo}, {Hathi},
  {Huang}, {van der Wel}, {Yan}, {Acquaviva}, {Alexander}, {Almaini}, {Ashby},
  {Barden}, {Bell}, {Bournaud}, {Brown}, {Caputi}, {Cassata}, {Challis},
  {Chary}, {Cheung}, {Cirasuolo}, {Conselice}, {Roshan Cooray}, {Croton},
  {Daddi}, {Dav{\'e}}, {de Mello}, {de Ravel}, {Dekel}, {Donley}, {Dunlop},
  {Dutton}, {Elbaz}, {Fazio}, {Filippenko}, {Finkelstein}, {Frazer}, {Gardner},
  {Garnavich}, {Gawiser}, {Gruetzbauch}, {Hartley}, {H{\"a}ussler},
  {Herrington}, {Hopkins}, {Huang}, {Jha}, {Johnson}, {Kartaltepe},
  {Khostovan}, {Kirshner}, {Lani}, {Lee}, {Li}, {Madau}, {McCarthy},
  {McIntosh}, {McLure}, {McPartland}, {Mobasher}, {Moreira}, {Mortlock},
  {Moustakas}, {Mozena}, {Nandra}, {Newman}, {Nielsen}, {Niemi}, {Noeske},
  {Papovich}, {Pentericci}, {Pope}, {Primack}, {Ravindranath}, {Reddy},
  {Renzini}, {Rix}, {Robaina}, {Rosario}, {Rosati}, {Salimbeni}, {Scarlata},
  {Siana}, {Simard}, {Smidt}, {Snyder}, {Somerville}, {Spinrad}, {Straughn},
  {Telford}, {Teplitz}, {Trump}, {Vargas}, {Villforth}, {Wagner}, {Wandro},
  {Wechsler}, {Weiner}, {Wiklind}, {Wild}, {Wilson}, {Wuyts}, \&
  {Yun}}]{koekemoer}
{Koekemoer}, A.~M., {Faber}, S.~M., {Ferguson}, H.~C., {et~al.} 2011, \apjs,
  197, 36

\bibitem[{{Kriek} {et~al.}(2009){Kriek}, {van Dokkum}, {Labb{\'e}}, {Franx},
  {Illingworth}, {Marchesini}, \& {Quadri}}]{kriek}
{Kriek}, M., {van Dokkum}, P.~G., {Labb{\'e}}, I., {et~al.} 2009, \apj, 700,
  221

\bibitem[{{Laidler} {et~al.}(2007){Laidler}, {Papovich}, {Grogin}, {Idzi},
  {Dickinson}, {Ferguson}, {Hilbert}, {Clubb}, \& {Ravindranath}}]{laidler}
{Laidler}, V.~G., {Papovich}, C., {Grogin}, N.~A., {et~al.} 2007, \pasp, 119,
  1325

\bibitem[{{Lee} {et~al.}(2012){Lee}, {Ferguson}, {Wiklind}, {Dahlen},
  {Dickinson}, {Giavalisco}, {Grogin}, {Papovich}, {Messias}, {Guo}, \&
  {Lin}}]{lee2012}
{Lee}, K.-S., {Ferguson}, H.~C., {Wiklind}, T., {et~al.} 2012, \apj, 752, 66

\bibitem[{{Lee} {et~al.}(2014){Lee}, {Ferguson}, {Somerville}, {Giavalisco},
  {Wiklind}, \& {Dahlen}}]{lee2014}
{Lee}, S.-K., {Ferguson}, H.~C., {Somerville}, R.~S., {et~al.} 2014, \apj, 783,
  81

\bibitem[{{Lee} {et~al.}(2009){Lee}, {Idzi}, {Ferguson}, {Somerville},
  {Wiklind}, \& {Giavalisco}}]{lee2009}
{Lee}, S.-K., {Idzi}, R., {Ferguson}, H.~C., {et~al.} 2009, \apjs, 184, 100

\bibitem[{{Longhetti} \& {Saracco}(2009)}]{long2012}
{Longhetti}, M., \& {Saracco}, P. 2009, \mnras, 394, 774

\bibitem[{{Lu} {et~al.}(2014){Lu}, {Mo}, \& {Wechsler}}]{lu}
{Lu}, Y., {Mo}, H.~J., \& {Wechsler}, R.~H. 2014, ArXiv e-prints,
  arXiv:1402.2036

\bibitem[{{Maraston}(2005)}]{maraston2005}
{Maraston}, C. 2005, \mnras, 362, 799

\bibitem[{{Maraston} {et~al.}(2010){Maraston}, {Pforr}, {Renzini}, {Daddi},
  {Dickinson}, {Cimatti}, \& {Tonini}}]{maraston2010}
{Maraston}, C., {Pforr}, J., {Renzini}, A., {et~al.} 2010, \mnras, 407, 830

\bibitem[{{McLure} {et~al.}(2013){McLure}, {Dunlop}, {Bowler}, {Curtis-Lake},
  {Schenker}, {Ellis}, {Robertson}, {Koekemoer}, {Rogers}, {Ono}, {Ouchi},
  {Charlot}, {Wild}, {Stark}, {Furlanetto}, {Cirasuolo}, \& {Targett}}]{mcLure}
{McLure}, R.~J., {Dunlop}, J.~S., {Bowler}, R.~A.~A., {et~al.} 2013, \mnras,
  432, 2696

\bibitem[{{Ouchi} {et~al.}(2009){Ouchi}, {Mobasher}, {Shimasaku}, {Ferguson},
  {Fall}, {Ono}, {Kashikawa}, {Morokuma}, {Nakajima}, {Okamura}, {Dickinson},
  {Giavalisco}, \& {Ohta}}]{Ouchi}
{Ouchi}, M., {Mobasher}, B., {Shimasaku}, K., {et~al.} 2009, \apj, 706, 1136

\bibitem[{{Pforr} {et~al.}(2012){Pforr}, {Maraston}, \& {Tonini}}]{pforr2012}
{Pforr}, J., {Maraston}, C., \& {Tonini}, C. 2012, \mnras, 422, 3285

\bibitem[{{Pforr} {et~al.}(2013){Pforr}, {Maraston}, \& {Tonini}}]{pforr2013}
---. 2013, \mnras, 435, 1389

\bibitem[{{Reddy} {et~al.}(2012){Reddy}, {Pettini}, {Steidel}, {Shapley},
  {Erb}, \& {Law}}]{reddy}
{Reddy}, N.~A., {Pettini}, M., {Steidel}, C.~C., {et~al.} 2012, \apj, 754, 25

\bibitem[{{Santini} {et~al.}(2015){Santini}, {Ferguson}, {Fontana}, {Mobasher},
  {Barro}, {Castellano}, {Finkelstein}, {Grazian}, {Hsu}, {Lee}, {Lee},
  {Pforr}, {Salvato}, {Wiklind}, {Wuyts}, {Almaini}, {Cooper}, {Galametz},
  {Weiner}, {Amorin}, {Boutsia}, {Conselice}, {Dahlen}, {Dickinson},
  {Giavalisco}, {Grogin}, {Guo}, {Hathi}, {Kocevski}, {Koekemoer},
  {Kurczynski}, {Merlin}, {Mortlock}, {Newman}, {Paris}, {Pentericci},
  {Simons}, \& {Willner}}]{santini2015}
{Santini}, P., {Ferguson}, H.~C., {Fontana}, A., {et~al.} 2015, \apj, 801, 97

\bibitem[{{Schenker} {et~al.}(2013{\natexlab{a}}){Schenker}, {Ellis},
  {Konidaris}, \& {Stark}}]{schenker13b}
{Schenker}, M.~A., {Ellis}, R.~S., {Konidaris}, N.~P., \& {Stark}, D.~P.
  2013{\natexlab{a}}, \apj, 777, 67

\bibitem[{{Schenker} {et~al.}(2013{\natexlab{b}}){Schenker}, {Robertson},
  {Ellis}, {Ono}, {McLure}, {Dunlop}, {Koekemoer}, {Bowler}, {Ouchi},
  {Curtis-Lake}, {Rogers}, {Schneider}, {Charlot}, {Stark}, {Furlanetto}, \&
  {Cirasuolo}}]{schenker13a}
{Schenker}, M.~A., {Robertson}, B.~E., {Ellis}, R.~S., {et~al.}
  2013{\natexlab{b}}, \apj, 768, 196

\bibitem[{{Scoville} {et~al.}(2007){Scoville}, {Abraham}, {Aussel}, {Barnes},
  {Benson}, {Blain}, {Calzetti}, {Comastri}, {Capak}, {Carilli}, {Carlstrom},
  {Carollo}, {Colbert}, {Daddi}, {Ellis}, {Elvis}, {Ewald}, {Fall},
  {Franceschini}, {Giavalisco}, {Green}, {Griffiths}, {Guzzo}, {Hasinger},
  {Impey}, {Kneib}, {Koda}, {Koekemoer}, {Lefevre}, {Lilly}, {Liu},
  {McCracken}, {Massey}, {Mellier}, {Miyazaki}, {Mobasher}, {Mould}, {Norman},
  {Refregier}, {Renzini}, {Rhodes}, {Rich}, {Sanders}, {Schiminovich},
  {Schinnerer}, {Scodeggio}, {Sheth}, {Shopbell}, {Taniguchi}, {Tyson}, {Urry},
  {Van Waerbeke}, {Vettolani}, {White}, \& {Yan}}]{scoville}
{Scoville}, N., {Abraham}, R.~G., {Aussel}, H., {et~al.} 2007, \apjs, 172, 38

\bibitem[{{Somerville} {et~al.}(2012){Somerville}, {Gilmore}, {Primack}, \&
  {Dom{\'{\i}}nguez}}]{somerville}
{Somerville}, R.~S., {Gilmore}, R.~C., {Primack}, J.~R., \& {Dom{\'{\i}}nguez},
  A. 2012, \mnras, 423, 1992

\bibitem[{{Somerville} {et~al.}(2008){Somerville}, {Hopkins}, {Cox},
  {Robertson}, \& {Hernquist}}]{somerville08}
{Somerville}, R.~S., {Hopkins}, P.~F., {Cox}, T.~J., {Robertson}, B.~E., \&
  {Hernquist}, L. 2008, \mnras, 391, 481

\bibitem[{{Wiklind} {et~al.}(2008){Wiklind}, {Dickinson}, {Ferguson},
  {Giavalisco}, {Mobasher}, {Grogin}, \& {Panagia}}]{wiklind}
{Wiklind}, T., {Dickinson}, M., {Ferguson}, H.~C., {et~al.} 2008, \apj, 676,
  781

\bibitem[{{Wuyts} {et~al.}(2009){Wuyts}, {Franx}, {Cox}, {F{\"o}rster
  Schreiber}, {Hayward}, {Hernquist}, {Hopkins}, {Labb{\'e}}, {Marchesini},
  {Robertson}, {Toft}, \& {van Dokkum}}]{wuyts2009}
{Wuyts}, S., {Franx}, M., {Cox}, T.~J., {et~al.} 2009, \apj, 700, 799

\end{thebibliography}

\noindent {\bf Appendix I: Definitions of Stellar Mass in Galaxies}
\\
There are three different definitions of the stellar mass commonly used 
in literature.
\\
1. Stellar mass is built up over time by star formation activity in galaxies, 
with stellar mass recycling ignored. If $\phi(t)\ dt$ is the stellar mass generated
in a galaxy between time $t$ and $t+\Delta t$ with a star formation rate $\phi(t)$, 
the stellar mass over the age of
the galaxy is $M_{int}=\int_0^{t_g} \phi(t)\ dt $, where $t_g$ is the current age of  the galaxy. 
In this case the stellar mass depends on the SFH of the galaxy. 
Assuming an exponentially declining SFR, $SFR(t) = SFR_0 e^{-t/\tau}$, the stellar mass is therefore calculated for each object as
$$ M_{int} = \tau SFR_0 (e^{t/\tau} - 1) $$
where $SFR_0$ is the SFR at $t=0$ and $\tau$ is the SFR time scale.  
\\
2. Stellar mass recycling is taken into account using 
the `instantaneous recycling approximation'. In this case a fixed fraction 
of the mass that goes into stars is 
returned to the Inter-Stellar Medium (ISM) immediately in each timestep 
to take into account the stellar mass loss in supernovae explosion or stellar winds. 
At any given time interval, $\Delta t$, the increment of the stellar mass is the star formation rate minus the mass fraction of the short-lived stars and stellar winds multiplied by the time interval. 
Therefore, the stellar mass of a galaxy at the age of $t_g$, assuming Instantaneous Re-cycling Approximation, $M_{ira}$, is estimated as 

$$ M_{ira} = \int_0^{t_g} (\phi(t) - \phi(t)R_{re})\ dt = (1-R_{re}) \int_0^{t_g} \phi(t)\ dt$$

where $\phi(t)$ is the SFR at time $t$ and $R_{re}$ is the recycling fraction, 
which is set to be a constant and depends on the IMF. Most SAMs adopt this prescription. 
\\
3. Stellar mass recycling is treated using detailed predictions of 
stellar population models for how much mass is returned from a stellar 
population of a given age in each timestep (e.g. \citealt{lu}). 
The stellar mass of a galaxy at the current age $t_g$ depends on the star formation history and the mass loss from all stars formed in the past and is estimated as 

$$ M_* = \int_0^{t_g} [\phi(t) - \phi(t) R_{re}(t_g-t)]\ dt$$

where $R_{re}(t_g-t)$ is the recycling fraction at time $t_g$ for the stellar mass formed at time $t$. 
The stellar mass of galaxies strongly depends on their SFH, with the recycled 
mass mainly depending on the IMF and age, with a secondary 
dependence on the metallicity of the stellar population. We show in section 6 that the stellar mass of galaxies weakly depends on the stellar mass loss.\\
\\
\\
\noindent {\bf Appendix II: TEST-1 and TEST-2 Simulated Catalogs}\\
\\

To generate the mock catalog to be as close as possible to the observed  
data, we first predict the observed 1-dimensional distributions of the expectation values for each of the main parameters (redshift, age, stellar mass and extinction). This is done by using a sample of galaxies in GOODS-S with available 
spectroscopic redshifts and by fitting their SEDs to model templates generated from BC03. For each parameter, we generated the 1-dimensional distribution for
the observed parameters and fitted them to analytic functions (i.e. Gaussian). We then drew a mock sample of 1000 galaxies from this distribution 
(with their associated multi-waveband photometry) and only retained those with
(a). ages between 10Myr and the age of the universe at the redshift of the galaxy and (b). with $0 < E(B-V) < 1$. The final sample selected for TEST-1 mock catalog satisfies these criteria, with the total number of galaxies adjusted to be similar to the spectroscopic sample in GOODS-S. This test therefore contains 559 simulated points. The distribution of the main parameters in TEST-1 catalog are 
presented in Figure 19. The redshift distribution here closely resembles the 
observed distribution for galaxies with spectroscopic redshifts in the GOODS-S field.

\begin{figure*}
\epsscale{0.6}
\plotone{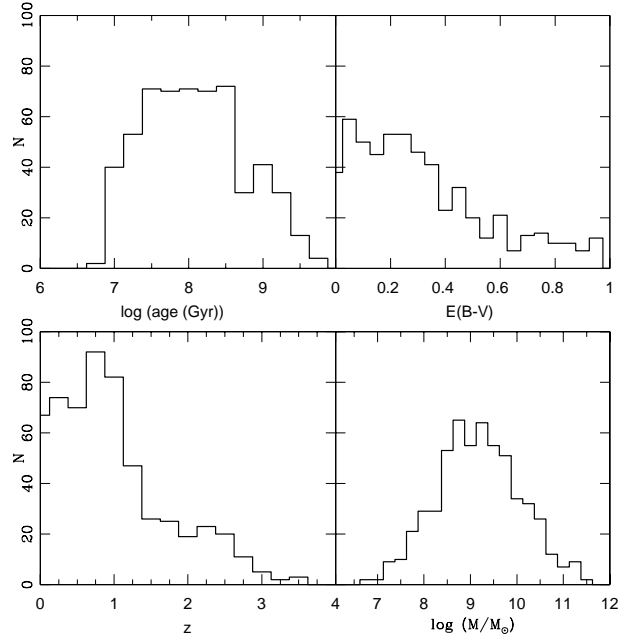}
\figcaption[fig.19.test1.hist.eps]{Distribution of physical parameters (redshift, stellar mass, age and extinction) for the TEST-1 mock catalog. The redshift distribution is taken to be the same as the observed distribution in the spectroscopic sample used for training the mock catalog.
\label{fig1}}
\end{figure*}

\begin{figure*}
\epsscale{0.6}
\plotone{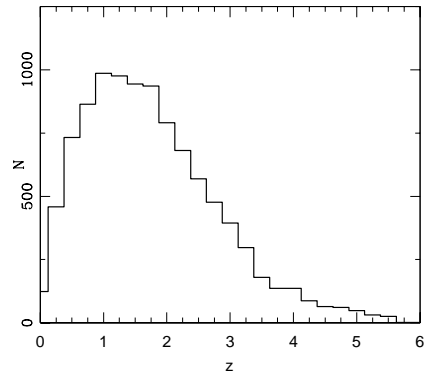}
\figcaption[fig.A2.test2A.hist.z.eps]{Input redshift distribution for TEST-2A mock catalog. 
\label{fig1}}
\end{figure*}

While a larger mock catalog (in terms of the number of galaxies generated) 
would reduce the poisson noise in the analysis, we aimed for a catalog which contains similar number of galaxies as those in the observed 
spectroscopic catalog. This allows a more realistic estimate of the
stellar mass calibration errors when applying the results from the mock data to the real data. 

For TEST-2, light cones were used to directly replicate CANDELS field geometry. The $N(z)$ for this model is generated to closely resemble the photometric redshift distribution for the GOODS-S field, as shown in Figure 20. In this test, stellar mass re-cycling is treated using the ``instantaneous''
recycling approximation in which a fixed fraction of mass that goes into stars is immediately returned into the ISM during each time step (\citealt{lu}).

%\noindent {\bf Acknowledgement} This study is based, in part, on observations from the Hubble Space Telescope under program number HST-GO-12060 provided by NASA through a grant from the Space Telescope Science Institute, which is operated by the Assiciation of Universities for Research in Astronomy, incorporated under NASA contract NAS5-26555.  

\end{document}